\documentclass[11pt]{article}
\usepackage{epsf}
\begin{document}

\renewcommand{\arraystretch}{1.3}
\renewcommand{\textfraction}{0.05}
\renewcommand{\topfraction}{0.9}
\renewcommand{\bottomfraction}{0.9}

\def\la{\;\raise0.3ex\hbox{$<$\kern-0.75em\raise-1.1ex\hbox{$\sim$}}\;}
\def\ga{\;\raise0.3ex\hbox{$>$\kern-0.75em\raise-1.1ex\hbox{$\sim$}}\;}
\def\lr{\;\raise0.3ex\hbox{$\rightarrow$\kern-1.0em\raise-1.1ex\hbox{$\leftarrow$}}\;}
\def\pFn{p_{\raise-0.3ex\hbox{{\scriptsize F$\!$\raise-0.03ex\hbox{\it n}}}} }  
\def\pFp{p_{\raise-0.3ex\hbox{{\scriptsize F$\!$\raise-0.03ex\hbox{\it p}}}} }  
\def\pFe{p_{\raise-0.3ex\hbox{{\scriptsize F$\!$\raise-0.03ex\hbox{\it e}}}} }  
%
\def\m@th{\mathsurround=0pt }     
\def\eqalign#1{\null\,\vcenter{\openup1\jot \m@th
   \ialign{\strut$\displaystyle{##}$&$\displaystyle{{}##}$\hfil
   \crcr#1\crcr}}\,}
\newcommand{\om}{\mbox{$\omega$}}              
\newcommand{\Om}{\mbox{$\Omega$}}              
\newcommand{\Th}{\mbox{$\Theta$}}              
\newcommand{\ph}{\mbox{$\varphi$}}             
\newcommand{\del}{\mbox{$\delta$}}             
\newcommand{\Del}{\mbox{$\Delta$}}             
\newcommand{\lam}{\mbox{$\lambda$}}            
\newcommand{\Lam}{\mbox{$\Lambda$}}            
\newcommand{\ep}{\mbox{$\varepsilon$}}         
\newcommand{\ka}{\mbox{$\kappa$}}              
\newcommand{\dd}{\mbox{d}}                     
\newcommand{\vect}[1]{\bf #1}                
\newcommand{\vtr}[1]{\mbox{\boldmath $#1$}}  
\newcommand{\vF}{\mbox{$v_{\mbox{\raisebox{-0.3ex}{\scriptsize F}}}$}}  
\newcommand{\pF}{\mbox{$p_{\mbox{\raisebox{-0.3ex}{\scriptsize F}}}$}}  
\newcommand{\kF}{\mbox{$k_{\rm F}$}}           
\newcommand{\kTF}{\mbox{$k_{\rm TF}$}}         
\newcommand{\kB}{\mbox{$k_{\rm B}$}}           
\newcommand{\tn}{\mbox{$T_{{\rm c}n}$}}        
\newcommand{\tp}{\mbox{$T_{{\rm c}p}$}}        
\newcommand{\te}{\mbox{$T_{eff}$}}             
\newcommand{\dash}{\mbox{--}}                  
\newcommand{\ex}{\mbox{\rm e}}                 
\newcommand{\rate}{\mbox{$\frac{ \mbox{erg}}{\mbox{cm$^3 \cdot $s}}$}}
\newcommand{\A}{\raisebox{-0.2ex}{\scriptsize{\it$\!$A}}}
\newcommand{\B}{\raisebox{-0.2ex}{\scriptsize{\it$\!$B}}}
\newcommand{\C}{\raisebox{-0.2ex}{\scriptsize{\it$\!$C}}}
\newcommand{\va}{v_{\A}}  
\newcommand{\vb}{v_{\B}}  
\newcommand{\vc}{v_{\C}}  
\newcommand{\RaC}{R_{\A}} 
\newcommand{\RbC}{R_{\B}} 
\newcommand{\RcC}{R_{\C}} 
\newcommand{\mur}{\raisebox{0.2ex}{\mbox{\scriptsize (M)}}} 
\newcommand{\Mn}{\raisebox{0.2ex}{\mbox{\scriptsize (M{\it n\/})}}}        %
\newcommand{\Mp}{\raisebox{0.2ex}{\mbox{\scriptsize (M{\it p\/})}}}        %
\newcommand{\MN}{\raisebox{0.2ex}{\mbox{\scriptsize (M{\it N\/})}}}        %
\newcommand{\RnpA}{\mbox{$R^{\Mn}_{p{\A}}$}}                               %
\newcommand{\RnnB}{\mbox{$R^{\Mn}_{n{\B}}$}}       
\newcommand{\RppA}{\mbox{$R^{\Mp}_{p{\A}}$}}                               %
\newcommand{\RpnB}{\mbox{$R^{\Mp}_{n{\B}}$}} 
\newcommand{\dur}{\raisebox{0.2ex}{\mbox{\scriptsize (D)}}  } 
\newcommand{\Ra}{R^{\dur}_{\A} }                                        %
\newcommand{\Rb}{R^{\dur}_{\B} }                   
\newcommand{\Rc}{R^{\dur}_{\C} } 
\newcommand{\lep}{\raisebox{-0.3ex}{$\mbox{\scriptsize{\it l}}$}}
\newcommand{\rec}{\raisebox{0.2ex}{\mbox{\scriptsize (CP)}}  } 
\newcommand{\rrr}{\rule{0cm}{0.4cm}}
\newcommand{\hh}{\rule{0.5cm}{0cm}}
\newcommand{\hb}{\rule{0.4cm}{0cm}}

\thispagestyle{empty}

\begin{center}
{\bf \Large Cooling Neutron Stars and Superfluidity in Their Interiors}\\[1ex]
D.G.\ Yakovlev, K.P.\ Levenfish, Yu.A.\ Shibanov \\[0.4ex]
{\small \it Ioffe Physical-Technical Institute}\\[0.4ex]
{\small \it Politekhnicheskaya 26,194021 St.-Petersburg, Russia}\\[0.4ex]
{\small e-mails: yak@astro.ioffe.rssi.ru ksen@astro.ioffe.rssi.ru
shib@stella.ioffe.rssi.ru}
\end{center}
\medskip

\begin{abstract}
We study the heat capacity and neutrino emission reactions
(direct and modified Urca processes, nucleon-nucleon bremsstrahlung,
Cooper pairing of nucleons) in matter of supranuclear density
of the neutron star cores with superfluid neutrons and
protons. Various superfluidity types are
analysed (singlet-state pairing and two types of triplet-state
pairing, without and with nodes of the gap at a
nucleon Fermi surface). The results are used for cooling simulations
of isolated neutron stars. Both, the standard cooling
and the cooling enhanced by the direct Urca process,
are strongly affected by nucleon superfluidity.
Comparison of cooling theory of isolated neutron stars
with observations of their thermal radiation may give
stringent constraints on the critical temperatures of the
neutron and proton superfluidities in the neutron star cores.
\end{abstract}

\tableofcontents
%
%


  \section{Introduction}
\noindent
Neutron stars (NSs) are unique astrophysical objects.
First, their observational manifestations are numerous
(radio and X-ray pulsars, X-ray bursters, X-ray transients, etc.;
Shapiro and Teukolsky, 1983).
Second, their structure and evolution are determined by
properties of matter under extreme conditions
which cannot be reproduced in laboratory
(supranuclear densities, superstrong magnetic fields,
superfluidity of baryon component of superdense matter, etc.).

It is widely accepted
(e.g.,
Imshennik and Nadyozhin, 1982;
Shapiro and Teukolsky, 1983)
that NSs are born at the final stage of evolution of
normal stars of mass
$M \ga 8 M_\odot$
in gravitational collapse of their cores
($M_\odot$ is the solar mass).
During collapse, matter of central layers
is compressed to nuclear densities and is enriched by neutrons.
As a result, a compact NS is created of mass
$M \sim M_\odot$ and radius ${\cal R} \sim 10$~km.
Its core consists mainly of neutrons (with some
admixture of protons, electrons, and -- possibly -- hyperons and other
particles). A NS is born hot, with the internal temperature about
$10^{11}$~K,
but cools down rapidly due to powerful neutrino emission
from the internal layers.

NSs possess strong gravitational fields; gravitational
acceleration on their surfaces is as large as
$\sim (2$--$3) \times 10^{14}$ cm~s$^{-2}$.
The stellar radius ${\cal R}$ is only 2--3 times
larger than the gravitational radius
${\cal R}_{g}=2GM/c^2$, where $G$ is the gravitational constant and
$c$ is the light speed. Therefore, the effects of
General Relativity are important in NS life.

In this review, we restrict ourselves by consideration
of structure and thermal evolution of isolated NSs of age
$t\la 10^6$ yr. Evolution of NSs in close binaries is
more complicated; it is described, for instance, by
Lewin et al.\ (1995).

Superfluidity and superconductivity of nucleons
in NS interiors
play important role in evolution of isolated NSs.
For instance,
sudden changes of spin periods (glitches) demonstrated by some
pulsars are commonly explained by interaction
of normal and superfluid components of matter in the
stellar crusts
(Pines, 1991).
Superfluidity affects the heat capacity and neutrino luminosity
of the star, and therefore its cooling. The effect of superfluidity
on the cooling is the main subject of the present review.

At a certain cooling stage ($10^2$--$10^5$~yr)
the NS internal temperature can strongly depend
on critical temperatures of transitions of nucleons into
superfluid state. This feature was first mentioned by
Page and Applegate (1992)
who called NSs ``thermometers" for measuring
the nucleon critical temperatures in asymmetric nuclear matter.
Microscopic calculations of such temperatures are complicated
owing to the absence of explicit manybody relativistic
quantum theory which would describe adequately strong interactions
of particles of various species. However, the critical temperatures
can be studied by the astrophysical method, by confronting
cooling simulations with observations of thermal radiation
of isolated NSs. This work is not finished yet but certain results
are already obtained. The aim of this review is to describe
the above method and main results obtained to-date
(by January 1999).


\section{Overall properties of neutron stars}
  \subsection{Structure}

If standard measurements are used,
neutron stars (NSs) are hot objects.
In several hundred thousand years after their birth their internal
temperature exceeds $10^7$~K. However, the equation of state
of internal NS layers is practically independent of temperature
since the main contribution into pressure comes from strongly
degenerate fermions of high energy
(Shapiro and Teukolsky, 1983).

\begin{figure}[t]
\begin{center}
\leavevmode
\epsfysize=8.5cm
\epsfbox[130 445 441 741]{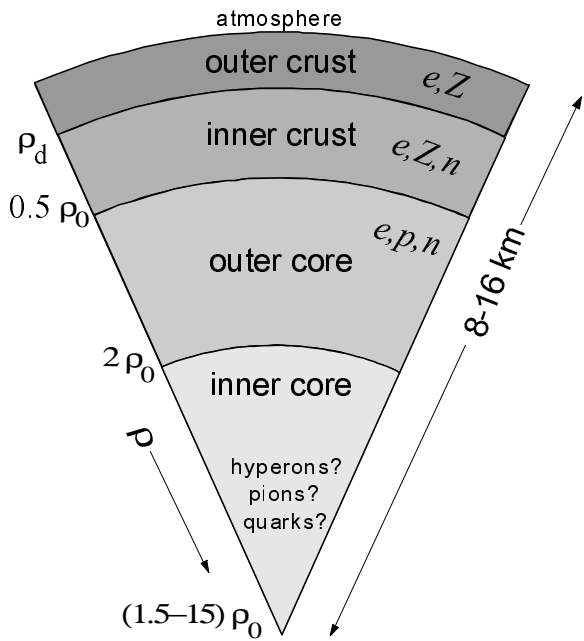}
\end{center}
\caption[]{\footnotesize Schematic cross section of a
$1.4 \, M_\odot$ neutron star.
Stellar parameters depend strongly on the
equation of state of internal layers.
}
\label{fig:sector}
\end{figure}

A NS can be subdivided into the {\it atmosphere}
and four internal regions: the {\it outer crust},
the {\it inner crust}, the {\it outer core}, and the {\it inner core}
as shown in Fig.\ \ref{fig:sector}.

{\bf The atmosphere} is a thin plasma layer, where the spectrum
of thermal electromagnetic NS radiation is formed.
In principle, this radiation contains valuable information
on stellar parameters (on temperature, gravitational acceleration
and chemical composition of the surface, on magnetic field, etc.,
see Sect.\ 8.1) and, as a result, on the internal structure.
Geometrical depth of the atmosphere varies from some ten centimeters
in a hot NS down to some millimeters in a cold one. Very cold NSs
may have no atmosphere at all but a solid surface.

NS atmospheres have been studied theoretically by many
authors (see, e.g., review articles by Pavlov et al., 1995;
Pavlov and Zavlin, 1998, and references therein).
Construction of the atmosphere models, especially for the cold NSs
(with the surface temperature $T_s \la 10^6$~K)
with strong magnetic fields 10$^{11}$--10$^{14}$~G,
is far from being complete owing to complexity of calculations
of the equation of state and spectral opacity of
the atmospheric plasma (Sect.\ 8.1.2).

{\bf The outer core (outer envelope)} extends from the
atmosphere bottom to the layer of the density
$\rho = \rho_d \approx
4.3 \times 10^{11}$ g~cm$^{-3}$ and has a depth about
some hundred meters
(Shapiro and Teukolsky 1983). Its matter consists of ions and
electrons. A very thin (no more than several meters in a hot NS)
surface layer contains a non-degenerate electron gas.
In deeper layers, electrons constitute a strongly degenerate,
almost ideal gas, which becomes relativistic at
$\rho \gg 10^6$ g~cm$^{-3}$.
For $\rho \ga 10^4$ g~cm$^{-3}$, atoms are fully ionized by
the electron pressure, being actually bare atomic nuclei.
The electron Fermi energy grows up with increasing $\rho$,
the nuclei suffer beta-decays and are enriched by neutrons
(see, e.g.,
Haensel and Pichon, 1994). At the base of the outer core
($\rho = \rho_d$) neutrons start to drip from nuclei
producing a free--neutron gas.

The depth of {\bf the inner crust (inner envelope)}
may be as large as several kilometers. The density
$\rho$ in the inner crust varies from $\rho_d$ at the upper boundary
to $\sim 0.5 \rho_0$ at the base. Here,
$\rho_0=2.8 \times 10^{14}$ g~cm$^{-3}$ is the saturation nuclear
matter density. Matter of the inner crust consists of
electrons, free neutrons and neutron--rich atomic nuclei
(Negele and Vautherin, 1973;
Pethick and Ravenhall, 1995).
The fraction of free neutrons increases with growing $\rho$.
At the core bottom (in the density range from
$10^{14}$ to about $1.5 \times 10^{14}$ g~cm$^{-3}$)
the nuclei may form clusters and have non-spherical shapes
(Lorenz et al., 1993;
Pethick and Ravenhall, 1995). The nuclei disappear
at the crust--core interface. Neutrons in the inner crust may
be in a superfluid state (Sect.\ 3.1).

{\bf The outer core} occupies the density range
$0.5 \rho_0 \la \rho \la 2 \rho_0$ and can be several kilometers in depth.
Its matter consists of neutrons $n$ with some
(several per cent by particle number)
admixture of protons $p$ and electrons $e$
(the so called {\it standard} nuclear composition).
Composition of matter is determined by the conditions
of electric neutrality and beta-equilibrium with respect to the
reactions
$ n \to p + e + \bar{\nu}_e$,
$\; p + e \to n + \nu_e $, where $\nu_e$ and $\bar{\nu}_e$ stand for
electron neutrino and antineutrino, respectively.
The electric neutrality requires equality of the
electron and proton number densities:
$n_p = n_e$.
The beta equilibrium establishes a relationship
between chemical potentials of the particles:
$ \mu_n = \mu_p + \mu_e $.
The neutrino chemical potential is omitted here
since immediately after the birth a
NS becomes fully transparent for neutrinos
which freely escape the star
(Friman and Maxwell, 1979;
Shapiro and Teukolsky, 1983).

All $npe$-plasma components are strongly degenerate.
The electrons form almost ideal relativistic Fermi gas.
The neutrons and protons, which interact via nuclear forces,
constitute a strongly non-ideal non-relativistic
Fermi-liquid. It can be described using
the formalism of quasi-particles
(Botermans and Malfliet, 1990; Baym and Pethick, 1991).
The Fermi momentum $p_{{\rm F}\!j}$
of particle species $j$ is related to their number density $n_j$ as
$p_{{\rm F}\!j} = \hbar \, (3\pi^2 \, n_j)^{1/3}$.
Therefore, the electric neutrality implies the equality
of the Fermi momenta of protons and electrons:
$\pFp=\pFe$.

Calculations show (e.g.,
Shapiro and Teukolsky, 1983) that
for $\rho \sim \rho_0$ the chemical potentials of neutrons and
electrons are
$\mu_n \approx \mu_e \approx 60$--$100$~MeV,
while the chemical potential of protons is much lower,
$\mu_p \approx 3$--$6$~MeV. With increasing density,
the Fermi energies of particles grow up, so that new particles
can be created. First of all, these are muons which,
like electrons, constitute a strongly degenerate, almost ideal
gas. The appearance of particles in the inner NS core
will be discussed below.

Let us emphasize that properties of the NS crust
($\rho \la 0.5 \rho_0$) are described by sufficiently
reliable microscopic theories. The situation with
matter of supranuclear density, $\rho \ga \rho_0$, is quite different.
Laboratory data on properties of this matter are
incomplete, and reliability of the theories decreases
with growing $\rho$. Exact selfconsistent quantum theory
of matter of supranuclear density has not been constructed yet.
Many theoretical equations of state have been proposed which
can be subdivided conventionally into the
{\it soft}, {\it moderate} and {\it stiff} ones with respect
to compressibility of matter.
These equations of state are considerably
different for $\rho > \rho_0$ . For instance, at
$\rho \approx 4 \rho_0$
a soft equation of state proposed by
Pandharipande (1971)
(based on the Reid, 1968, nucleon--nucleon soft--core
potential)
gives the pressure which is about one order of magnitude lower
than that given by a stiff equation of state constructed by
Pandharipande and Smith (1975)
using the mean field approach.

Almost all microscopic theories predict
the appearance of neutron and proton superfluidity
in the outer NS core (Sect.\ 3.1). The proton superfluidity is
accompanied by superconductivity (most likely, of second type)
and affects evolution of the internal magnetic fields
(see, e.g.,
Ruderman, 1991).

The outer core of a low-massive NS extends to the stellar center.
More massive NSs possess also {\bf the inner core}.
Its radius may reach several kilometers,
and its central density may be as high as $(10$--$15) \rho_0$.
Composition and equation of state of the inner core
are poorly known. It should be emphasize that
these properties are most crucial for determining
structure and evolution of massive NSs, and they constitute
the main NS ``mystery". Several hypotheses have been put
forward and are being discussed
in the literature; it is impossible to reject any of them
at present:

(1) Hyperonization of matter ---  appearance of
$\Sigma$- and $\Lambda$-hyperons
(see, e.g.,
Shapiro and Teukolsky, 1983). The fractions of
$p$ and $e$ may be so high that the powerful direct Urca
process of neutrino emission becomes allowed
(Lattimer et al., 1991) as well as similar reactions
involving hyperons
(Prakash et al., 1992). In these cases, the neutrino luminosity
of the star is enhanced by 5--6 orders of magnitude
(see, e.g.,
Pethick, 1992) as compared to the standard neutrino
luminosity of the outer core produced mainly by
the modified Urca process. This accelerates considerably NS cooling
(Sects.\ 2.2 and 7).

(2) The second hypothesis,
proposed in different forms by
Bahcall and Wolf (1965a, b), Migdal (1971), Sawyer (1972),
and Scalapino (1972), assumes the appearance of
pion condensation. The condensates soften the equation of state
and enhance the neutrino luminosity by allowing
the reactions of the direct Urca type
(Bahcall and Wolf, 1965a, b;
Maxwell et al., 1977;
Muto and Tatsumi, 1988).
However, many modern microscopic
theories of dense matter predict weakly polarized pionic degrees
of freedom which are not in favor of pion condensation
(Pethick, 1992).

(3) The third hypothesis predicts a
phase transition to the strange quark matter composed
of almost free $u$, $d$ and $s$ quarks with small admixture of
electrons (see, e.g.,
Haensel, 1987;
Schaab et al., 1996). In these cases the neutrino
luminosity is thought to be considerably higher than the standard
luminosity due to switching on the direct Urca processes
involving quarks
(Iwamoto, 1980;
Burrows, 1980;
Pethick, 1992).
However, in some models (e.g., Schaab et al., 1997a)
the presence of quarks does not enhance the neutrino
luminosity.

(4) Following
Kaplan and Nelson (1986), Nelson and Kaplan (1987)
and Brown et al.\ (1988), several authors
considered the hypothesis on kaon condensation in dense matter.
Kaon condensates may also enhance the neutrino
luminosity by several orders of magnitude
(Pethick, 1992).
A critical analysis of contemporary theories of kaon condensation
was given by Pandharipande et al.\ (1995).

For any equation of state, one can build a set of NS models
with different central densities $\rho_c$.
This can be done by solving numerically the equation
of hydrostatic equilibrium with account for the effects
of General Relativity (the Oppenheimer--Volkoff equation, see, e.g.,
Shapiro and Teukolsky, 1983). With increasing
$\rho_c$, the stellar mass $M$ usually grows up, and the radius ${\cal R}$
decreases (a star becomes more compact).
As a rule, the mass $M$ reaches maximum at some $\rho_c$, which
corresponds to the most compact stable stellar configuration.
The mass $M_c$ of this configuration is the
{\it maximum} NS mass, for a given equation of state.
Stellar models with higher central density are usually
unstable (with respect to collapsing into a black hole)
and cannot exist for a long time. Using a set of NS models, one
can construct a ``mass-radius" diagram which depends strongly
on the equation of state in the NS core. The softer the equation of state,
the more compact NS, and the lower the maximum mass.
For instance, this mass ranges from $\sim$1.4 to $\sim$1.6 $M_\odot$
for different soft equations of state; from
$\sim$1.6 to $\sim$1.8 $M_\odot$ for moderate equations of state,
and from $\sim$1.8 to $\sim$3 $M_\odot$ for stiff equations of state.
The problem of maximum NS mass is crucial
for identification of black holes in binary systems
(Cherepashchuk, 1996).

The equation of state in the NS cores can be studied
by confronting theory with observations in different ways.
The majority of methods are based on determination (constraint)
of NS mass and/or radius and comparison
of these observational results with the
mass-radius diagrams for different
equations of state (see, e.g.,
Shapiro and Teukolsky, 1983).
Unfortunately, no absolutely decisive arguments
have been given so far in favor
of the stiff, moderate or soft equations of state.
One can definitely rule out
only the ultra-soft equations of state which give the
maximum NS mass lower than $1.44 \, M_\odot$,
the mass of the Hulse--Taylor pulsar (PSR 1913+16),
which is the most massive known star
in close double neutron--star binaries
(where NS masses are determined
very accurately). In this review, we will discuss another
method to explore properties of superdense matter ---
by NS cooling.

  \subsection{Neutrino emission}

In $\sim 20$~s after its birth, a NS becomes transparent
for neutrinos
(Shapiro and Teukolsky, 1983)
generated in its interiors.
Leaving the star, neutrinos carry away energy and cool the star.
Therefore, a study of neutrino reactions is important
for the NS cooling theories. The most powerful neutrino
emission is produced in the NS core. Thus we will not
discuss the neutrino reactions in the NS crust. Their detailed
description can be found, for instance, in the review article
by
Imshennik and Nadyozhin (1988)
as well as in the articles by
Soyer and Brown (1979),
Itoh et al.\ (1984, 1996),
Pethick and Thorsson (1994),
and Kaminker et al.\ (1997, 1999a).
Typical neutrino
energies are much higher than their assumed rest mass energies;
therefore, neutrinos can be considered usually as massless particles.

There are many neutrino reactions in the NS core
(see, e.g.,
Shapiro and Teukolsky, 1983;
Pethick, 1992).
Their efficiency is determined by the equation of state
of matter which is poorly known (see above).
For certainty, we will not consider exotic models
of strange matter, pion or kaon condensates, but
outline the main neutrino reactions in
$npe$ matter. A more detailed analysis of these reactions,
particularly in the presence of nucleon superfluidity,
is given in Sects.\ 4--6. \\[0.5ex]

{\bf Modified Urca process.}
Starting from the classical article by
Chiu and Salpeter (1964)
this processes has been treated as the main
neutrino generation process for the {\it standard}
NS cooling. Cooling is called {\it standard}
provided very powerful neutrino reactions are absent,
like the direct Urca process or analogous processes in
exotic models of dense matter.

The modified Urca reaction is similar to the familiar
reactions of beta decay and beta capture, but involves
additional spectacular nucleon. In the
$npe$-matter, the reaction can go through two channels
\begin{eqnarray}
    n + n &\rightarrow& n + p + e + \bar{\nu}_e  \, , \;\;\;\;
    n + p + e \rightarrow n + n  + \nu_e \; ;
\label{eq:Murca_N} \\
    n + p &\rightarrow& p + p + e + \bar{\nu}_e  \, , \;\;\;\;
    p + p + e \rightarrow n + p  + \nu_e \; ,
\label{eq:Murca_P}
\end{eqnarray}
which we define as the {\it neutron} and {\it proton}
branches of the modified Urca process, respectively.

The spectacular nucleon is needed to conserve
momentum of reacting particles.
The familiar beta-decay and beta-capture reactions
(called the direct Urca process discussed below)
are forbidden (strongly suppressed) in the outer NS core.
The suppression is due to relatively low
fractions of electrons and protons, owing to which
the Fermi momenta of $n$, $p$, and $e$
do not obey the ``triangle rule",
$\pFn \leq \pFe + \pFp$, required for momentum conservation
(the momenta of emitting neutrinos $p_\nu \sim k_{\rm B}T/c$
can be neglected: they are determined by temperature of matter,
$T$, and are much smaller than the momenta of other particles;
here $k_{\rm B}$ is the Boltzmann constant).
The neutrino energy emission rate
(emissivity) $Q^{\Mn}$ in the neutron
branch (\ref{eq:Murca_N}) of the modified Urca process
for $\rho \sim \rho_0$ can be estimated as
(Friman and Maxwell, 1979)
$ Q^{\Mn} \sim 10^{22} \: T^8_9\; $ erg$ \; $cm$^{-3} \;$s$^{-1}$,
where $T_9 = T/10^9$ K. The proton branch is nearly as efficient:
$Q^{\Mp} \sim Q^{\Mn}$
(Sect.\ 5.1). \\[0.5ex]

{\bf Neutrino bremsstrahlung due to nucleon-nucleon scattering.}
The standard neutrino luminosity is determined also
by the processes of neutrino bremsstrahlung radiation in
nucleon--nucleon collisions
\begin{equation}
   n+n   \rightarrow   n+n + \nu + \bar{\nu} \, , \;\;\;\;\;
   n+p   \rightarrow   n+p + \nu + \bar{\nu} \, , \;\;\;\;\;
   p+p   \rightarrow   p+p + \nu + \bar{\nu},
\label{eq:Brems}
\end{equation}
allowed in the entire NS core.
Here $\nu$ means neutrino of any flavor
($\nu_e$, $\nu_{\mu}$, or $\nu_{\tau}$). In normal (non-superfluid) matter
the bremsstrahlung processes
(Friman and Maxwell, 1979)
are weaker than the modified Urca process:
$Q^{(NN)} \sim (10^{19}$--$10^{20}) \; T_9^8\,$ erg$\;$cm$^{-3}\;$s$^{-1}$.
However, they can be important in
superfluid matter (Sect.\ 5). \\[0.5ex]

{\bf Direct Urca process.}
As mentioned above, a sequence of beta-decay and beta-capture
reactions,
\begin{equation}
    n  \rightarrow p + e + \bar{\nu}_e  \, , \;\;\;\;
    p + e \rightarrow n  + \nu_e \, ,
\label{eq:Durca_Nucleon}
\end{equation}
called the {\it direct Urca process},
is forbidden in the outer NS core due to insufficiently high
fractions of $e$ and $p$. It has been thought for a long time
that the process is forbidden also in the inner NS core.

However, the process (\ref{eq:Durca_Nucleon})
becomes allowed
(Lattimer et al., 1991),
if the fraction of protons
(among all baryons) $x_p = n_p/n_b$ exceeds some critical value
$x_p=x_{\rm c}$. In
$npe$-matter, this happens for
$\pFn \leq 2 \pFp$,
which gives $x_{\rm c} = 1/9= 0.111$. If muons are present
for the same number density of baryons $n_b$, the proton fraction
appears to be slightly higher than in the $npe$ matter, and
the electron fraction slightly lower. In this case,
$x_{\rm c}$ is higher and reaches
0.148 for relativistic muons
(Lattimer et al., 1991).

In the simplest model of superdense matter as a gas of
non-interacting Fermi particles
(Shapiro and Teukolsky, 1983)
the proton fraction is not high:
$x_p < x_{\rm c}$ for any density. However, this may be not so
for realistic equations of state. This circumstance was first
outlined by Boguta (1981)
but his article has been unnoticed for a long time.

It was the paper by Lattimer et al.\ (1991)
which initiated wide discussion of the direct Urca process.
The authors showed that, for many realistic models of matter,
$x_p$ exceeded slightly
$x_{\rm c}$ at densities several times higher than the
standard nuclear matter density. Therefore, the nucleon
direct Urca process can be allowed in the inner cores of
rather massive NSs. Moreover, practically all equations
of state in the inner stellar core, which predict appearance of
hyperons, open direct Urca reactions involving hyperons
(Prakash et al., 1992).

According to Lattimer et al.\ (1991)
the neutrino emissivity in the reaction
(\ref{eq:Durca_Nucleon}) is
$Q^{\dur} \sim  10^{27}\; T^6_9 \,$ erg$ \;$cm$^{-3} \;$s$^{-1}$.
For $T \sim 10^9$~K, the direct Urca process is about 5--6 orders
of magnitude more efficient than the modified Urca.
Therefore, sufficiently massive NSs suffer
{\it the enhanced cooling} (Sect.\ 7).

In a series of papers initiated by Voskresensky
and Senatorov (1984, 1986)
the neutrino reactions
of the direct Urca type have been studied
for the models of dense matter with
highly polarized pion degrees of freedom.
Pion condensation in this matter takes place at
$\rho \sim \rho_0$ and is very efficient. Even for lower
$\rho$, before the condensation occurs,
the neutrino emissivity appears to be much higher
than the standard one due to polarizability of pion vacuum.
Cooling of NSs with this equation of state has been simulated
recently by Schaab et al.\ (1997b).
We will not consider these models. \\[0.5ex]

{\bf Neutrino emission due to Cooper pairing of nucleons.}
Onset of nucleon superfluidity switches on a new neutrino
generation mechanism concerned with creation of Cooper pairs.
The process had been proposed and calculated in the pioneering
article by Flowers et al.\ (1976)
and rediscovered independently by
Voskresensky and Senatorov (1986, 1987) ten years later.
Until recently, the process has been forgotten due to unknown reasons
and has not been included into NS cooling simulations.
It was Page (1998a) who ``remembered"  the process and introduced
it into the cooling theory.
Cooling simulations including this process have  been
performed recently by a number of authors, in particular, by
Schaab et al.\ (1997b), Levenfish et al.\ (1998, 1999) and
Yakovlev et al.\ (1998, 1999).
The process in question represents actually
(Yakovlev et al., 1999) neutrino pair emission (any neutrino flavor)
by a nucleon $N$ (neutron or proton) whose dispersion relation
contains an energy gap:
\begin{equation}
    N  \to N + \nu + \bar{\nu}  \, .
\label{eq:Recomb}
\end{equation}
The reaction cannot occur without superfluidity:
the emission of a neutrino pair by a free nucleon is forbidden by
energy--momentum conservation.
According to Yakovlev et al.\ (1999)
(see also Sect.\ 6.1),
the neutrino emissivity due to pairing of neutrons is
$Q^{\rec} \sim 10^{21} T^7_9 \, F(\tau)$
erg$ \;$cm$^{-3} \;$s$^{-1}$, where
$\tau=T/T_c$,
$T_c$ is the critical temperature of superfluidity onset,
and $F(\tau)$ is a function, which has maximum
$F \sim 1$ at $\tau \sim 0.4$. The main neutrino energy release takes place
in the temperature range
$0.2 \, T_c \la T \la T_c$. The efficiency of the process
in this range can be compared to or even larger than
(Sect.\ 6.2) the efficiency of the modified or even the direct
Urca processes suppressed partly by the superfluidity.
This determines the importance of ``Cooper" neutrinos for
NS cooling. Neutrino emission due to pairing of protons
appears to be much weaker owing to the smallness of the vector constant
of weak neutral current involving protons.

  \subsection{Cooling}

Initially a NS cools mainly via the neutrino emission
from its core. However, direct detection of the neutrino flux
is possible only during a NS birth in supernova explosion.
So far it was done only
once, from the Supernova 1987A in the Large
Magellanic Cloud (see, e.g.,
Imshennik and Nadyozhin, 1988).
A powerful neutrino outburst in a supernova explosion
lasts for about 20 s. Afterwards the neutrino flux decreases
rapidly in time.

Cooling of a NS is accompanied by the loss of its
thermal energy which is mainly stored in the stellar core.
The energy is carried away through two channels: first,
by neutrino emission from the entire stellar body
(mostly from the core), where the most powerful neutrino reactions
take place, and second, by heat conduction through the internal
stellar layers towards the surface, and further, by thermal emission of
photons from the surface. The neutrino cooling dominates at the
initial cooling stage (see below), while the photon cooling
dominates later, when the neutrino luminosity becomes weak
(it fades faster than the photon luminosity with decreasing temperature).

From mathematical point of view, cooling simulation reduces
to solving the heat diffusion equation within the star
(Thorne, 1977)
with account for the volume (neutrino emission) and surface
(photon emission) energy sinks. As a rule, one considers
one-dimensional diffusion along radial coordinate in a
spherically--symmetric star. In about
$t \ga 10^2$--$10^3$ yr, owing to the high thermal conductivity
of the internal layers, a wide, almost isothermal region
is formed within the entire core and
the main fraction of the crust. In this case, cooling simulations are
considerably simplified being reduced to solving the global
thermal--balance equation: the loss of the thermal energy
(determined by the total heat capacity) is governed
by the neutrino and photon luminosities of the star.
This approach is described and used in
Sect.\ 7.2. Accordingly the main elements of the cooling theory
are: (1) NS heat capacity, (2) neutrino luminosity,
(3) dependence of the photon luminosity on
internal temperature (determined by the thermal conductivity of
the outermost stellar layers). The first two elements
are discussed in detail in this review.

Character of cooling depends on many parameters:
equation of state of internal layers, superfluidity
of nucleons in the stellar core, NS mass, magnetic field,
chemical composition of the surface layers, etc.
Confronting the cooling theory with observations enables one,
in principle, to constrain these parameters. In this review
we discuss the constraints on
the critical temperatures of neutron and proton superfluidities
in the NS cores.

\begin{figure}[t]
\begin{center}
\leavevmode
\epsfysize=8.5cm
\epsfbox[140 20 470 405]{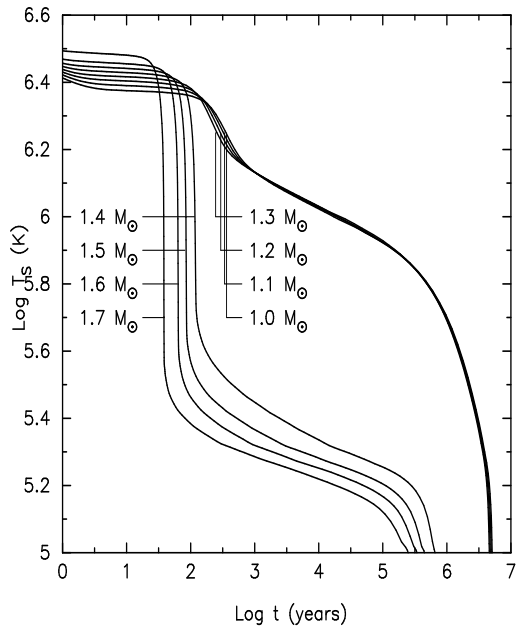}
\end{center}
\caption[]{\footnotesize
Redshifted surface temperature $T_s^\infty$
versus stellar age $t$
according to
Page and Applegate (1992) for non-superfluid NS models
of different mass.}
\label{fig:PA}
\end{figure}
					     
While simulating NS cooling one calculates the
{\it cooling curves}, the dependence of the effective
stellar surface temperature $T_s$ on stellar age $t$.
The effective temperature defines the photon luminosity:
$L_\gamma= 4 \pi {\cal R}^2 \sigma T^{\, 4}_s\,$,
where $\sigma$ is the Stefan--Boltzmann constant.
The luminosity $L_\gamma^\infty$, as detected by a distant observer
with account for the effects of General Relativity,
is related to the local luminosity $L_\gamma$ as
$L_\gamma^\infty = L_\gamma \, (1-{\cal R}_g/{\cal R})$.
Here ${\cal R}_g = 2GM/c^2$ is the gravitational NS radius.
The effective temperature of thermal radiation
$T_s^\infty$, detected by a distant observer,
is related to the local temperature $T_s$ by
$T_s^\infty = T_s \sqrt{1-{\cal R}_g/{\cal R}}$.

Figure \ref{fig:PA} shows typical cooling curves
for (non-superfluid) NSs of different masses
(from
Page and Applegate, 1992, with permission of the authors).
The NS models are based on the equation of state
which opens the direct Urca process for
$\, M > 1.35 \, M_\odot$. The stars with lower $M$ undergo the standard
cooling.

If the direct Urca is allowed in the stellar core,
the star cools much faster than that with the standard
neutrino luminosity. Typical cooling times
$t \sim T/\dot{T}$ can be estimated from thermal balance
of a plasma element in the stellar core: $t \sim CT/Q_\nu$.
Here, $Q_\nu$ is the neutrino emissivity, and
$C$ is the specific heat capacity. Using the heat capacity
(Sect.\ 3.3) of an ideal degenerate gas of neutrons at
$\rho=\rho_0$ and appropriate neutrino emissivity
from Sect.~2.2, one obtains
(see, e.g.,
Shapiro and Teukolsky, 1983;
Haensel, 1994):
$ t^{\dur} \sim \mbox{1~min} \, / \, T_9^4 \,$,
$\;\; t^{\mur} \sim \mbox{1~yr}\, / \, T_9^6 \,$.
It is seen that the direct Urca process decreases the internal
stellar temperature down to
$10^9$~K in $t^{\dur}\sim 1$ min, and down to $10^8$~K in several days.
For the standard neutrino emissivity, the
cooling time
$t^{\mur}$ is about 1~yr and $10^6$~yr, respectively.

These estimates describe temperature variation in the stellar core
at the initial cooling stage. The crust cools initially much slower since
its own neutrino energy losses are relatively small.
Sufficiently low thermal conductivity of the crust
produces thermal insulation between the crust and the core.
Thus, the core and the crust cool independently at the initial stage,
and the NS surface radiation carries no information
on the thermal state of the core
(almost flat initial parts of the cooling curves in Fig.\ \ref{fig:PA}).

The cooling wave from the rapidly cooling core reaches the stellar
surface in 10--1000 yr
(see, e.g.,
Nomoto and Tsuruta, 1987;
Lattimer et al., 1994), and the surface temperature $T_s$
falls down sharply
(Fig.\ \ref{fig:PA}). In the course of the enhanced cooling
$T_s$ drops by almost one order of magnitude in several years,
which decreases the photon luminosity
by a factor of $\sim 10^4$. For the standard cooling, the temperature
fall is not so significant and lasts $\sim 10^2$~yr.

After the thermal relaxation in the NS interiors is over
the neutrino luminosity from the core remains higher than
the photon surface luminosity. However, at this stage
the core controls the surface temperature. This is
the second relatively flat part of the cooling curves.
The internal temperature exceeds the surface temperature
by more than one order of magnitude; the main temperature
gradient takes place in the outermost layers of the outer crust.

Finally, a rather old star appears to be at the {\it photon} cooling stage.
The neutrino luminosity becomes much smaller than
the photon one, and the cooling curves turn out to be much steeper
(Fig.\ \ref{fig:PA}). The change of the cooling regime
is associated with different temperature dependences
of the neutrino and photon cooling rates.
Let us remind that the neutrino emissivity is
$\, Q^{\dur} \propto T^6 \,$ for the direct Urca process,
and $Q^{\Mn} \propto T^8\,$ for the modified Urca process (Sect.\ 2.2).
The photon luminosity falls down much weaker with decreasing temperature,
as $\sim T^{2.2}$
(Gudmundsson et al., 1983). Simple estimates show that the time
$t_\nu$ of transition to the photon cooling stage is mainly determined by
the NS heat capacity and by the temperature dependence of the neutrino
energy losses; typically, $t_\nu \sim 10^5$ yr. The internal temperature
of the star of age $t_\nu$ is mainly determined by the efficiency
of the neutrino energy losses and by their temperature dependence.

The appearance of superfluidity in the NS core changes drastically the
cooling process. This is discussed in subsequent Sects.\ 3 -- 8.
We will study systematically the effects of
superfluidity on the heat capacity and
neutrino luminosity of NSs. This
will allow us to ``calibrate" theoretical
cooling curves and use cooling NSs as ``thermometers"
for measuring nucleon critical temperatures in their cores.
We will see that the nucleon superfluidity can either enhance or
slow down cooling (depending on parameters) and it strongly
reduces difference between the enhanced and standard cooling.
We will analyse observations of thermal radiation from
cooling NSs and show that the observational data are difficult to explain
without assuming existence of superfluidity
in the NS cores.

%
%
%

\section{Superfluidity and heat capacity 
         of neutron star cores}

\subsection{Nucleon superfluidity}

The theory of electron superconductivity in metals was developed by
Bardeen, Cooper and Schrieffer (1957; BCS).
Superconductivity is explained by Cooper pairing of
electrons caused by a weak attraction due to electron--phonon interaction.
Superconducting state appears with decreasing temperature
as a result of the second--order phase transition;
typical critical temperatures are
$T_c \sim$ (1--10)~K. For $T<T_c$, the dispersion relation of
electrons contains an energy gap $\Delta$, with $\Delta \sim \kB T_c$
at $T \ll T_c$.

One year after the publication of the BCS theory
Bohr et al.\ (1958)
suggested that the phenomenon
much in common with superconductivity
could appear in the systems of nucleons in atomic nuclei.
Cooper pairing of nucleons could occur due to nuclear attraction.
It was expected that the gap in the nucleon spectrum,
$\Delta \sim 1$~MeV ($T_c \sim 10^{10}$~K), was many orders of magnitude
larger than for electrons in metals.
Later the presence of pair correlations of
nucleons in atomic nuclei and associated energy gap in the nucleon
spectrum has been investigated theoretically
and confirmed experimentally (Nobel Prize
of Bohr, Mottelson and Reynoter in 1975).

Migdal (1959) was one of the first who applied the BCS theory
to atomic nuclei.
He noticed also that neutron superfluidity caused by nuclear forces
could occur in neutron matter of the inner NS layers where
critical temperatures $T_c \sim10^{10}$~K could be expected.

BCS equations which describe symmetric nuclear matter in atomic nuclei
and asymmetric neutron--rich NS matter have much in common but have also
some differences. For instance, pairing in atomic nuclei takes
place in the singlet-state of a nucleon pair. In this case, the energy gap
is isotropic, independent of orientation of nucleon momentum.
On the other hand, one can expect triplet-state
pairing in NS matter (see below) which leads to the anisotropic gap.

Calculations of the energy gap in symmetric nuclear matter
have been carrying out since 1959, starting from the classical
papers by Cooper et al.\ (1959) and Migdal (1959).
Without pretending to give complete description of this activity
(see, e.g., Chen et al., 1993, for a review)
let us mention the early articles by
Ishihara et al.\ (1963),
Henley and Wilets (1964),
Kennedy et al.\ (1964),
and Kennedy (1968)
who studied the gap in the nucleon spectrum for different model potentials
(Tamagaki, 1968)
of nucleon--nucleon interaction. It was shown by the middle of 1960s
that the gap was extremely sensitive to the repulsive part
of the potential and to the effective masses of nucleons
in nuclear matter. These conclusions are qualitatively
correct for NSs (see below).

It is well known that the BCS theory is used also
to describe superfluidity in liquid  $^3$He.
The foundation of the theory was built
in the paper by Cooper et al.\ (1959) mentioned above.
Superfluidity in $^3$He (caused by interatomic attraction)
is quite different from that in nucleon matter;
the critical temperature in
$^3$He at normal pressure is as small as 2.6 mK. However, there
is one important similarity: pairing can occur in the
triplet state of interacting particles (with orbital momentum $l$=1),
which leads to anisotropic gap. The BCS equations are similar
in these cases. The first article devoted to the triplet-state
pairing was written by
Anderson and Morel (1961). While deriving the equation for
the anisotropic gap the authors overlooked contribution
from one of three triplet spin states. This inaccuracy was
corrected by
Balian and Werthamer (1963).

Let us outline microscopic theories of superfluidity in NSs.
Five years after
Migdal (1959),
Ginzburg and Kirzhnits (1964) published a brief article
where they estimated the gap produced by the singlet-state pairing
of neutrons at density $\rho = 10^{13}$--$10^{15}$
g cm$^{-3}$ and obtained $\Delta \sim$(5--20) MeV.
Very serious step was made by
Wolf (1966).
He showed that the singlet-state neutron pairing takes place
in the inner NS crust
($\rho \la \rho_0$), but disappears in the core,
since the singlet-state
$nn$-interaction becomes repulsive in high-density matter.
The number density of protons in the NS core is much smaller
than the number density of neutrons
(Sect.\ 2.1). Accordingly the single-state $pp$-interaction
is attractive there and leads to the proton pairing.

Before the discovery of pulsars in 1967
the theory of superfluidity in NSs was developing rather slowly.
A detailed review of ``pre-pulsar" articles was given by
Ginzburg (1969).
The discovery of pulsars initiated a great interest to the theory.
Baym et al.\ (1969) analyzed macroscopic consequences
of the neutron superfluidity (rotation of superfluid component in the
form of quantized vortices) and proton superfluidity -- superconductivity
(splitting of the internal stellar magnetic field into
Abrikosov magnetic flux tubes).
This paper made foundation of the modern theory which explained
pulsar glitches by interaction of normal and superfluid
components of matter inside NSs
(see, e.g.,
Pines, 1991).

Very important contribution to the theory was made by
Hoffberg et al.\ (1970).
They showed that the triplet-state $^3{\rm P}_2$ interaction
of neutrons at
$\rho \ga \rho_0$ was attractive. Thus, the triplet-state
neutron superfluidity with anisotropic gap can occur in
the NS core. The authors performed the first calculations
of the critical temperature of the triplet-state
neutron superfluidity in NSs.

The article by Hoffberg et al.\ (1970) was followed by many others,
where the theory was developed and nucleon critical temperatures
in NSs were calculated. Superfluidities of various types have
been considered using different model potentials of nucleon--nucleon
interaction. Some authors have made use of the potentials
renormalized with account for polarization properties of matter
on the basis of various many--body theories.
A comparative analysis of different approaches has been done, for instance,
by
Amundsen and {\O}stgaard (1985a, b), Wambach et al.\ (1991),
Baldo et al.\ (1992), and Chen et al.\ (1993).
A detailed review has been written by
Takatsuka and Tamagaki (1993) not long ago.
Many articles have been devoted to the triplet-state
superfluidity of neutrons in the NS cores
(see, e.g.,
Hoffberg et al., 1970;
Tamagaki, 1970;
Takatsuka and Tamagaki, 1971, 1993, 1997a, b, c;
Takatsuka, 1972a, 1972b;
Amundsen and {\O}stgaard, 1985b;
Baldo et al., 1992;
Elgar{\o}y et al., 1996a, c).
Let us mention model calculations by Muzikar et al.\ (1980)
of the triplet-state neutron pairing in magnetized
NS cores.
According to these calculations, the magnetic field $B \ga 10^{16}$ G
makes the triplet-state superfluidity with the nodes at
the Fermi-surface energetically more preferable than
the familiar triplet-state superfluidity without nodes
(Sect.\ 3.2).
Singlet-state proton superfluidity in the NS cores
has been considered thoroughly by many authors
(e.g.,
Chao et al., 1972;
Takatsuka, 1973;
Amundsen and {\O}stgaard, 1985a;
Baldo et al., 1990, 1992;
Chen et al., 1993;
Takatsuka and Tamagaki, 1993, 1997c;
Elgar{\o}y et al., 1996a, b).
Elgar{\o}y and Hjorth-Jensen (1998)
have studied a singlet-state
proton superfluidity in the NS cores and estimated maximum densities
at which this superfluidity disappears.
Let us mention also the articles by
Sedrakian et al.\ (1997) and
Civitareze et al.\ (1997),
where the possibility of neutron-proton Cooper pairing in
uniform nucleon matter has been analyzed.
This pairing is possible in symmetric nuclear matter
but does not occur in NSs due to large difference of
neutron and proton Fermi momenta.
Many authors (e.g.,
Hoffberg et al., 1970;
Krotscheck, 1972;
Clark et al., 1976;
Takatsuka, 1984;
Amundsen and {\O}stgaard, 1985a;
Chen et al., 1986, 1993;
Ainsworth et al., 1989;
Baldo et al., 1990, 1992;
Wambach et al., 1991, 1993;
Takatsuka and Tamagaki, 1993;
Broglia et al., 1994;
Elgar{\o}y et al., 1996b, 1996d;
Pethick and Ravenhall, 1998)
have considered the singlet-state pairing of free neutrons
in the inner NS crust.
Nucleons within atomic nuclei in the inner crust
can also suffer pairing which is, however, weaker
than for free neutrons
(see, e.g.,
Takatsuka, 1984;
De Blasio and Lazzari, 1995;
Elgar{\o}y et al., 1996d;
Barranco et al., 1998;
Pethick and Ravenhall, 1998).

Although we will not consider exotic models of NS cores
(Sect.\ 2.1), let us mention that superfluidity is
possible in these models as well. For instance,
Takatsuka and Tamagaki (1993)
have reviewed calculations of neutron and proton superfluid gaps
in pion condensed matter. They have obtained also
new results in this field (Takatsuka and Tamagaki, 1997a, b, c).
The same authors
(Takatsuka and Tamagaki, 1995)
have studied the nucleon superfluidity in the presence of kaon
condensation. The pion or kaon condensates mix strongly
neutron and proton states which may induce triplet-state
pairing of quasi-protons. Some authors have discussed
superfluidity in quark matter
(e.g.,
Balian and Love, 1984;
Iwasaki, 1995;
Schaab et al., 1996, 1997a).
If hyperons appear in
$npe$-matter,
they can also be in a superfluid state
(Balberg and Barnea, 1998).
Electrons and muons, which interact via Coulomb forces,
can be, in principle, superfluid as well.
However, corresponding critical temperatures are too
low to be of practical interest. For instance,
according to an estimate by
Ginzburg (1969) the critical temperature of
degenerate electrons at
$\rho \ga 100$ g cm$^{-3}$ does not exceed
1~K, while the internal temperature of a cooling NS of age
$t \la 10^6$ does not fall below $10^6$ (Sect.\ 7).

Let us summarize properties of nucleon superfluidity
in the cores and crusts of NSs with the standard
($npe$) composition.

(1) Singlet-state neutron superfluidity exists in the inner
NS crust and disappears at the density
$\rho \sim \rho_0$, at which an effective neutron--neutron
singlet-state attraction transforms into repulsion.
Density dependence of the critical temperature
$T_{cn}$ has maximum at a subnuclear density.
Maximum values of
$T_{cn}$ range from $10^8$~K to $10^{11}$~K, for different
models of dense matter.

(2) Triplet-state neutron superfluidity appears
in the NS core at
$\rho \ga \rho_0$ owing to an effective attraction between neutrons in the
triplet state. Density dependence of the appropriate critical
temperature, as a rule, has maximum at supranuclear
density. Maximum values
of $T_{cn}$
vary from $10^8$~K to $10^{10}$~K for different microscopic models.

(3) Protons in the NS cores can suffer singlet-state pairing.
The dependence $T_{cp}(\rho)$ has usually maximum at
a supranuclear density, and the maximum values of
$T_{cp}$ range from $10^8$~K to $10^{10}$~K for different
models of matter.

(4) The critical temperature is very sensitive
to the strength of repulsive core of nucleon--nucleon interaction
(see, e.g.,
Takatsuka and Tamagaki, 1971, 1993;
Krotscheck, 1972).
$T_c$ increases strongly with core softening,
i.e., with increasing attraction between nucleons.
Even a weak additional attraction
(e.g., due to inclusion of coupling between
the $^3{\rm P}_2$ and $^3{\rm F}_2$ states for the triplet-state
pairing of neutrons: see
Takatsuka, 1972a;
Takatsuka and Tamagaki, 1993) may increase
$T_c$ by several orders of magnitude. Generally, the superfluidity is
stronger for softer equation of state.

(5) The critical temperature falls down rapidly
with decreasing the effective mass of nucleons in dense matter
(see, e.g.,
Takatsuka and Tamagaki, 1971, 1993;
Chao et al., 1993),
i.e., with decreasing density of state of nucleons at the Fermi surface.
If the effective mass is sufficiently low
($m^\ast_N \la 0.5 \, m_N$) the superfluidity may entirely disappear.

(6) The critical temperature depends strongly on
the method of description of many-body (in-medium) effects.
For instance, the in-medium effects
for neutrons in the NS crusts
decrease $\tn$ typically by several times
(e.g.,
Clark et al., 1976;
Wambach et al., 1991, 1993;
Takatsuka and Tamagaki, 1993).

\begin{figure}[t]                         
\begin{center}
\leavevmode
\epsfysize=8.5cm
\epsfbox[0 20 325 350]{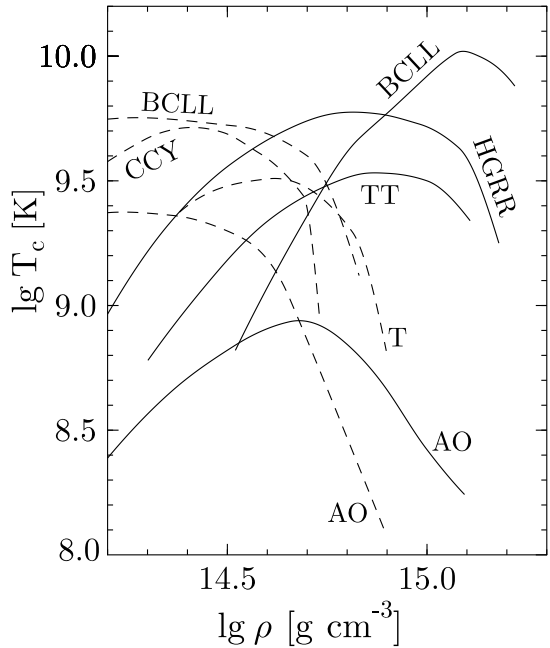}
\end{center}
\caption[]{\footnotesize
     Critical temperatures of triplet-state neutron
     superfluidity (solid lines) and singlet-state proton
     superfluidity (dashed lined) versus density
     according to different authors:
     HGRR --- Hoffberg et al.\ (1970),
     TT --- Takatsuka and Tamagaki (1971),
     CCY --- Chao et al.\ (1972),
     T --- Takatsuka (1973),
     AO --- Amundsen and {\O}stgaard (1985a, b),
     BCLL --- Baldo et al.\ (1992)
     (see text for details).
     }
\label{fig:Gaps}
\end{figure}

For illustration, in Fig.\ \ref{fig:Gaps} we present
the critical temperature of the triplet-state neutron
superfluidity and the singlet-state proton superfluidity
obtained by different authors.

The HGRR curve has been calculated by
Hoffberg et al.\ (1970)
for neutron superfluidity using the Tabakin (1964) model of
nucleon--nucleon potential
which reproduces quite well experimental phases of
nucleon scattering at energies
$\leq$ 320 keV.
Hoffberg et al.\ (1970) have neglected the in-medium effects;
the effective neutron mass has been set equal to its bare mass.
The TT--curve has been calculated by
Takatsuka and Tamagaki (1971) for the neutron superfluidity
under the same assumptions but using
the one-pion-exchange model of nucleon interaction
(OPEG $^3$0-1) with somewhat harder core.
This lowers $T_{cn}$.
The solid AO--curve has been obtained by
Amundsen and {\O}stgaard (1985b) for the neutron superfluidity
using similar one-pion-exchange approach
(OPEG), but the effective neutron mass has been determined
self-consistently and has appeared to be lower
than the bare neutron mass. This additionally lowers
$T_{cn}$. The BCLL--curve is a result of recent calculations
by Baldo et al.\ (1992)
using an empirical nucleon potential
(Argonne $v_{14}$) which reproduces accurately laboratory
data on nucleon scattering. The potential and neutron mass have not been
renormalized to account for the in-medium effects. As a result,
the values of $T_{cn}$ are high again.

The dashed CCY curve shows the results
by Chao et al.\ (1972)
for proton superfluidity using the simplified
Serber potential which describes satisfactorily
the experimental phases of singlet-state nucleon-nucleon scattering.
The effective proton mass has been determined self-consistently.
The T curve presents the results by
Takatsuka (1973)
obtained under the same assumption but using a more realistic
OPEG potential.
The dashed AO curve has been calculated by
Amundsen and {\O}stgaard (1985a)
using the same one-pion-exchange approach but
another method of evaluation of the effective
proton mass which leads to noticeably different critical temperatures
$T_{cp}$. Finally, the dashed
BCLL curve
has been derived by Baldo et al.\ (1992)
under the same assumptions as the solid
BCLL curve. A neglect of polarization effects leads to
the high critical temperature $T_{cp}$.

Let us emphasize large scatter of
$T_{cn}$ and $T_{cp}$. For instance, the BCLL curves
(Baldo et al., 1992) mentioned above
give very high critical temperatures in matter of supranuclear
density while, for instance, calculations by
Takatsuka and Tamagaki (1997b)
indicate significant decrease of critical temperatures
at densities of several
$\rho_0$, at which the direct Urca process is open.
Under these conditions, while analysing the
effects of superfluidity,
we will not use any specific
microscopic results but will treat
$T_{cn}$ and $T_{cp}$ as free parameters.

\subsection{Energy gaps and critical temperatures}

Thus, the cases of the $^1{\rm S}_0$ or $^3{\rm P}_2$ pairing
in the NS cores are of special interest.
The $^3{\rm P}_2$ pairing in $npe$ matter
occurs mainly in the system of neutrons.
While studying this pairing one should take into account
the states with different projections $m_J$ of
the total pair momentum onto quantization axis:
$|m_J|=0$, 1, 2. The actual (energetically favourable)
state may be a superposition of states with different
$m_J$ (see, e.g.,
Amundsen and {\O}stgaard, 1985b;
Baldo et al., 1992). Owing to uncertainties of
microscopic theories this state is still unknown;
it depends possibly on density and temperature.
In simulations of NS cooling, one usually considers
the triplet-state pairing with $m_J=0$
(excluding the recent article by
Schaab et al., 1998b). Below we will consider the
$^3{\rm P}_2$ superfluidities
with $m_J=0$ and $|m_J|=2$, since their effects on the
heat capacity and NS neutrino luminosity
are qualitatively different.

\begin{table}[t]
\caption{Three superfluidity types of study}
\begin{center}
  \begin{tabular}{||c|cccc||}
  \hline \hline
         & Superfluidity type  & $\lambda$ &   $F(\vartheta)$
         & $\kB T_c/\Delta(0)$  \\
  \hline
  {\bf A} & $^1{\rm S}_0$  &     1     &        1
         &   0.5669   \\
  {\bf B} & $^3{\rm P}_2\ (m_J =0)$ & $1/2$
          & $(1+3\cos^2\vartheta)$  & 0.8416  \\
  {\bf C} & $^3{\rm P}_2\ (|m_J| =2)$ & $3/2$
          & $\sin^2 \vartheta$      & 0.4926  \\
  \hline \hline
\end{tabular}
\label{etab:ABC}
\end{center}
\end{table}

For certainty, we will analyze the BCS superfluidity
for an ensemble of almost free (qua\-si-)\-nuc\-le\-ons.
The superfluidity types
$^1{\rm S}_0$,
$^3{\rm P}_2$ ($m_J=0$) and  $^3{\rm P}_2$ ($m_J=2$)
under study will be denoted by
{\bf A}, {\bf B} and {\bf C}, respectively (Table \ref{etab:ABC}).
Superfluidity onset is accompanied by the appearance
of the energy gap $\delta$ in
momentum dependence of the nucleon energy
$\ep({\vect{p}})$. Near the Fermi surface
($|p- \pF| \ll \pF$) one has
(e.g., Lifshitz and Pitaevskii, 1980)
\begin{equation}
   \ep = \mu - \sqrt{\del^2 + \eta^2} \;\;
                     \mbox{at} \;\; p<\pF; \;\;\;\;\;\;
   \ep = \mu + \sqrt{\del^2 + \eta^2} \;\;
                     \mbox{at} \;\; p\geq\pF .
\label{eq:Dispers}
\end{equation}
Here, $\eta=\vF (p-\pF)$, $\vF$ and $\pF$ are the nucleon
Fermi velocity and momentum, respectively, $\mu$ is the
chemical potential; it is assumed that $\delta \ll \mu$.
In the cases of study, $\delta^2 = \Delta^2 (T) F(\vartheta)$, where
$\Delta(T)$ is an amplitude, which determines
temperature dependence of the gap,
$F(\vartheta)$ describes dependence of
$\delta$ on angle $\vartheta$ between the quantization axis
and particle momentum $\vect{p}$. The quantities $\Del$ and $F$
are determined by the superfluidity type
(Table \ref{etab:ABC}).
In case {\bf A} the gap is isotropic,
$\del=\Del$. In cases {\bf B} and {\bf C}
the gap is anisotropic (depends on $\vartheta$).
Let us notice that in case {\bf C} the gap
vanishes at the Fermi-sphere poles at any temperature
(since $F(0)=F(\pi)=0$), i.e., the superfluidity does not
affect the nucleons which move along the quantization axis.

The gap amplitude $\Del(T)$ is determined by the BCS equation
(see, e.g.,
Lifshitz and Pitaevskii, 1980;
Tamagaki 1970) which can be written as
\begin{equation}
     \ln \left[ \frac{\Delta_0}{\Delta(T)} \right] =
     2 \lambda \int {\dd \Omega \over 4 \pi}
     \int^\infty_0 \, \frac{\dd x}{z} \, f \, F(\vartheta),
\label{eq:BCS}
\end{equation}
where $\Delta_0=\Del(0)$,
$\dd \Omega$ is a solid angle element in the direction of ${\vect{p}}$,
$f= (1 + {\rm e}^z)^{-1}$
is the Fermi--Dirac distribution,
$\lambda$ is a numeric coefficient (Table \ref{etab:ABC}),
\begin{equation}
    z=\frac{\ep-\mu}{\kB T}={\rm sign}(x)\sqrt{x^2+y^2}, \;\;\;
    x=\frac{\eta}{\kB T}, \;\;\;  y=\frac{\del}{\kB T}\, .
\label{eq:DimLessVar}
\end{equation}
Using Eq.\ (\ref{eq:BCS}) one can easily obtain the values of
$\kB T_c / \Delta_0$ presented in Table \ref{etab:ABC}.
It is convenient to introduce the variables
\begin{equation}
   v = \frac{\Delta(T)}{\kB T} , \;\;\;\;
   \tau = \frac{T}{T_c}.
\label{eq:DefGap}
\end{equation}
The dimensionless gap amplitude $v$ describes the temperature
dependence of the gap. It is determined by the superfluidity type
and the dimensionless temperature
$\tau$. In case {\bf A} the amplitude $v$ corresponds to the
isotropic gap, in case {\bf B} it corresponds to the minimum
and in case {\bf C} to the maximum gap at the Fermi surface.
In these notations, the dimensionless gap $y$ has the form:
\begin{equation}
  y_{\A} = \va, \;\;\;\;
  y_{\B} = \vb \, \sqrt{1+3 \cos^2 \vartheta},  \;\;\;\;\;
  y_{\C} = \vc \, \sin \vartheta.
\label{eq:y_v}
\end{equation}

Using Eq.~(\ref{eq:BCS}) one can obtain the asymptotes of the
gap amplitude near the critical temperature and in the
limit of the so called strong superfluidity ($T \ll T_c$).
For instance, at $T \to T_c$, $T < T_c$  ($\tau \to 1$)
we have (see, e.g.,
Lifshitz and Pitaevskii, 1980;
Levenfish and Yakovlev, 1994a):
$\,v=\beta \sqrt{1-\tau}$, where
$\beta_A= 3.063$,
$\beta_B= 1.977$,
$\beta_C= 3.425$.
For $T \ll T_c$ we obtain
$v=\Delta_0/(k_{\rm B} T_c \, \tau)$.
Levenfish and Yakovlev (1993, 1994a)
calculated $v=v(\tau)$ for intermediate $\tau$
and proposed the analytic fits of the numeric data:
\begin{eqnarray}
\eqalign{
  \va  =  \sqrt{1-\tau}
          \left( 1.456 - \frac{0.157}{\sqrt{\tau}} + \frac{1.764}{\tau}
          \right),
\cr
  \vb  =  \sqrt{1-\tau} \left( 0.7893 + \frac{1.188}{\tau} \right),
\cr
  \vc  =  \frac{\sqrt{1-\tau^4}}{\tau}
          \left(2.030 - 0.4903 \tau^4 + 0.1727 \tau^8 \right).
}
\label{eq:FitGaps}
\end{eqnarray}
These fits will be useful for evaluating the heat capacity
and neutrino luminosity in the superfluid NS cores.

Analytic fits presented here and below
reproduce numeric data with mean errors about
1--2\%, while the maximum error does not exceed 5\%.
This fit accuracy is more than sufficient for
implications to NS cooling. Our fits reproduce
also various asymptotes of the quantities of study.

Notice that for the triplet-state superfluidity with $|m_J|=1$ or
for a triplet-state superfluidity described by
superposition of states with different $m_J$, the anisotropic
gap $\delta$ depends not only on $\vartheta$, but also on
azimuthal angle $\phi$ of nucleon momentum $\vect{p}$.
A study of the effects of such superfluidity on the heat capacity and
neutrino luminosity is complicated and has not been performed yet.

\subsection{Heat capacity of superfluid neutron star cores}

Consider $npe$ matter of the NS cores.
The specific (per unit volume) heat capacity is equal to
the sum of partial heat capacities
of all particle species $j$: $ C= \sum_{\mbox{\tiny j}} C_j$, where
$j = n, \, p, \, e$.
Since the particles are strongly degenerate, the heat capacities at
constant volume and pressure are nearly the same
(Lifshitz and Pitaevskii, 1980)
and we make no difference between them.
A partial heat capacity is determined by the
standard thermodynamic formula
(Lifshitz and Pitaevskii, 1980)
\begin{equation}
       C_j = \frac{2}{(2 \pi \hbar)^3}
       \int \dd {\vect{p}}_j \; (\ep_j - \mu_j)
        \frac{\dd f_j}{\dd T},
\label{eq:DefC}
\end{equation}
where $\ep_j$ and ${\vect{p}}_j$
stand for the particle energy and momentum, respectively,
$\mu_j$ is the chemical potential, and
$f_j$ is the Fermi-Dirac distribution.

In our case, the electrons constitute an almost ideal,
strongly degenerate, ultra-relativistic gas.
Accordingly,
\begin{equation}
       C_e = \frac{m_e^\ast \pFe k_{\rm B}^2 T}{3 \hbar^3}  \approx
           5.67 \times 10^{19}  \left( n_e \over n_0 \right)^{2/3} \, T_9
          \;\;\;\; \frac{\mbox{erg}}{\mbox{cm$^3\,$K}},
\label{eq:Ce}
\end{equation}
where $m^\ast_e=\mu_e/c^2 \approx \pFe/c$; $\pFe$ and $n_e$
denote, respectively, the electron Fermi momentum and number density,
$n_0=0.16$~fm$^{-3}$ is the standard nucleon number density
in atomic nuclei. In the presence of muons, one should add
their partial heat capacity $C_\mu$ to the total heat capacity;
$C_\mu$ is similar to $C_e$
with the only difference that muons may be non-relativistic.

Neutrons and protons ($j = N= n$ and $p$)
in the NS cores constitute a non-relativistic, strongly non-ideal
Fermi-liquid
(Sect.\ 2.1). The partial heat capacity (\ref{eq:DefC}) of normal
nucleons is:
\begin{equation}
 C_{N 0} = \frac{m_N^* \, \pF_{\!N} \,
                    k_{\rm B}^2 \, T}{3 \hbar^3} \approx
 1.61 \times 10^{20} \; \frac{m_N^\ast}{m_N}
          \: \left(n_N \over  n_0 \right)^{1/3} \, T_9 \;\;\;\; \;\;
        \frac{ \mbox{erg}}{\mbox{cm$^3\,$K}}.
\label{eq:Cnuc}
\end{equation}
Here, $\pF_{\!N}$ is the nucleon Fermi momentum
determined by the nucleon number density $n_N$, and
$m_N^\ast$ is the effective mass of the
(quasi-)nucleon in dense matter. Notice that the main contribution
into the heat capacity comes from nucleons with energies
near the Fermi level, $ |\ep_N - \mu_N |
\la \kB T$, which can participate in processes
of energy exchange $\sim \kB T$.

For $T<T_c$, Eq.\ (\ref{eq:DefC}) should
include the energy gap
(\ref{eq:Dispers}). Generally, the nucleon heat capacity
can be written as
\begin{equation}
     C_N =C_{N 0} \, R(T),
\label{eq:reductionC}
\end{equation}
where $C_{N 0}$ is the partial heat capacity of normal nucleons
(\ref{eq:Cnuc}), and the factor
$R$ describes variation of the heat capacity by the
superfluidity.
$R$ depends on superfluidity type and
on dimensionless temperature
$\tau$. Clearly, $R(T)=1$ for $T>T_c$ ($\tau>1$).

A general expression for
$R$ is obtained from Eq.\ (\ref{eq:DefC})
by introducing the dispersion relation
(\ref{eq:Dispers}), dimensionless variables
(\ref{eq:DimLessVar}) and by taking into account Eq.\ (\ref{eq:Cnuc}):
\begin{equation}
     R = \frac{3}{2 \pi^3} \int \! \! \dd \Om \;
        \int_0^{\infty} \! \dd x\, z\, T\, \frac{\dd f(z)}{\dd T}.
\label{eq:DefR}
\end{equation}
Notice that the Fermi-Dirac distribution
$f$ depends on $T$ directly as well as functionally,
through the dispersion relation (\ref{eq:Dispers}).
Therefore, $\dd f /\dd T$ contains derivative of the
gap amplitude $\Delta(T)$ with respect to $T$.

The heat capacity of superfluid Fermi-systems is well known
in the physical literature (especially for the singlet-state
pairing) but not in the astrophysics of NSs. The heat capacity
$C_{A}(T)$ for case {\bf A} has been calculated by
M\"{u}hlschlegel (1959)
and is described in textbooks
(Lifshitz and Pitaevskii, 1980).
Maxwell (1979)
proposed a fit of $C_A(T)$
in the temperature range $0.2 \, T_c \le T \le T_c$.
Anderson and Morel (1961)
derived an asymptote of $C_C(T)$ at $T \ll T_c$
for superfluidity of type {\bf C}.
Simple expressions for
$R_A$, $R_B$ and $R_C$ convenient for evaluating
the nucleon heat capacity in the NS cores were obtained
by Levenfish and Yakovlev (1994a)
and are given below.

When temperature falls below the critical value,
the heat capacity suffers a jump produced by latent heat release
at the phase transition.
In case {\bf A} the jump is given by
$R_A(T_c) = 2.426$
(Lifshitz and Pitaevskii, 1980),
while in cases {\bf B} and {\bf C} the jump is $R_B(T_c) = R_C(T_c) =
2.188$. These jumps affect NS cooling (Sect.\ 7.2).

\begin{figure}[t]                         
\begin{center}
\leavevmode
\epsfysize=8.5cm
\epsfbox[65 30 280 280]{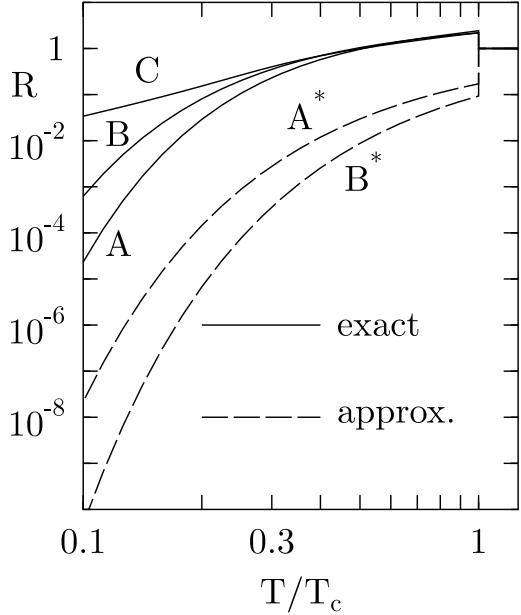}
\end{center}
\caption[]{\footnotesize
   Factors $R$, which describe variation of the heat capacity
   by superfluidity, versus $T/T_c$.
   Letters near the curves show superfluidity type.
   Dashed curves {\bf A}$^\ast$ and
   {\bf B}$^\ast$ present simplified factors $R^\ast$
   for superfluidities {\bf A} and {\bf B}, which were
   commonly used in applications earlier (see text).
}
\label{fig:Capacity}
\end{figure}

The energy gap in the nucleon spectrum strongly reduces the heat capacity
for temperatures much below the critical one
($T \ll T_c$, $v \to \infty$).
Let us remind, that the main contribution into
the heat capacity in non-superfluid matter
would come from nucleons with energies
$|\ep - \mu| \la \kB T$. However, in the low-temperature limit
$T \ll T_c$ for all nucleons in cases
{\bf A} or {\bf B} and almost for all nucleons
(excluding a small fraction of particles near the poles of the
Fermi sphere) in case {\bf C}
the gap $\delta$
is much larger than $\kB T$. Therefore, the ``heat-capacious" nucleons
with energies $|\ep - \mu| \la \kB T$ are either absent or
almost absent, which suppresses the heat capacity.
Accordingly, we have the asymptotes:
\begin{eqnarray}
\eqalign{
 \RaC = \frac{3 \sqrt{2}}{ \pi^{3/2} } \, v^{5/2} \ex^{-v} =
        {3.149 \over \tau^{5/2}} \,
        \exp \left( -{1.764 \over \tau} \right) ,
\cr
 \RbC = \frac{\sqrt{3}}{\pi} \, v^2 \, \ex^{-v} =
        \frac{0.7781}{\tau^2} \,
        \exp \left( - {1.188 \over \tau} \right) ,
\cr
  \RcC = \frac{7 \pi^2}{5 \, v^2} = 3.353 \, \tau^2 \, .
} 
\label{eq:RcCasy}
\end{eqnarray}
Here and hereafter
we set $v=v_A$, $v=v_B$ or $v=v_C$
in the factors
$R$, labeled by indices $A$, $B$ or $C$,
respectively.
With decreasing temperature,
$R_A$ and $R_B$ are reduced exponentially,
while $R_C$ is reduced much slower, proportionally to
$T^2$. The exponential reduction is associated with
exponentially small ``efficiency of excitation"
of quasi-nucleons near the Fermi surface in the presence of a large gap.
In the case of the heat capacity, this effect strongly reduces
``heat exchange", but, generally, all nucleon reactions
(transitions) are reduced.
A small exponent is absent in case
{\bf C} because the gap
$\delta_C=\delta_C(T,\vartheta)$ vanishes for nucleons at the Fermi--sphere
poles. The superfluidity disappears near the poles, and
these nucleons have ``normal" heat capacity. However the
fraction of these nucleons decreases proportionally to
$T^2$, which leads to a power-law reduction of the
heat capacity of the nucleon liquid as a whole.

The results of numeric calculations of
$R$, as well as the asymptotes for $v \to 0$ and
$v \to \infty$ are described by the simple fit expressions
(Levenfish and Yakovlev, 1994a)
\begin{eqnarray}
\eqalign{
  R_{\A} = \left( 0.4186 + \sqrt{(1.007)^2
          +(0.5010\,v)^2} \right)^{2.5}
          \exp \left(1.456- \sqrt{(1.456)^2 + v^2} \right),
\cr 
 R_{\B} = \left( 0.6893 + \sqrt{(0.790)^2
          +(0.2824\,v)^2} \right)^2
          \exp \left(1.934 - \sqrt{(1.934)^2 +  v^2} \right),
\cr 
 R_{\C} = \frac{2.188 - (9.537 \times 10^{-5} \, v)^2
         + (0.1491 \, v)^4}{ 1 - (0.2846 \, v)^2
         + (0.01335 \,v)^4 + (0.1815 \, v)^6}.
}
\label{eq:RcCfit}
\end{eqnarray}
Equations (\ref{eq:RcCfit}), combined with
(\ref{eq:FitGaps}), enable one to evaluate easily
the factor $R$ as a function of the dimensionless temperature $\tau$.

\begin{figure}[t]                         
\begin{center}
\leavevmode
\epsfysize=8.5cm
\epsfbox[45 30 260 280]{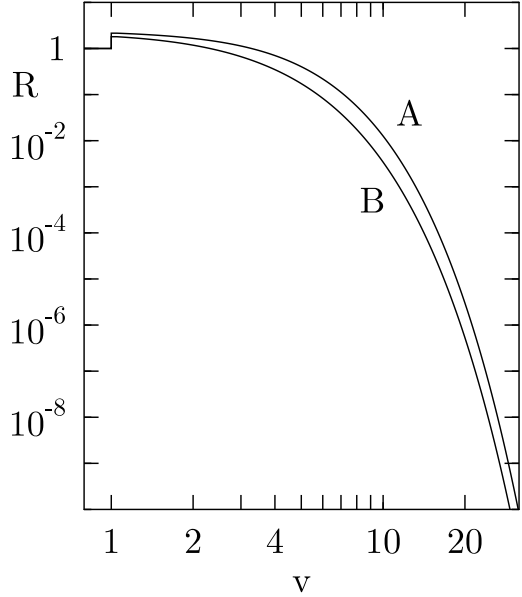}
\end{center}
\caption[]{\footnotesize
   Factors $\RaC$ (curve {\bf A}) and $\RbC$ (curve {\bf B}),
   as in Fig.\ \ref{fig:Capacity}, but versus
   dimensionless gap amplitude $v$.}
\label{fig:Capacity_v}
\end{figure}

The dependence of $R$ on $\tau$ for superfluidity types
{\bf A}, {\bf B}, {\bf C}
is shown in Fig.~\ref{fig:Capacity}.
The factors $R_A$ and $R_B$ as functions of the dimensionless
gap amplitude $v$ are close which is illustrated in Fig.\
\ref{fig:Capacity_v}.

In simulations of the NS cooling before publication of
the articles by
Levenfish and Yakovlev (1993, 1994a),
as far as we know, one considered the superfluidity
of {\bf A} or {\bf B} type and used the
simplified factors of the form
$R^\ast=\exp(-v^\ast)$, where $v_A^\ast =\delta_A/(\kB T)=1.764/\tau$,
$v_B^\ast=\delta_{\rm max}/(\kB T)=2.376/\tau$, $\delta_{\rm max}=
2 \Delta_B(0)$ is the maximum energy gap
$\delta_{\B} (0,\vartheta)$ as a function of $\vartheta$
at $T=0$. These factors were introduced from physical
consideration (see, e.g.,
Page and Applegate, 1992)
basing on the suggestion by
Maxwell (1979).
As seen from Fig.\ \ref{fig:Capacity}, the simplified factors do
not reproduce the heat capacity jump at $T=T_c$ and
strongly overestimate reduction of the heat capacity at
$T \ll T_c$. Indeed, the correct asymptotes in the limit $T \ll T_c$
contain large pre-exponents omitted in the simplified expressions.
In addition, the factors $R_B$ and $R_B^\ast$
differ by exponent arguments. Instead of the correct argument
$v_B=\delta_{\rm min}/(\kB T)$
(see Eq.\ (\ref{eq:RcCasy})) the simplified factor
contains the argument
$v_B^\ast=\delta_{\rm max}/(\kB T)=2 \delta_{\rm min}/(\kB T)$,
which is twice larger. As a result, for instance, at $T=0.1 \, T_c$
the exact factor $R_A$ exceeds the simplified factor $R_A^\ast$
by about three orders of magnitude, while the exact factor $R_B$
exceeds $R_B^\ast$ by seven orders of magnitude.

Although the local heat capacities of particles
are determined by a given equation of state,
the relative contribution of $n$, $p$ and $e$
into the total heat capacity of a non-superfluid NS core
does not depend strongly on the equation of state
(Page, 1994).
The neutron heat capacity is about
$3/4$ of the total one, the proton heat capacity is about
$1/4$ of the total one, and the electron contribution is only about
$5\%$. Therefore, a strong superfluidity of $n$
(with normal $p$) reduces the NS heat capacity by a factor of $\sim 4$.
In the opposite case (strong proton superfluidity and
normal neutrons) the total heat capacity is reduced only by $\sim 25\%$.
If neutrons and protons are strongly superfluid at once,
only $\sim 5\%$ of the heat capacity of the normal core survives. In
Sect.\ 7.2 we will see that the asymmetry of the nucleon composition
of the NS cores affects the NS cooling.

%
%
%

\section{Reduction of direct Urca process by nucleon superfluidity}
\subsection{Direct Urca process in non-superfluid matter}

Neutrino cooling of a NS with the standard nuclear composition
is determined by the direct Urca process (\ref{eq:Durca_Nucleon})
involving neutrons and protons; the process
is allowed (Lattimer et al., 1991)
for many realistic equations of state in the inner NS core.
In the outer core, the process is forbidden by
momentum conservation of reacting particles (Sect.\ 2.2).

The process (\ref{eq:Durca_Nucleon}) consists of the direct and
inverse reactions. Under beta-equilibrium condition,
their rates are equal, and it is sufficient to calculate
the neutrino emissivity in the first reaction and
double the result. The total emissivity
will be labeled by the upperscript
(D); in a normal (non-superfluid) matter,
according to Lattimer et al.\ (1991), it is given by
($\hbar\! = \! c \! = \! \kB \! = \!1$):
\begin{equation}
    Q^{\dur} \!\!  =  \!
            2 \! \int \!
            \left[ \prod_{j=1}^2 { \dd^3 p_j \over (2 \pi)^3} \right] \!
            { \dd^3 p_e \over 2 \ep_e (2 \pi)^3} \;
            { \dd^3 p_\nu \over 2 \ep_\nu (2 \pi)^3}
            (2 \pi)^4 \, \delta(E_f-E_i) \,
            \delta( {\vect{P}}_f - {\vect{P}}_i ) \;
            \ep_\nu \, {\cal L}
            \sum_{\rm spins} | M |^2 \; \Th \, ,
\label{eq:QdDef}
\end{equation}
where ${\vect{p}}_j$ is a nucleon momentum
($j$=1 corresponds to neutron and $j$=2 to proton),
${\vect{p}}_e$ and $\ep_e$ are electron momentum and energy, respectively,
${\vect{p}}_\nu$ and $\ep_\nu$ are antineutrino momentum and energy,
the delta function $ \delta(E_f-E_i) $
describes energy conservation,
and $\delta({\vect{P}}_f - {\vect{P}}_i )$
describes conservation of momentum of particles in the initial
({\it i\/}) and final ({\it f\/}) states. Furthermore,
${\cal L}$ stands for the product of Fermi--Dirac distributions
or corresponding blocking factors of nucleons and electron,
$| M |^2 $ is the squared reaction amplitude.
Summation is over initial and final spin states.
The step function $\Th$ forbids the reaction in matter of not too high density
(Sect.\ 2.2):  $\; \Th=1$ if
$\pF_n, \, \pF_p, \, \pF_e$
satisfy the ``triangle condition'' and $\Th=0$ otherwise.

Nucleons and electrons in the NS core are strongly
degenerate and the main contribution into the integral
(\ref{eq:QdDef}) comes from narrow regions of momentum space
near the corresponding Fermi surfaces. Thus one can set
$p = \pF$ in all smooth functions under the integral.
For non-relativistic nucleons, the quantity
$|M|^2$ is independent of particle momenta and can be
taken out of the integral. The remaining integral is
evaluated by the standard method of decomposition of
integration over directions and magnitudes of particle momenta.
Introducing the dimensionless variables
in accordance with Eq.\ (\ref{eq:DimLessVar}),
the neutrino emissivity (\ref{eq:QdDef}) can be written as
(e.g., Levenfish and Yakovlev, 1994b)
\begin{eqnarray}
&&    Q^{\dur}  =  {4 \over (2 \pi)^{8}} T^6
            AIS \, \overline{| M |^2} \; \Th,
\label{eq:DecompDur} \\
&&   A = \left[ \prod_{j=1}^4 \int \dd \Om_j \; \right]
            \delta \left( \sum_{j=1}^4 {\vect{p}}_j  \right),
\label{eq:Adur} \\
&&   I = \int_0^\infty \dd x_\nu \; x_\nu^3
       \left[ \prod_{j=1}^3 \int_{-\infty}^{+\infty}
       \dd x_j \; f_j \right]
       \delta \left( \sum_{j=1}^3 x_j-x_\nu \right),
\label{eq:Idur} \\
&&   S = \prod_{j=1}^3 \pF_{\! j} m_j^\ast.
\label{eq:Sdur}
\end{eqnarray}
The quantity $A$ contains integrals over orientations of
particle momenta
($j=$ 1, 2, 3, and 4  corresponds to $n$, $p$, $e$, and $\nu$, respectively);
$ \dd \Om_j $ is solid angle element in direction of ${\vect{p}}_j$.
All vector lengths ${\vect{p}}_j$ in the delta-function
must be set equal to the corresponding Fermi momenta.
The quantity $I$, given by Eq.\ (\ref{eq:Idur}),
includes integration over dimensionless energies of neutrino
$x_\nu = p_{\nu}/T =\ep_\nu /T$
and other particles $x_j = \vF_{\!j}(p-\pF_{\!j})/T$
(see Eq.\ (\ref{eq:DimLessVar}));
$f_j= [ \exp(x_j)+1]^{-1}$ is the Fermi--Dirac distribution.
For particles $j$=2 and 3, we have transformed
$\left[1-f(x) \right] \to f(x)$ by replacing $x \to -x$. Finally,
$S$ contains products of the density states at the Fermi surfaces
of $n$, $p$ and $e$;
$m_j^\ast$ is an effective particle mass, with
$m_e^\ast = \mu_e/c^2$.
As shown, for instance, by
Prakash et al.\ (1992)
the quantity $\overline{|M|^2}$,
summed over spins of final particles
and averaged over spin of initial neutron and over directions
of electron and neutrino momenta is
$G^2_{\rm F} \, \cos^2 \! \theta_{\rm C} \, (f_V^2+3g_A^2)$,
where $G_{\rm F}=1.436 \times 10^{-49}$ erg$\;$cm$^3$
is the Fermi constant of weak interaction; $f_V=1$
and $g_A=1.26$ are, respectively, the vector and axial-vector constants
for the reaction of study, and
$\theta_{\rm C}$ is the Cabibbo angle
($\sin  \theta_{\rm C} =0.231$).

In the absence of superfluidity, the integrals
$A$ and $I$ are standard (e.g.,
Shapiro and Teukolsky, 1983):
$  A_0 = 32 \pi^3 /\,( \pFn \, \pFp \, \pFe)$ and
$\,  I_0 = 457 \pi^6 / 5040$.
Thus, the neutrino emissivity of the direct Urca process
in non-superfluid matter (in the standard physical units) is
(e.g., Prakash et al., 1992)
\begin{eqnarray}
\eqalign{
 Q^{\dur}_0 = {457\, \pi \over 10080} \,
              G^2_{\rm F} \, \cos^2 \theta_{\rm C} \, (f_V^2+3 g_A^2) \;
              {m_n^\ast \, m_p^\ast \, m_e^\ast \over \hbar^{10} c^3} \;
              (\kB T)^6 \, \Th
\cr  
\phantom{ Q^{\dur}_0}
       \approx 4.00 \times 10^{27}
                \left( {n_e \over n_0} \right)^{\!1/3} \;
                 {m_n^\ast \, m_p^\ast \over m_n^2} \,
                 T^6_9 \, \Th
                \;\;\;\;\;\;\; \rate \,.
}
\label{eq:Qdur0}
\end{eqnarray}
Here, as before, $n_0=0.16$~fm$^{-3}$.

Similar formula is valid for the direct Urca processes involving
hyperons. Notice that
$f_V$ and $g_A$ can be renormalized under the action of
in-medium effects. Exact calculation of these
effects is complicated. Here and below, for certainty, we will
use non-renormalized values. Notice also
that the direct Urca process is affected by the strong magnetic
field, $B \ga 10^{16}$~G
(Baiko and Yakovlev, 1999).

\subsection{Direct Urca process in superfluid matter}

The main contribution into neutrino generation in non-superfluid matter
comes from nucleons with energies $|\ep - \mu| \la \kB T$.
The nucleon spectrum in superfluid matter contains
the energy gap which suppresses the reaction. The essence of
suppression is the same as in the case of suppression
of heat capacity by strong superfluidity (Sect.\ 3.3). In contrast
to the heat capacity, even a weak superfluidity usually suppresses
the neutrino reactions (excluding the neutrino emission due to
Cooper pairing of nucleons, Sects.\ 2.2, 6.1).

The expression for the neutrino emissivity in the direct Urca
process can be generalized to the case of superfluid matter
by introducing the energy gap into the
$\ep({\vect{p}}_j)$ dependence
(see Eq.\ (\ref{eq:Dispers})) for nucleons.
Let protons ($j=2$) suffer Cooper pairing of type {\bf A}, while
neutrons ($j=1$) suffer pairing of types {\bf A},
{\bf B} or {\bf C}. In order to incorporate superfluidity
in Eq.\ (\ref{eq:Idur}) it is sufficient to
replace $x_j \to z_j$ (see Eq.\ (\ref{eq:DimLessVar})) and introduce
averaging over orientations of
${\vect{p}}_1$. Then the emissivity can be written as
\begin{equation}
    Q^{\dur} \! = \! Q^{\dur}_0 \, R^{\dur}, \;\;
    R^{\dur}(v_1,v_2) =
                  {I \over I_0} \!= \!
                    \int \frac{\dd \Om}{4\pi} \,J(y_1,y_2)  = \!
                    \int_0^{\pi/2} \!\! \dd \vartheta \, \sin \!\vartheta \,
                    J(y_1,y_2),
\label{eq:RdurcaDef}
\end{equation}
where $y_2 \equiv v_2$, $y_1$ is given by Eq.\ (\ref{eq:y_v}),
\begin{equation}
 J(y_1,y_2) \! = \!  { 1 \over I_0}
                \int_0^{+\infty}         \!\!\! \dd x_{\nu}\, x_{\nu}^3 \,
                \int_{-\infty}^{+\infty} \!\!\! \dd x_1 \,  f(z_1) \,
                \int_{-\infty}^{+\infty} \!\!\! \dd x_2 \,  f(z_2) \,
                \int_{-\infty}^{+\infty} \!\!\! \dd x_e \, f(x_e) \,
                 \delta (x_{\nu}\!-\! z_1\! -\! z_2\! -\!x_e),
\label{eq:JDef}
\end{equation}
d$\Omega$ is solid angle element along ${\vect{p}}_1$,
$\vartheta$ is an angle between ${\vect{p}}_1$ and the
quantization axis; $\; Q^{\dur}_0$  is the emissivity
(\ref{eq:Qdur0}) in non-superfluid matter; and
$R^{\dur}$ is the factor which describes superfluid suppression of
the reaction. One formally has
$R^{\dur} =1$ for normal nucleons, while
$R^{\dur} <1$ in the presence of superfluidity.
The integrals (\ref{eq:RdurcaDef}) and (\ref{eq:JDef})
have been calculated by
Levenfish and Yakovlev (1993, 1994b)
for different combinations of neutron and proton superfluidities.
The results are discussed below.

\subsection{Superfluidity of neutrons or protons}

Consideration of the cases of superfluid
neutrons or protons is similar. For instance,
let neutrons be superfluid. Then we can set $z_2=x_2$ in Eqs.\
(\ref{eq:RdurcaDef}) and (\ref{eq:JDef}), and
$R^{\dur}$ will depend on one argument
$v_1=v$ and on superfluidity type.
For $\tau=T/T_{c} \ge 1$, as mentioned above,
$R^{\dur}=1$. For a strong superfluidity ($\tau \ll 1$, $v \gg 1$)
the neutrino emission is greatly suppressed
($R^{\dur} \ll 1$).
The asymptotes of $R^{\dur}$ in the cases of strong superfluidity
of types {\bf A}, {\bf B} and {\bf C} read:
\begin{eqnarray}
\eqalign{
 \Ra \!\! = \!\! \frac{252}{457 \,\pi^6} \, \sqrt{\frac{\pi}{2}} \,
                v^{5.5}\, \exp(-v)
                =  \frac{0.0163}{\tau^{5.5}} \;
                            \exp \left(- \frac{1.764}{\tau} \right),
\cr   
 \Rb \!\! = \!\! \frac{126}{457\, \pi^5 \sqrt{3}} \, v^5 \exp(-v)
               = \frac{0.00123}{\tau^5} \,
                \exp \left(- \frac{1.188}{\tau}  \right),
\cr   
 \Rc \!\! = \!\! \frac{6029\, \pi^2}{5484 \, v^2}
              = 2.634 \, \tau^2.
}
\label{eq:RcAsy}
\end{eqnarray}
The asymptotes of $\Ra$ and $\Rb$ are seen to be exponential while
the asymptote of $\Rc$ is power-law. The latter circumstance is
associated with the nodes of the gap at
$\vartheta=0$ and $\pi$. Comparing with the results of Sect.\ 3.3,
we see that these asymptotes are similar to
the asymptotes of the nucleon heat capacity in the limit of strong
superfluidity.

\begin{figure}[t]                         
\begin{center}
\leavevmode
\epsfysize=8.5cm
\epsfbox[135 495 410 785]{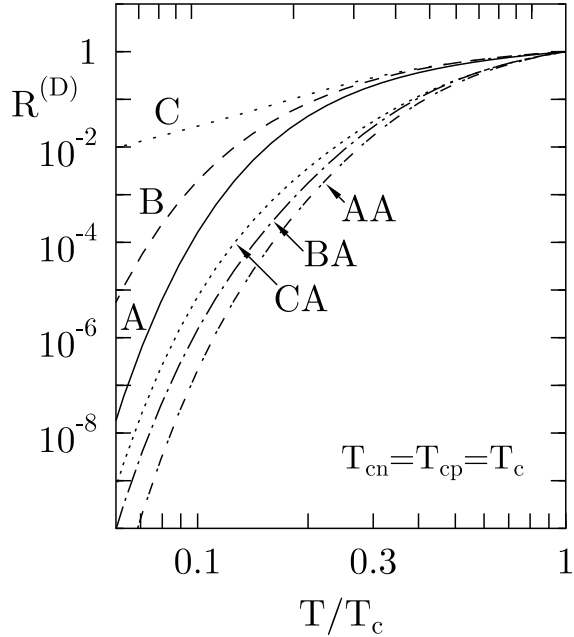}
\end{center}
\caption[]{\footnotesize
          Superfluid reduction factors of the direct Urca process
          versus $\tau=T/T_c$.
          Letters near the curves indicate superfluidity type
          (see Table 1): {\bf A}, {\bf B} and {\bf C} mean superfluidity
	  of neutrons or protons, while {\bf AA}, {\bf BA} and {\bf CA}
	  mean superfluidity of neutrons and protons.}
\label{fig:Rone}
\end{figure}

The asymptotes (\ref{eq:RcAsy}) and numeric values of the integrals
(\ref{eq:RdurcaDef}) and (\ref{eq:JDef})
for intermediate $v$ can be fitted by:
\begin{eqnarray}
\eqalign{
 \Ra \!\! = \!\! \left[ 0.2312 + \sqrt{ (0.7688)^2+(0.1438\,v)^2}
                   \right]^{5.5} \,
         \exp \! \left( 3.427 - \sqrt{ (3.427)^2+v^2 } \right), \hspace{1cm}
\cr   
 \Rb \!\! = \!\! \left[ 0.2546 + \sqrt{ (0.7454)^2+(0.1284\,v)^2}
                   \right]^5 \,
         \exp \! \left( 2.701 - \sqrt{ (2.701)^2+v^2 } \right),
\cr   
 \Rc \!\! = \!\! \frac{ 0.5 + (0.09226\,v)^2}{1 + (0.1821\, v)^2
                             + (0.16736\, v)^4} \,
       + \, \frac{1}{2} \,\exp \! \left( 1 - \sqrt{1+(0.4129\,v)^2} \right).
}
\label{eq:RcFit}
\end{eqnarray}
Equations (\ref{eq:RcFit}), with account for (\ref{eq:FitGaps}),
enable one to calculate $R^{\dur}$ for any $\tau$.
The results are presented in Fig.~\ref{fig:Rone}.

In simulations of NS cooling prior to publication
of the above results
(Levenfish and Yakovlev, 1993, 1994b)
one usually used
(e.g., Page and Applegate 1992)
the simplified factors of reduction of the direct Urca process
by the superfluidity of type {\bf A} or
{\bf B}. These factors were proposed by
Maxwell (1979)
from simple (and inadequate) consideration. Comparison
of the accurate and simplified factors shows that the latter
factors strongly overestimate the direct
Urca reduction. For instance, for
$T=0.1 \, T_c$ the accurate factor
$R_A^{\rm (D)}$ appears to be about four orders of magnitude
larger than the simplified one,
and the accurate factor $R_B^{\rm (D)}$ is more
than seven orders of magnitude larger. The difference has the same
nature as the difference of accurate and simplified
heat-capacity factors discussed in Sect.~3.3.

\subsection{Superfluidity of neutrons and protons}

Let us consider reduction of the direct Urca process by
combined actions of superfluidities of neutrons and protons of types
{\bf (AA)}, {\bf (BA)} and {\bf (CA)}.
In these cases, the reduction factor
$R^{\dur}$ depends on two arguments,
$v_1=v_n$ and $v_2=v_p$.

Calculation of $R^{\dur}$ in the presence of
the neutron and proton superfluidities
is complicated; let us consider the factor
$R^{\dur}_{AA}$, as an example.
According to Eqs.\ (\ref{eq:y_v}) and (\ref{eq:RdurcaDef}),
for the singlet-state pairing
$R^{\dur}_{AA}(v_1,v_2)$ = $ J(v_1,v_2) = J(v_2,v_1)$,
where $y_1=v_1$, $\, y_2 = v_2$. Clearly,
$R^{\dur}_{AA}(0,0)=1$. If both superfluidities are strong
($v_1 \gg 1$ and $v_2 \gg 1$) and
$v_2-v_1  \gg \sqrt{v_2}$,
the asymptote of the reduction factor is:
\begin{eqnarray}
 &&    R^{\dur}_{AA} \!\! = \!\! J(v_1,v_2) = {1 \over I_0}
              \left( \frac{\pi}{2} \, v_2 \right)^{1/2}
              \exp(-v_2) \, K,
\label{eq:RaaAsy} \\
 &&  K = \frac{s}{120} \,
      \left( 6v_2^4+83v_2^2v_1^2 +16v_1^4 \right)
       -  \frac{1}{8}\,v_2v_1^2
          \left( 3v_1^2+4v_2^2 \right)
          \ln \left( \frac{v_2 + s}{v_1} \right),
\label{eq:KaaAsy}
\end{eqnarray}
where $s \! = \! \sqrt{v_2^2 - v_1^2}$.
In the limit $v_1 \ll v_2$ Eq.\ (\ref{eq:KaaAsy}) gives
$K=v_2^5/20$, which corresponds to the asymptote
(\ref{eq:RcAsy}) of $R^{\dur}_{A}$.
In another limit
$ \sqrt{v_2} \ll v_2-v_1 \ll v_2$
we obtain $ K=(2/315) \, s^9/v_2^4$.
The asymptote (\ref{eq:KaaAsy}) fails
in the range $ \mid v_2-v_1 \mid \la \sqrt{v_2}$.
One can show
(Levenfish and Yakovlev, 1994b)
that $K \sim \sqrt{v_2}$ for $v_1=v_2$.

Let us present also the fit expression which reproduces the asymptote
(\ref{eq:RaaAsy}) and numeric values of
$R^{\dur}_{AA}$ calculated in a wide range of arguments
($\sqrt{v_1^2+v_2^2\,} \la 50$):
\begin{equation}
      R^{\dur}_{AA}= J(v_1,v_2) = \frac{u}{u+0.9163}\, S + D.
   \label{eq:RaaFit}
\end{equation}
In this case
\begin{eqnarray}
&&  S = {1 \over I_0} \;
         (K_0+K_1+0.42232\,K_2)
           \,\left( \frac{\pi}{2} \right)^{1/2}
        p_s^{1/4} \exp(- \sqrt{p_e}),
\nonumber  \\
&& K_0 = \frac{ \sqrt{p-q}}{120} \; (6p^2+83pq+16q^2) -
        \sqrt{p} \, \frac{q}{8}\,(4p+3q)\;
        \ln \left( \frac{ \sqrt{p}+ \sqrt{p-q}}{ \sqrt{q}} \right),
\nonumber  \\
&& K_1 = \frac{\pi^2 \sqrt{p-q}}{6} \, (p+2q) \; - \; \frac{\pi^2}{2} \, q
      \sqrt{p}\,
      \ln \left( \frac{ \sqrt{p}+ \sqrt{p-q}}{ \sqrt{q}} \right),
\nonumber  \\
&& K_2 = \; \frac{7 \, \pi^4}{60} \, \sqrt{p-q},
\nonumber \\
&& 2p  =  u+12.421 + \sqrt{w^2+16.350 \, u+45.171},
\nonumber \\
&& 2q  =  u+12.421 - \sqrt{w^2+16.350 \, u+45.171},
\nonumber \\
&& 2p_s  =  u + \sqrt{w^2+5524.8\,u+6.7737},
\nonumber \\
&&  2p_e  =  u + 0.43847 + \sqrt{w^2+8.3680\,u+491.32},
\nonumber \\
&&  D  =  1.52 \, \mbox{\rule{0cm}{0.8cm}} (u_1 u_2)^{3/2} \; (u_1^2+u_2^2)\,
    \exp(-u_1-u_2),
\nonumber \\
&&  u_1  =  1.8091 + \sqrt{v_1^2+(2.2476)^2},
\nonumber \\
&&  u_2 =  1.8091 + \sqrt{v_2^2+(2.2476)^2},
\label{eq:RaaFit_}
\end{eqnarray}
with $u = v_1^2+v_2^2$ and $w = v_2^2-v_1^2$.
For $v_2=0$, the factor $R^{\dur}_{AA} (v_1,0)$ agrees with
the factor $R^{\dur}_A (v_1)$ given by Eq.\ (\ref{eq:RcFit}).

\begin{figure}[t]                         
\begin{center}
\leavevmode
\epsfysize=17.5cm
\epsfbox[40 70 310 775]{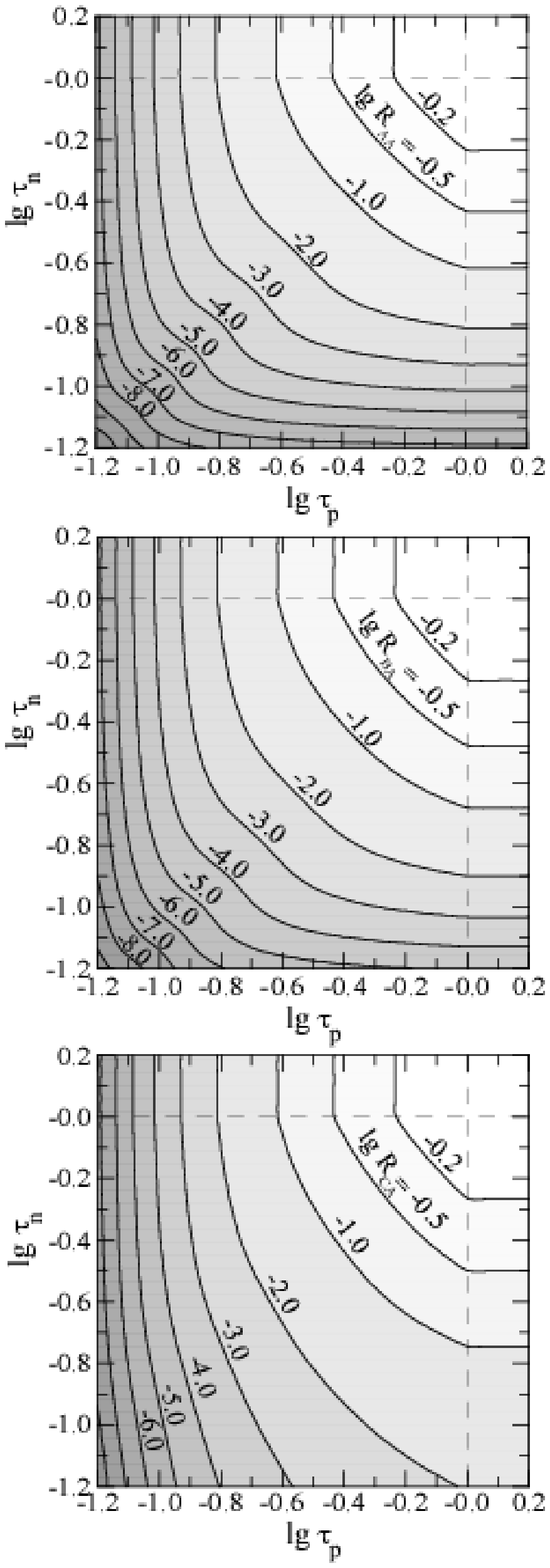}
\end{center}
\caption[]{\footnotesize
          Lines of constant reduction factors
          $R^{\rm (D)}$ for the direct Urca
          in the presence of neutron and proton superfluidities
          {\bf (AA)}, {\bf (BA)} and {\bf (CA)}.
          The curves are labeled by the values $\lg R^{\rm (D)}$.
          In the region $\tau_n \ge 1$ and $\tau_p \ge 1$
          neutrons and protons are normal and $R^{\rm (D)}=1$.
          In the region $\tau_n < 1$, $\tau_p \ge 1$,
          where only neutrons are superfluid, and in the region
          $\tau_n \ge 1$, $\tau_p < 1$, where only protons are
          superfluid,
          $R^{\rm (D)}$ depends on the only parameter
          (Sect.\ 4.3). In the region
          $\tau_n < 1,\; \tau_p < 1$ both, neutron and protons,
          are superfluid at once.
          }
\label{fig:Rboth}
\end{figure}

Figure \ref{fig:Rboth} shows lines of
$R^{\dur}_{AA}\!=\,$const versus
$\tau_1=T/T_{c1}$ and $ \tau_2=T/T_{c2}$.
The lines exhibit superfluid reduction of the direct Urca
for any $T$, $T_{c1}$ and $T_{c2}$.
The behaviour of $R^{\dur}_{AA}$ at $\tau_1^2+ \tau_2^2 \ll 1$
(both superfluidities are strong) is of special interest.
In this case one can obtain an approximate relationship
(Lattimer et al., 1991;
Levenfish and Yakovlev, 1994b)
\begin{equation}
      R^{\dur}_{12} \sim
           \min\left( R^{\dur}_1 \! , \, R^{\dur}_2 \right)\, ,
\label{eq:Estimation}
\end{equation}
where $R^{\dur}_1$ and $R^{\dur}_2$ are the reduction factors
for superfluidities of one type.
According to (\ref{eq:Estimation}), the factor
$R^{\dur}_{12}$ is mainly determined by a stronger
superfluidity (1 or 2). The presence of the second,
weaker superfluidity reduces naturally
$R^{\dur}_{12}$, but not to a great extent.
This is confirmed by the asymptote
(\ref{eq:RaaAsy}). A transition from the regime in which
$R^{\dur}_{12} \sim R^{\dur}_1$ to the regime in which
$R^{\dur}_{12} \sim R^{\dur}_2$ takes place in the region $v_2 \sim v_1$,
where the asymptote $R^{\dur}_{12}$ requires special consideration.

Using Eq.\ (\ref{eq:RdurcaDef}), it is not difficult to evaluate
$R^{\dur}$ for cases {\bf (BA)} and {\bf (CA)}, in which
the protons suffer singlet-state pairing
while the neutrons suffer triplet-state pairing.
The evaluation is reduced to one-dimensional integration in Eq.\
(\ref{eq:RdurcaDef}) of the quantity
$J(y_1,y_2)$, fitted by Eqs.\ (\ref{eq:RaaFit})
and (\ref{eq:RaaFit_}).
The results of calculations of
$R^{\dur}_{BA}$ and $R^{\dur}_{CA}$
for any $T$, $T_{c1}$ and $T_{c2}$
are shown in Fig.~\ref{fig:Rboth}.
The dependence of $R^{\dur}_{BA} $ and
$R^{\dur}_{CA} $ on $\tau_1$ and $\tau_2$ is similar to the
dependence of $R^{\dur}_{AA}$, shown in the same figure,
but now $R^{\dur} (\tau_1,\tau_2) \neq R^{\dur} (\tau_2,\tau_1)$.
The approximate expression (\ref{eq:Estimation})
in these cases is valid as well as confirmed by
the asymptotes
(Levenfish and Yakovlev, 1994b).
Since the superfluidity of type
{\bf C} reduces the neutrino emission much weaker
than that of type {\bf A} or {\bf B};
the transition from one dominating superfluidity to the other
in case {\bf (CA)} for $v_1 \gg 1$ takes place at
$v_2 \sim \ln v_1 $. Moreover, for
$v_1 \ga v_2 \gg 1$ the factor
$R^{\dur}_{CA}$ appears to be much larger than $R^{\dur}_{AA}$
or $R^{\dur}_{BA}$ (Fig.\ \ref{fig:Rone}).

For not very strong superfluidities {\bf (BA)}
($\sqrt{v_1^2 + v_2^2\,} \la 5 $ or
$[(1-\tau_2)/0.65]^4+$ $[(1-\tau_1)/0.76]^4 $ $\la 1$),
the reduction factor is fitted by
%
\begin{equation}
R^{\dur}_{BA} = \frac{10^4-2.839 \, v_2^4 - 5.022
     \, v_1^4}{ 10^4 + 757.0 \, v_2^2
    + 1494 \, v_1^2 + 211.1 \, v_2^2v_1^2 + 0.4832\,v_2^4v_1^4}.
                                                           \label{eq:RabFit}
\end{equation}
In the case of not too strong superfluidities
({\bf CA}) ($\sqrt{v_1^2 + v_2^2\,} \la 10$ or
$[(1-\tau_2)/0.825]^6+[(1-\tau_1)/0.8]^6 \la 1$)
$R^{\dur}_{CA}$ is fitted by the expression:
%
\begin{eqnarray}
  R^{\dur}_{CA} &=& 10^4 \times \left( 10^4 + 793.9\,v_2^2 + 457.3 \, v_1^2
       + 66.07 \, v_2^2v_1^2 +2.093 \, v_1^4 + \right. \nonumber \\
       &+& \left. 0.3112 \, v_2^6 + 1.068 \, v_2^4v_1^2 +
     0.01536 \, v_2^4v_1^4 +0.006312 \, v_2^6v_1^2 \right)^{-1} \, .
\label{eq:RacFit}
\end{eqnarray}
If $v_1=0$ or $v_2=0$, the above fits agree with Eqs.\ (\ref{eq:RcFit}).
The tables of $R^{\dur}_{BA}$
and $R^{\dur}_{CA}$, as well as the asymptotes of these factors
in the limit of strong superfluidity are given in
Levenfish and Yakovlev (1994b).

Before publication of the above results
(Levenfish and Yakovlev, 1993, 1994b),
as far as we know, simulation of the enhanced NS cooling
with account of the superfluidity of neutrons and protons
was carried out in the only article by
Van Riper and Lattimer (1993).
The authors assumed the simplified reduction factor
$R^{\dur}_{12} = R^{\dur}_1 R^{\dur}_2$.
The actual factor, Eq.\ (\ref{eq:Estimation}), is quite different,
i.e., the simplified factor strongly overestimates the
effect of superfluidity of neutrons and protons.

As shown by Prakash et al.\ (1992), practically all
equations of state in the inner NS cores,
which predict appearance of hyperons, open
the direct Urca reactions involving hyperons:
\begin{equation}
         B_1 \to B_2 + l + \bar{\nu} \, , \quad
         B_2 + l \to B_1 + \nu \, ,
\label{eq:DurcaHyperon}
\end{equation}
where $B_1$ and $B_2$ are baryons (nucleons, $\Lambda$-, $\Sigma$- or
$\Xi$-hyperons, $\Del^-$-resonance),
$l$ is a lepton (electron or $\mu$-meson).
The hyperonic processes (\ref{eq:DurcaHyperon}) do not conserve
strangeness; the neutrino emissivities
in these processes are somewhat lower than in the nucleon direct Urca
process (\ref{eq:Durca_Nucleon}).
The emissivities are given by the expressions similar to
(\ref{eq:Qdur0}), but with different numerical coefficients
(Prakash et al., 1992).

Hyperons, like nucleons, may be in a superfluid state
(Balberg and Barnea, 1998).
Their superfluidity is most likely
to be of singlet-state type due to their relatively small
number density. The reduction factors for
the direct Urca processes involving hyperons are
similar to those for the nucleon direct Urca.

%
%

\section{Modified Urca process and neutrino bremsstrahlung 
         due to nucleon-nucleon scattering}

\subsection{Two branches of the modified Urca process}

In this section we will analyse the {\it standard}
neutrino energy loss rates in the processes
(\ref{eq:Murca_N})--(\ref{eq:Brems})
(Sect.\ 2.2). In the absence of superfluidity, these processes
were studied by
Bahcall and Wolf (1965a, b),
Flowers et al.\ (1975),
Friman and Maxwell (1979),
as well as by Maxwell (1987). The latter author considered also
the processes involving hyperons. The most detailed article
seems to be that
by Friman and Maxwell (1979),
where, however, the proton branch of the modified Urca-process
was neglected. The proton branch was analysed and
the superfluid reduction factors for the standard process
were calculated by
Yakovlev and Levenfish (1995).
Below in will mainly follow consideration of these authors.

The main standard neutrino energy loss mechanism
in a non-superfluid matter is the modified Urca process.
It will be labeled by upperscripts (M{\it N\/}),
where $N=n$ indicates the neutron branch
(\ref{eq:Murca_N}) of the process, while
$N=p$ indicates the proton branch (\ref{eq:Murca_P}).
Both branches consist of direct and inverse reactions
and are described by similar Feynman diagrams.
If beta-equilibrium is established, the rates of the
direct and inverse branches are equal; it is sufficient to
calculate the rate of any reaction and double the result.
The general expressions for the neutron and proton
branches of the modified Urca process can be written as
($\hbar = c = \kB = 1$):
\begin{equation}
    Q^{\MN} \! =  2  \int
              \left[ \prod_{j=1}^4 { \dd^3 p_j \over (2 \pi)^3} \right]
              { \dd^3 p_e \over 2 \ep_e (2 \pi)^3} \;
              { \dd^3 p_\nu \over 2 \ep_\nu (2 \pi)^3} \;  \ep_\nu \,
       (2 \pi)^4 \del(E_f-E_i)\del({\vect{P}}_f - {\vect{P}}_i ) \,
               { {\cal L}\over 2} \, \sum_{\rm spins} | M |^2 ,
  \label{eq:Q_mur}
\end{equation}
where ${\vect{p}}_j$ is a nucleon momentum
($j=1$,~2,~3,~4), ${\vect{p}}_e$ and
$\ep_e$ are, respectively, the electron momentum and energy; while
${\vect{p}}_\nu$ and $\ep_\nu$ are,
respectively, the neutrino momentum and energy.
The delta function $ \del(E_f-E_i) $ describes energy conservation,
and $\del( {\vect{P}}_f - {\vect{P}}_i )$ describes momentum conservation;
subscripts $i$ and $f$ refer to the initial and final particle
states, respectively. Furthermore,
${\cal L}$ means product of the Fermi--Dirac functions or corresponding
blocking factors for nucleons and electron;
$| M |^2$ is the squared reaction amplitude, the sum is over all
particle spins. The factor
2 in the denominator before the summation sign is introduced
to avoid double counting of the same collisions of identical particles.

Introducing dimensionless variables,
the emissivity (\ref{eq:Q_mur}) can be written
(Yakovlev and Levenfish, 1995) in the form
similar to the direct Urca emissivity
(see Eq.\ (\ref{eq:DecompDur}) and explanations afterwards):
\begin{eqnarray}
&&    Q^{\MN}  =  {1 \over 4 \, (2 \pi)^{14} }\; T^8
            AIS \sum_{\mbox{\scriptsize spins}} | M |^2 ,
\label{eq:DecompMur} \\
&&   A = 4 \pi \;  \left[ \prod_{j=1}^5 \int \dd \Om_j \; \right]
            \del \left( \sum_{j=1}^5 {\vect{p}}_j  \right),
\label{eq:A} \\
&&   I = \int_0^\infty \dd x_\nu \; x_\nu^3
       \left[ \prod_{j=1}^5 \int_{-\infty}^{+\infty}
       \dd x_j \; f_j \right]
       \del \left( \sum_{j=1}^5 x_j-x_\nu \right),
\label{eq:I} \\
&&   S = \prod_{j=1}^5 \pF_{\!j} m_j^\ast.
\label{eq:S}
\end{eqnarray}
The quantity
$A$ contains integrals over orientations of particle momenta
($j$=5 corresponds to electron);
all lengths of the momenta ${\vect{p}}_j$ of nucleons and electron
in the delta function should be set equal to the Fermi momenta.
Typical neutrino momentum $p_\nu \sim T$ is much smaller
than the momenta $\pF$ of other particles. Thus, it can be neglected
in momentum conservation; the integration over orientations of the
neutrino momentum in $A$ immediately yields $4 \pi$.
The quantity $I$, given by Eq.\ (\ref{eq:I}),
includes integrals over dimensionless energies of neutrino
$x_\nu = p_{\nu}/T =\ep_\nu /T$
and other particles $x_j = \vF_{\!j} (p-\pF_{\!j})/T$
(see Eq.\ (\ref{eq:DimLessVar}));
$\; f_j= [ \exp(x_j)+1]^{-1}$. Finally, the quantity
$S$ is the product of density of states of the
particle species $1 \le j \le 5$ (with effective masses
$m_j^\ast$) at the Fermi surfaces.

Owing to the similarity of the Feynman diagrams
for the neutron and proton branches of the
modified Urca process,
(\ref{eq:Murca_N}) and (\ref{eq:Murca_P}),
the reaction amplitudes
$M$ and phase integrals $I$ are practically the same. The reactions
differ by the products of the density of states
$S$ and by the angular integrals $A$.

In a non-superfluid matter, one has $I=I_0= 11513 \, \pi^8\,/ 120960$
(see, e.g.,
Shapiro and Teukolsky, 1983).
The amplitude $M$ for the neutron branch of the
modified Urca process was calculated by
Friman and Maxwell (1979)
using the Weinberg--Salam--Glashaw theory of electroweak interaction.
Long-range (small momentum transfer) part of the nucleon--nucleon interaction
was described in the one--pion--exchange approach,
while the short-range (large momentum transfer) part
was described in the frame of the Landau Fermi--liquid theory
(e.g.,
Baym and Pethick, 1991).
The squared amplitude $| M |^2 $ summed over spin states
is given by Eq.\ (39) in
Friman and Maxwell (1979).

The angular integrals (\ref{eq:A})
for the processes
(\ref{eq:Murca_N}) and (\ref{eq:Murca_P}) differ due to the
difference of neutron and proton Fermi momenta
(Sect.\ 2.1). The proton reaction branch involves three protons
with rather small Fermi momentum and the only neutron with
larger momentum, while the neutron branch involves three
neutrons with large momentum and the only proton with small momentum.
Direct calculation of the angular integrals for these processes
in the absence of superfluidity gives
(Shapiro and Teukolsky, 1983;
Yakovlev and Levenfish, 1995):
\begin{equation}
     A_{n0} = {2 \pi \, (4 \pi)^4 \over \pFn^3} \, , \;\;\;\;\;\;
     A_{p0}  =  {2 \pi \, (4 \pi)^4 \over \pFp^2 \, \pFn}
         \left( 1 -{\pFe \over 4 \pFp} \right)  \Theta,
\label{eq:Anp}
\end{equation}
where $\Theta=1$ if the proton branch is allowed by momentum
conservation, and
$\Theta=0$ otherwise. Let us remind that
in the outer NS core the Fermi momenta
$\pFe$ and $\pFp$ are much smaller than
$\pFn$ (Sect.\ 2.1).
Thus $\Theta = 1$ for
$\pFn <$ $3 \pFp + \pFe$. Notice that the expression for
$A_{n0}$ is obtained from Eq.\
(\ref{eq:A}) in case $\pFn > \pFe + \pFp$. In the opposite case of
$\pFn \leq \pFe + \pFp$,
$A_{n0}$ is given by Eq.\ (13)
in Yakovlev and Levenfish (1995),
but in this case the direct Urca process dominates and the
modified Urca processes are insignificant.

Using the above expressions for $I_0$ and
$A_{n0}$,
Friman and Maxwell (1979)
calculated the neutrino emissivity in the neutron
branch of the modified Urca process
(in the standard physical units):
\begin{eqnarray}
     Q^{\Mn}_0 & = & {11513 \over 30240} \,
               {G_{\rm F}^2 g_A^2 m_n^{\ast 3} m_p^\ast \over 2 \pi}
               \left( {f^\pi \over m_\pi } \right)^4
               {\pFe (\kB T)^8 \over \hbar^{10} c^8} \,
               \alpha_n \beta_n
      \nonumber \\
      & \approx & 8.55 \times 10^{21}
                \left( {m_n^\ast \over m_n } \right)^3
                \left( {m_p^\ast \over m_p } \right)
                \left( {n_e  \over n_0 } \right)^{1/3} \!
                T_9^8 \, \alpha_n \beta_n \; \; \;        \rate,
    \label{eq:Qn0}
\end{eqnarray}
where
$g_A = 1.26$ is the axial--vector constant of weak hadron current,
$f^\pi \approx 1$ is the
$\pi N$-interaction constant in the $p$-state
in the one--pion--exchange approximation, and
$m_\pi$ is the pion mass ($\pi^\pm$). The factor
$\alpha_n$ describes momentum dependence of the squared reaction amplitude
in the Born approximation, and
$\beta_n$ contains different corrections.
According to Eq.\ (62)
in Friman and Maxwell (1979)
$\alpha_n \approx 1.76 - 0.63 \, (n_0 /n_n)^{2/3}$, where $n_n$ is
the number density of neutrons.
In their final Eq.\ (65c) for $Q^{\Mn}_0$
Friman and Maxwell (1979)
used the value $\alpha_n = 1.13$, calculated for $\rho = \rho_0$,
and set $\beta_n$ = 0.68 (to account for the correlation effects).

The expression for the neutrino emissivity in the
proton reaction was derived by
Yakovlev and Levenfish (1995):
\begin{eqnarray}
     Q^{\Mp}_0 & = & {11513 \over 30240} \,
               {G_{\rm F}^2 g_A^2 m_p^{\ast 3} m_n^\ast \over 2 \pi}
               \left( { f^\pi \over m_\pi } \right)^4
               {\pFe (\kB T)^8 \over \hbar^{10} c^8}
               \alpha_p \beta_p
               \left( 1 - {\pFe \over 4\pFp} \right) \Theta
      \nonumber \\
      &\approx&  8.53 \times 10^{21}
                \left( {m_p^\ast \over m_p } \right)^3
                \left( {m_n^\ast \over m_n } \right)
                \left( {n_e  \over n_0 } \right)^{1/3}
                T_9^8 \alpha_p \beta_p
                \left(1 - {\pFe \over 4 \pFp} \right) \Theta \;\;\;
                \rate. \hspace{1.0cm}
    \label{eq:Qp0}
\end{eqnarray}
Taking into account all uncertainties concerned with the
calculation of the reaction amplitude
by Friman and Maxwell (1979)
we set $\alpha_p= \alpha_n$ and $\beta_p = \beta_n$.
It is easy to see that the emissivities
in the neutron (\ref{eq:Qn0}) and proton
(\ref{eq:Qp0}) branches of the process are similar.
The main difference of the proton branch is in its threshold
character: the reaction is allowed for
$\pFn < 3\pFp + \pFe$. In $npe$-matter, this inequality is
equivalent to $\pFn < 4\pFe$, i.e., to $n_e > n_n/64$.
The latter condition is realized almost everywhere in the
NS core. It can be violated only for ultra-soft equations
of state at $\rho \la \rho_0$. For these equations of state,
the proton branch can be forbidden in the outermost
part of the NS core. Similar but much more stringent
threshold conditions can be formulated for the direct Urca process
(\ref{eq:Durca_Nucleon}) (Sects.\ 2.2 and 4).

Comparing $Q^{\Mp}_0$ and $Q^{\Mn}_0$, we find:
$ Q^{\Mp}_0 / Q^{\Mn}_0  = ( m_p^\ast / m_n^\ast )^2
 [ 1 - \pFe / (4 \pFp)]$.
For instance, at  $m_n^\ast= m_p^\ast$ and $\pFe \!= \pFp$ we have
$Q^{\Mp}_0 = 0.75 \, Q^{\Mn}_0$.
Therefore the proton branch of the process is nearly as efficient
as the neutron branch.

The potential efficiency of the proton branch was outlined by
Itoh and Tsuneto (1972) who, however, did not calculate $Q^{\Mp}_0$.
Later the neutrino emissivity $Q^{\Mp}_0$ was calculated by
Maxwell (1987)
who found it negligibly small as compared to $Q^{\Mn}_0$.
This conclusion is erroneous due to several inaccuracies made by
Maxwell (1987)
(and analysed by
Yakovlev and Levenfish, 1995).
In particular, Maxwell incorrectly neglected the electron momentum
in momentum conservation.

%
  \subsection{Modified Urca processes in superfluid matter}

Nucleon superfluidity reduces the modified Urca processes
(see Sect.\ 4.2). Let us analyse this reduction.
We will adopt the traditional assumption
(Sect.\ 3.1) that the proton superfluidity is of type {\bf A},
while the neutron superfluidity is of type {\bf B}.
Our results will also be valid for the neutron pairing of type {\bf A}
(at $\rho \la \rho_0$), as will be mentioned later.
Reduction of the modified Urca by the neutron pairing of type
{\bf C} has not been considered so far.

Superfluidity affects the nucleon dispersion relations
under the integral (\ref{eq:Q_mur})
in accordance with Eq.\ (\ref{eq:Dispers}). The neutrino emissivity
of the neutron and proton branches of the process can be presented
in the form
\begin{equation}
    Q^{\Mn}\! = Q^{\Mn}_0 R^{\Mn}, \;\;\;\; Q^{\Mp}\! = Q^{\Mp}_0 R^{\Mp} \, ,
\label{MurcaReduc}
\end{equation}
where $Q^{\Mn}_0$ and $Q^{\Mp}_0$ are the emissivities
(\ref{eq:Qn0}) and (\ref{eq:Qp0}) in non-superfluid matter,
while $R^{\Mn}$ and $R^{\Mp}$ are the factors, which describe
superfluid reduction of the reactions
($R^{\MN} \le 1$). Generally,
$R^{\MN} = J_N / (I_{N0} A_{N0})$, where
\begin{equation}
   J_N   =  4\pi \int \prod_{j=1}^5 \dd \Om_j \;
          \int_0^{\infty} \dd x_\nu \, x_\nu^3
          \left[ \prod_{j=1}^5 \int_{-\infty}^{+\infty}
          \dd x_j \, f(z_j) \right]
         \del \left( x_\nu - \sum_{j=1}^5 z_j \right)
          \del \left( \sum_{j=1}^5 {\vect{p}}_j \right),
\label{eq:JN}
\end{equation}
$\;f(z)=1/({\rm e}^z+1)$;
$z_j$ for nucleons ($j \leq 4$)
is defined by Eq.\ (\ref{eq:DimLessVar});
one should set $z_5 =x_5 = x_e$ for an electron.

Equation (\ref{eq:JN}) enables one to calculate the reduction factors
$R^{\Mn}$ and $R^{\Mp}$ as a function of $T$, $\tn$ and $\tp$.
Below we present the results
(Yakovlev and Levenfish, 1995)
for the proton superfluidity of type {\bf A} and
normal neutrons as well as for
the neutron superfluidity of type {\bf B} and normal protons.
The behaviour of $R^{\Mn}$ and $R^{\Mp}$
in the presence of the joint superfluidity of nucleons
is described in Sect.\ 5.4.\\[0.3ex]

{\bf Singlet-state proton pairing.} 
Since the singlet-state gap is isotropic the angular
and energy integrations in Eq.\ (\ref{eq:JN})
are decomposed, and the angular integration remains the same
as in non-superfluid matter.

Just as in the case of the direct Urca,
the reduction factors of the neutron
and proton branches of the modified Urca process,
$R_{pA}^{\Mn}$ and $R_{pA}^{\Mp}$, can be expressed in terms of
the dimensionless energy gap
$v_p= v_A$ (see Eq.\ (\ref{eq:FitGaps}); here and hereafter
the subscripts in
$R^{\MN}$ indicate the superfluid particle species and
superfluidity type). Clearly,
$R_{pA}^{\Mn}\! = \! 1 $  and
$R_{pA}^{\Mp}\! = \! 1 $  for $T\! \ge \! \tp$ ($\va\! = \! 0$).
In the case of strong superfluidity
($T \! \ll \! \tp$, $\va \! \to \! \infty$),
the asymptotes of the both factors are
(Yakovlev and Levenfish, 1995)
\begin{eqnarray}
 R_{pA}^{\Mn} &=&  {72 \sqrt{2 \pi} \over 11513 \pi^8} \,
                               \va^{7.5} \exp (-\va)
           =  {1.166  \times 10^{-4} \over \tau_p^{7.5}}
            \exp \left(- {1.764 \over \tau_p} \right) \, ,
\label{eq:Rn1asy} \\
    R_{pA}^{\Mp} &=& {120960 \over 11513 \pi^8} \, \xi \va^7 \exp(-2\va)
    = {0.00764 \over \tau_p^7}
    \exp \left( - {3.528 \over \tau_p} \right),
\label{eq:Rp1asy} \\
     \xi &=& {3 \pi \over 320} \left[ 339 \sqrt{3} -
    {885 \over 2} \ln ( \sqrt{3} +2) \right] \approx 0.130 \, ,
\label{eq:xi}
\end{eqnarray}
where $\tau_p=T/\tp$.

Yakovlev and Levenfish (1995) calculated
$R_{pA}^{\Mn}$ and $R_{pA}^{\Mp}$
for intermediate $\va$ and fitted them by the analytic expressions,
which reproduced also the asymptotes
in the limit of $v \to \infty$ and obeyed the condition
$R^{\MN}(0)=1$:
\begin{eqnarray}
    R_{pA}^{\Mn} \! &=& \! { a^{7.5} + b^{5.5} \over 2}
            \exp \left( 3.4370 - \sqrt{ (3.4370)^2 + \va^2} \, \right),
\label{eq:Rn1fit} \\
       a \! &=& \! 0.1477 + \sqrt{ (0.8523)^2 + (0.1175\va)^2}, \;\;\;
     b = 0.1477 + \sqrt{ (0.8523)^2 + (0.1297\va)^2}; \hspace{0.5cm}
\nonumber \\
   R_{pA}^{\Mp} \! &=& \!
           \left[0.2414 \! + \!
           \sqrt{ (0.7586)^2 \! + \! (0.1318\va)^2} \right]^7
             \exp \left( 5.339 \! - \! \sqrt{ (5.339)^2 + (2\va)^2} \, \right) .
\label{eq:Rp1fit}
\end{eqnarray}
Equations (\ref{eq:Rn1fit}) and
(\ref{eq:Rp1fit}), together with (\ref{eq:FitGaps}),
fully determine the dependence of
$R_{pA}^{\Mn}$ and $R_{pA}^{\Mp}$ on $\tau_p$.

The above results
(Yakovlev and Levenfish, 1995)
are valid also for the singlet-state superfluidity of neutrons.
Evidently, in that case one should set
$v_n = v_A $ and
\begin{equation}
     R_{nA}^{\Mp}(\va)=R_{pA}^{\Mn}(\va), \quad
     R_{nA}^{\Mn}(\va)=R_{pA}^{\Mp}(\va).
\label{NewR}
\end{equation}
Then Eq.\ (\ref{eq:Rn1fit}) describes reduction of the
proton branch of the process and Eq.\
(\ref{eq:Rp1fit}) describes reduction of the neutron branch.

Wolf (1966) (Fig.\ 2 of his article)
as well as Itoh and Tsuneto (1972) were the first
who considered the reduction factor
$R_{nA}^{\Mn}$ of the neutron branch of the modified Urca process
by the singlet-state neutron superfluidity; this factor is analogous
to the factor $R_{pA}^{\Mp}$.
Notice that Itoh and Tsuneto (1972) analysed only the asymptote
(\ref{eq:Rp1asy}).
In both articles the same asymptote
(\ref{eq:Rp1asy}) was obtained but with different
numerical factors $\xi$.
Wolf (1966) got $\xi =0.123$, while
Itoh and Tsuneto (1972)
obtained $\xi =\pi /15 \approx 0.209$.
Recently $R_{nA}^{\Mn}$ has been calculated independently by
Pizzochero (1998)
under artificial assumption that the superfluid gap
is temperature independent. His results are described
by the factor $R_{nA}^{\Mn}(\va)$ presented above
with $\va=1.764/\tau$.
\\[0.3ex]

{\bf Triplet-state neutron pairing.} 
In this case the neutron gap is anisotropic.
The proton branch of the modified Urca process is
analysed easily since the only one superfluid particle is involved.
The expression for the reduction factor $R_{nB}^{\Mp}$
reduces to a one-dimensional integral over angle
$\vartheta_n$
(between neutron momentum and quantization axis) of the factor
$R_{pA}^{\Mn}(v)$ fitted by the expression (\ref{eq:Rn1fit}).
An argument $v_A$ in the latter expression should be formally
replaced by $y_B$ in accordance with Eq.\ (\ref{eq:y_v})
(Sect.\ 3.2).
It is evident that $ R_{nB}^{\Mp}(v)=1$ for $v=\vb=0$.
In the limit $\vb \to \infty$ one can use the asymptote
(\ref{eq:Rn1asy}) in the integrand.
Then for $T \ll T_{cn}$ ($\vb \gg 1$) 
according to Yakovlev and Levenfish (1995)
\begin{equation}
  R_{nB}^{\Mp} = {72 \over 11513 \, \pi^7 \sqrt{3} }\, \vb^{7} \exp(-\vb) =
      {3.99 \times 10^{-6} \over \tau_n^7}
      \exp \left(- {1.188 \over \tau_n} \right) \, ,
\label{eq:Rp2asy}
\end{equation}
where $\tau_n=T/T_{cn}$.
Yakovlev and Levenfish (1995)
calculated also $R_{nB}^{\Mp}$ for intermediate values of $\vb$
and fitted the results by the analytic expression:
\begin{eqnarray}
\eqalign{
   R_{nB}^{\Mp}  =  { a^7 + b^5 \over 2} \exp \left(2.398 -
                \sqrt{ (2.398)^2 +\vb^2} \, \right),
\cr
 a = 0.1612 + \sqrt{ (0.8388)^2 + (0.1117 \vb)^2},
   \;\;\;\; b = 0.1612 + \sqrt{ (0.8388)^2 + (0.1274 \vb)^2}.
}
\label{eq:Rp2fit}
\end{eqnarray}
%

\begin{figure}[t]                         
\begin{center}
\leavevmode
\epsfysize=8.5cm
\epsfbox[40 30 250 280]{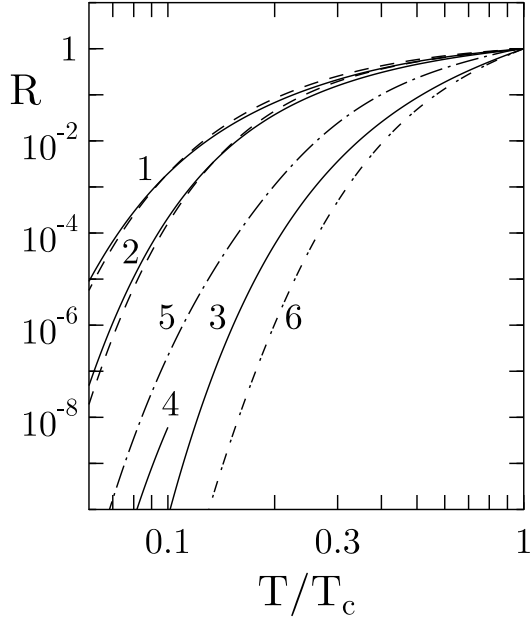}
\end{center}
\caption[]{\footnotesize
         Reduction factors of various neutrino emission
         processes by superfluidity of neutrons or protons versus $T/T_c$.
         Curves {\bf 1} show reduction of the
         $p$-branch of the modified Urca process (solid line)
         and the direct Urca process (dashed line)
         by neutron superfluidity of type {\bf B}.
         Curves {\bf 2} correspond to reduction of the
         $n$-branch of the modified Urca (solid line)
         and the direct Urca (dashed line) by proton
         superfluidity of type {\bf A}.
         Dot-and-dash lines {\bf 5}, {\bf 6} and the solid line
         {\bf 3} refer to the $np$, $pp$-scattering and $p$-branch
         of the modified Urca processes, respectively,
         for the same superfluidity. Solid line
         {\bf 4} is the asymptote of the reduction factor for
         the $n$-branch of the modified Urca process
         due to $n$ superfluidity of type {\bf B}.
}
\label{fig:Rstandard_tau}
\end{figure}

Exact calculation of the reduction factor $R_{nB}^{\Mn}$
of the neutron branch of the modified Urca process
by the triplet-state neutron superfluidity
for intermediate values of $\vb$ is complicated;
an approximate expression will be given in Sect.\ 5.4.
Here we present only the asymptote of $R_{nB}^{\Mn}$
in the limit $\tau_n \ll 1$
(Yakovlev and Levenfish, 1995)
\begin{equation}
  R_{nB}^{\Mn} \! = \! {120960 \over 11513 \pi^8} \times {2 \over 3 \sqrt{3}} \;
    \xi  \, \vb^6 \exp(-2\vb)
    =  {1.56 \times 10^{-4} \over \tau_n^6}
    \exp \left( - {2.376 \over \tau_n} \, \right).
\label{eq:Rn2asy}
\end{equation}
In this case, as for the proton reaction involving superfluid protons
(\ref{eq:Rp1asy}), the effect of superfluidity appears to be
very strong: the exponent argument in
$R_{nB}^{\Mn}$ contains the doubled gap
(three neutrons which participate in the reaction
belong to superfluid component of matter).

\begin{figure}[t]                         
\begin{center}
\leavevmode
\epsfysize=8.5cm
\epsfbox[50 35 255 275]{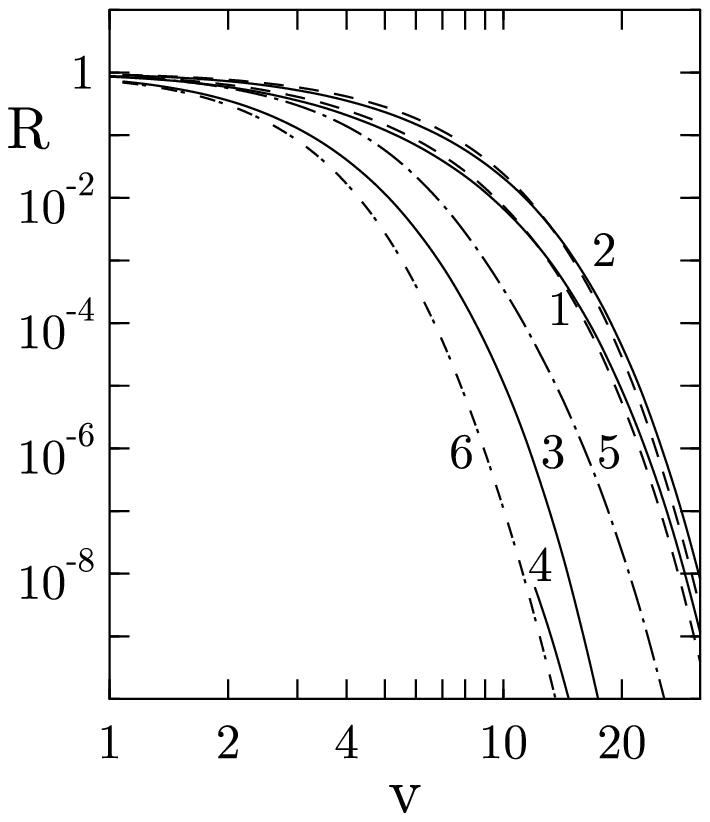}
\end{center}
\caption[]{\footnotesize
           Same as in Fig.~\protect{\ref{fig:Rstandard_tau}}, but
           versus dimensionless gap parameter.}
\label{fig:Rstandard_v}
\end{figure}

The dependence of the reduction factors
$R_{pA}^{\Mn}$, $R_{pA}^{\Mp}$, $R_{nB}^{\Mp}$, and
$R_{nB}^{\Mn}$ on dimensionless temperature
$\tau$ and dimensionless gap parameter $v$
is plotted in Figs.\ \ref{fig:Rstandard_tau} and \ref{fig:Rstandard_v}.
For comparison, we present also the reduction factors
of the direct Urca process, $\Ra$ and $\Rb$.
Let us mention that in the case of the strong
neutron superfluidity ($T \ll \tn$) and normal protons
the proton branch of the modified Urca process
becomes much more efficient than the neutron branch.
However in this case the main neutrino emission comes from
the neutrino bremsstrahlung due to $pp$-scattering,
which is not affected by the neutron superfluidity.

%
\subsection{Neutrino bremsstrahlung due to nucleon-nucleon
            scattering in superfluid matter}

Let us consider the effect of superfluidity on generation
of neutrino pairs in
$nn$, $np$ and $pp$-scattering (\ref{eq:Brems}); these processes are often
referred to as
{\it neutrino nucleon-nucleon bremsstrahlung}.
In a non-superfluid matter, the neutrino bremsstrahlung is about two
orders of magnitude less efficient than the modified
Urca process. Corresponding neutrino emissivities
(in the standard physical units) are
(Friman and Maxwell, 1979;
Yakovlev and Levenfish, 1995)
\begin{eqnarray}
   Q^{\,(nn)}_0 & = & {41 \over 14175} \, {G_F^2 g_A^2 m_n^{\ast4} \over
             2 \pi \hbar^{10} c^8} \left( {f^\pi \over m_\pi} \right)^4
             \pFn \alpha_{nn} \beta_{nn} (\kB T)^8 {\cal N}_{\nu}
\nonumber \\
         & \approx & 7.4 \times 10^{19}
             \left( {m_n^\ast \over m_n} \right)^4
             \left({ n_n \over n_0} \right)^{1/3}
             \alpha_{nn} \beta_{nn} {\cal N}_{\nu} T_9^8
             \; \; \; \rate,
\label{eq:Qnn0} \\
   Q^{\,(np)}_0 & = & {82 \over 14175} \,{ G_F^2 g_A^2 m_n^{\ast2}
             m_p^{\ast2} \over
             2 \pi \hbar^{10} c^8} \left( {f^\pi \over m_\pi} \right)^4
             \pFe \alpha_{np} \beta_{np} (\kB T)^8 {\cal N}_{\nu}
             \rule{0em}{5.5ex}
\nonumber \\
         & \approx & 1.5 \times 10^{20} \left(
             {m_n^\ast \over m_n}\, {m_p^\ast \over m_p} \right)^2
             \left({ n_e \over n_0} \right)^{1/3}
             \alpha_{np} \beta_{np} {\cal N}_{\nu} T_9^8
             \; \; \; \rate,
\label{eq:Qnp0}   \\
   Q^{\,(pp)}_0 & = & {41 \over 14175} \, { G_F^2 g_A^2 m_p^{\ast4} \over
             2 \pi \hbar^{10} c^8} \left( {f^\pi \over m_\pi} \right)^4
             \pFp \alpha_{pp} \beta_{pp} (\kB T)^8 {\cal N}_{\nu}
             \rule{0em}{5.5ex}
\nonumber \\
         & \approx & 7.4 \times 10^{19}
             \left( {m_p^\ast \over m_p} \right)^4
             \left({ n_p \over n_0} \right)^{1/3}
             \alpha_{pp} \beta_{pp} {\cal N}_{\nu} T_9^8
             \; \; \; \rate,
\label{eq:Qpp0}
\end{eqnarray}
where $m_{\pi}$ is the $\pi^0$ mass, and
${\cal N}_{\nu}$ is the number of neutrino flavors.
$\alpha_{NN}$ and $\beta_{NN}$ have the same meaning as in
Eqs.\ (\ref{eq:Qn0}) and (\ref{eq:Qp0}); they are
slowly varying functions of nucleon Fermi momenta, i.e., density.
Friman and Maxwell (1979) assumed the factors
$\alpha_{NN}$ to be equal to their values at $\rho = \rho_0$:
$\alpha_{nn} = 0.59$, $\alpha_{np} =1.06$, $\alpha_{pp}= 0.11$, and
the correction factors (line $d=0.7$~fm in their Table 1) $\beta_{NN}$
to be $\beta_{nn} = 0.56$, $\beta_{np} = 0.66$.
They did not calculate $\beta_{pp}$.
The presented values of $\beta_{nn}$ and $\beta_{np}$
indicate that it is reasonably to set $\beta_{pp} \approx 0.7$.
Hereafter we will use ${\cal N}_{\nu}=3$
in Eqs.\ (\ref{eq:Qnn0})--(\ref{eq:Qpp0}) while
Friman and Maxwell (1979)
took into account two neutrino flavors.

In analogy with the Urca processes let us introduce the factors
$R^{(NN)}$, which describe superfluid reduction of the neutrino
bremsstrahlung due to $NN$ scattering:
\begin{equation}
        Q^{(NN)} = Q^{(NN)}_0 R^{(NN)}.
\label{eq:Redubre}
\end{equation}
Consider a singlet--state pairing
{\bf A} of neutrons or protons. Then the modification of the
dispersion relation by superfluidity affects only
the phase integral. For a non-superfluid matter,
the latter integral is given by Eq.\ (46)
in Friman and Maxwell (1979)
(cf.\ with Eq.~(\ref{eq:I})):
\begin{equation}
   I_{NN}^{(0)}  =  \int_0^\infty \dd x_\nu \; x_\nu^4
       \left[ \prod_{j=1}^4 \int_{-\infty}^{+\infty}
       \dd x_j \; f(x_j) \right]  \,
       \del \left( \sum_{j=1}^4 x_j - x_\nu \right)  =
       {164 \pi^8 \over 945},
\label{eq:INN}
\end{equation}
where in this case $x_\nu=\ep_\nu / T$ determines the total
energy $\ep_\nu$ of a neutrino pair. Accordingly the
reduction factors take the form:
\begin{equation}
   R^{(NN)}  =  {945 \over 164 \pi^8}
       \int_0^\infty \dd x_\nu \; x_\nu^4
       \left[ \prod_{j=1}^4 \int_{-\infty}^{+\infty}
       \dd x_j \; f(z_j) \right]
       \del \left( \sum_{j=1}^4 z_j - x_\nu \right),
\label{eq:RNN}
\end{equation}
where the dimensionless quantity $z_j$ is defined
in Eq.\ (\ref{eq:DimLessVar}).
It is clear that $R^{(NN)} \!= \!1$ for $\tau \! \ge \! 1$.

In the case of the singlet-state proton pairing
the reduction factors for the
$np$ and $pp$ processes,
$R^{\,(np)}_{pA}$ and $ R^{\,(pp)}_{pA}$, are
reduced to two dimensional integrals which can be calculated
numerically
(Yakovlev and Levenfish, 1995).
In the limit of strong superfluidity
($\tau_p \ll 1$, $\va \to \infty$) the asymptotes of these factors are:
\begin{eqnarray}
     R^{\,(np)}_{pA} & = & {945 \over 164 \pi^8} \, \xi_1 \va \exp(-\va)
                      =
           {0.910 \over \tau_p} \exp \left( - { 1.764 \over \tau_p} \right),
\label{eq:Rnpasy} \\
     R^{\,(pp)}_{pA} &=& { 8505 \over 41 \pi^6} \; \va^2 \exp(-2\va)
            =
      {0.671 \over \tau_p^2} \exp \left( - {3.528 \over \tau_p} \right),
\label{eq:Rppasy}
\end{eqnarray}
where $\xi_1 \approx 849$.
The asymptotes (\ref{eq:Rnpasy}) and (\ref{eq:Rppasy}), as well as
numeric values of
$R^{\,(np)}_{pA}$ and  $R^{\,(pp)}_{pA}$
calculated for intermediate
values of $\va$ can be fitted by the expressions:
\begin{eqnarray}
&\eqalign{
      R^{\,(np)}_{pA} = {1 \over 2.732}
      \left[ a \exp \left( 1.306 - \sqrt{ (1.306)^2 + \va^2} \right) \right.
\cr
       \phantom{R^{\,(np)}_{pA}}
       \left. + \; 1.732 \, b^7
          \exp \left( 3.303 - \sqrt{(3.303)^2 + 4\va^2} \right) \right],
\cr
      a = 0.9982 + \sqrt{(0.0018)^2 + (0.3815 \va)^2}, \;\;\;
      b = 0.3949 + \sqrt{(0.6051)^2 + (0.2666 \va)^2} \, ;
}
\label{eq:Rnpfit}
\end{eqnarray}
\begin{eqnarray}
\eqalign{
      R^{\,(pp)}_{pA} = {1 \over 2}
             \left[ c^2
             \exp \left( 4.228 - \sqrt{ (4.228)^2 + (2 \va)^2} \right) \right.
\cr
       \phantom{R^{\,(pp)}_{pA} }
       +  \left. d^{7.5} \exp
             \left( 7.762 - \sqrt{ (7.762)^2 + (3\va)^2} \right) \right],
\cr
      c  =  0.1747 + \sqrt{ (0.8253)^2 + (0.07933 \va)^2}, \;\;\;
      d  =  0.7333 + \sqrt{ (0.2667)^2 + (0.1678 \va)^2}.
}
\label{eq:Rppfit}
\end{eqnarray}
For the singlet-state neutron pairing, we evidently have
\begin{equation}
  R^{\,(np)}_{nA}(\va)=  R^{\,(np)}_{pA}(\va), \;\;\;\;
  R^{\,(nn)}_{nA}(\va)=  R^{\,(pp)}_{pA}(\va).
\label{New1}
\end{equation}

The reduction factors considered above are shown in
Fig.\ \ref{fig:Rstandard_tau} versus
$T/T_c$ and in Fig.~\ref{fig:Rstandard_v} versus
dimensionless gap parameter $v$.

Summarizing the results of Sects.\ 5.2 and 5.3 we note that the exponent
argument in the asymptote of a reduction factor
contains a single gap if one or two reacting particles
belong to superfluid component of matter.
Under the number of reacting particles we mean
total number of particles (in the initial and final states
of a bremsstrahlung process, in the initial and final state
of direct or inverse reaction of an Urca process) belonging
to superfluid component. The gap in the exponent argument
is doubled if three or four superfluid particles are
involved, etc. As seen from Fig.\ \ref{fig:Rstandard_v}, the factor
$R^{\,(np)}_{pA}$ (two superfluid particles)
falls down with increasing the superfluidity strength
(with increasing $v$)
much more rapidly than the factors
$R_{pA}^{\Mn}$ or $R_{nB}^{\Mp}$ (one superfluid particle).
Accordingly, $R^{\,(pp)}_{pA}$ (four superfluid particles)
falls down faster than
$R_{pA}^{\Mp}$ (three superfluid particles).
Let us stress that we discuss the neutrino reactions considered
in Sects.\ 4 and 5. The formulated rule is invalid for the
neutrino emission due to Cooper pairing of nucleons
(Sect.\ 6.1).

\subsection{Neutrino reactions in the presence of neutron and proton
            superfluidity}

If neutrons and protons are superfluid at once
calculations of multi-dimensional integrals
(\ref{eq:JN}) and (\ref{eq:RNN}),
which determine reduction of the standard neutrino reactions,
becomes very complicated. However, if a very high accuracy
is not required, one can avoid calculation by noticing
that the reduction factors for the processes involving
one superfluid particle
$( \Ra, \, R_{pA}^{\Mn}, \, \Rb \; \mbox{and} \; R_{nB}^{\Mp} ) $
are close to one another as functions of the dimensionless
parameter $v$ (Fig.\ \ref{fig:Rstandard_v}).
This enabled
Levenfish and Yakovlev (1996)
to formulate approximate {\it similarity criteria}
for different reduction factors.
Using these criteria we have constructed the approximate
reduction factors for the proton and neutron branches
of the modified Urca process in the presence of neutron and proton
superfluidity:
\begin{eqnarray}
R^{\Mp}_{BA}(v_n,v_p) &\approx& { R^{\dur}_{BA}(v_n,2v_p) \over
                              \Rb(v_n) } \, R_{nB}^{\Mp}(v_n) \, ,
                                                       \label{eq:RpAB} \\
R^{\Mn}_{BA}(v_n,v_p) &\approx& { R^{\dur}_{BA}(2v_n,v_p) \over
                              \Ra(v_p) } \, R_{pA}^{\Mn}(v_p).
                                                       \label{eq:RnAB}
\end{eqnarray}
We expect that these factors, as functions of corresponding
parameters $v$
(corrected due to the number of superfluid particles)
do not differ strongly from the reduction factor
$R^{\dur}_{BA}(v_n,v_p)$ for the direct Urca process.
If protons are normal ($v_p=0$), then the expression for
$R^{\Mp}_{BA} (v_n,v_p)$ becomes exact; if neutron are normal
($v_n=0$), the factor $R^{\Mn}_{BA} (v_n,v_p)$ is exact.
In addition, the approximate factors satisfy the relationship
analogous to (\ref{eq:Estimation}).

One can also expect similarity
of the reduction factors for other neutrino reactions.
For instance, the
reduction of the neutron branch of the modified Urca
process, $R_{nB}^{\Mn}$, by a moderate neutron superfluidity
($v \la 10$) should not deviate strongly from
the reduction of the proton branch of the modified Urca,
$R_{pA}^{\Mp}$, by the proton superfluidity:
\begin{equation}
    R_{nB}^{\Mn} \approx R_{pA}^{\Mp}(v_n).
\label{RmnnB}
\end{equation}
The approximate reduction factor
$R^{\,(nn)}_{nB}$ for the $nn$ scattering
in the presence of $n$ superfluidity
and the approximate reduction factor
$R^{\,(np)}_{BA}$ of the
$np$ scattering by the neutron and proton superfluidities
can be written as
\begin{eqnarray}
      R^{\,(nn)}_{nB} &\approx& R^{\,(pp)}_{pA}(v_n)\, ,
\label{eq:Rnn_B} \\
     R^{\,(np)}_{BA} &\approx&  {R^{\dur}_{BA}(v_n,v_p) \over
                                \Ra(v_p) } \, R^{\,(np)}_{pA} (v_p) \, .
\label{eq:Rnp_BA}
\end{eqnarray}
In the absence of the neutron superfluidity, Eq.\ (\ref{eq:Rnp_BA})
becomes exact.

The modified Urca and neutrino bremsstrahlung
processes can involve hyperons. After minor modifications,
the above reduction factors can be valid
for these reactions as well.

%
%
%
\section{Neutrino emission due to Cooper pairing of nucleons}

\subsection{Neutrino emissivity produced by nucleon pairing}

In contrast to the neutrino emission processes considered
in Sects.\ 4 and 5, this process is allowed only in the presence
of superfluidity (Sect.\ 2.2):
the superfluidity distorts the nucleon dispersion relation
near the Fermi surface and opens the reaction (\ref{eq:Recomb}).
Actually the process consists in emission of a neutrino pair
by a nucleon whose dispersion relation contains
an energy gap; however, in theoretical studies, it is
convenient to use the formalism of quasi-particles
and treat it
(Flowers et al., 1976)
as annihilation of two quasi-nucleons
$\tilde{N}$ into a neutrino pair:
\begin{equation}
 \tilde{N} + \tilde{N}  \to \nu + \bar{\nu}.
\label{eq:Rec}
\end{equation}
The reaction goes via weak neutral currents and
produces neutrinos of all flavors. Following Yakovlev et al.\ (1999)
we will outline derivation of the
neutrino emissivity due to singlet-state or triplet-state
pairing of non-relativistic nucleons. The reaction is described
by the Hamiltonian ($\hbar = c = \kB = 1$)
\begin{equation}
     \hat{H} = -{ G_{\rm F} \over 2 \sqrt{2} } \,
     \left( c_V \, J_0 l_0 - c_A \, {\vect{J}} {\vect{l}} \right),
\label{H_w}
\end{equation}
where $G_{\rm F}$ is the Fermi constant, and
$c_V$ and $c_A$ are, respectively, the vector and axial-vector
constants of neutral hadron currents.
For neutron currents, we have
(see, e.g.,
Okun', 1990) 
$c_V=1$, $c_A=g_A=1.26$, while for proton currents
$c_V=4 \, \sin^2 \Theta_{\rm W} -1 \approx -0.08$, $c_A=-g_A$, where
$\Theta_{\rm W}$ is the Weinberg angle, $\sin^2 \Theta_{\rm W} = 0.23$.
Strong difference of $c_V$ for neutrons and protons
comes from different quark structure of these particles.
Furthermore,
\begin{equation}
      J^\mu = \left(J^0,{\vect{J}} \right) =
            \left(\hat{\Psi}^+ \hat{\Psi},
                  \hat{\Psi}^+ \vtr{\sigma} \hat{\Psi}
            \right),
    \quad
    l^\mu = \overline{\psi}_\nu \gamma^\mu (1 + \gamma^5) \psi_\nu
\label{l}
\end{equation}
are 4-vectors of neutral currents of
quasi-nucleons and neutrino
($\mu$=0,1,2,3), respectively;
$\psi_\nu$ is the neutrino wave function,
upper bar denotes Dirac conjugate;
$\gamma^\mu$ and $\gamma^5$ are Dirac gamma-matrices,
$\vtr{\sigma}$ is a vector Pauli matrix;
$\hat{\Psi}$ is a second--quantized quasi-nucleon wave function.
The function $\hat{\Psi}$ is derived using the Bogoliubov transformation.
Its description for the singlet-state and triplet-state
pairing is given, for instance, in
Lifshitz and Pitaevskii (1980)
and
Tamagaki (1970).
In the both cases
\begin{equation}
  \hat{\Psi} = \sum_{ \vect{p} \sigma \eta} \, \chi_\sigma \left[
               \ex^{ -i \epsilon t+i{\bf pr} } \,
               U_{\sigma \eta}(\vect{p}) \, \hat{\alpha}_{\vect{p} \eta}+
               \ex^{i \epsilon t-i {\bf pr} } \,
               V_{\sigma \eta}(- \vect{p}) \, \hat{\alpha}_{\vect{p} \eta}^+
               \right],
\label{Psi}
\end{equation}
where $\vect{p}$ and
$ \epsilon =\sqrt{v^2_{\rm F}  (p- \pF)^2 + \del_{\vect{p}}^2 } \, $
are, respectively, the quasiparticle momentum and energy
(with respect to the Fermi level).
A basic spinor $\chi_\sigma$
describes a nucleon state with fixed spin projection
($\sigma\! =\! \pm 1$)
onto the quantization axis (axis $z$);
$\eta $ enumerates quasi-nucleon spin states;
$\del_{\vect{p}}$ is the energy gap in the quasi-particle spectrum,
$\vF$ is the Fermi velocity,
$\; \hat{\alpha}^+_{{\vect{p}}\eta}$ and $ \hat{\alpha}_{{\vect{p}}\eta}$
are, respectively, creation and annihilation operators
for a quasiparticle in a
$({\vect{p}} \eta)$ state; $\hat{U}(\vect{p})$ and
$\hat{V}({\vect{p}})$ are the operators of the
Bogoliubov transformation.
For $|p-\pF| \ll \pF$, their matrix elements obey the relationships
$U_{\sigma \eta}({\vect{p}}) = u_{\vect{p}} \, \del_{\sigma \eta}\;$
and
$
\sum_{\sigma \eta} |V_{\sigma \eta}(\vect{p})|^2 =2 \,v_{\vect{p}}^2$,
where
\begin{equation}
   u_{\vect{p}} = \left[ {1 \over 2}
           \left( 1 + { \vF  (p- \pF) \over
           \epsilon} \right) \right]^{1/2},~~~
   v_{\vect{p}} = \left[ {1 \over 2}
           \left( 1 - { \vF  (p- \pF) \over
           \epsilon } \right) \right]^{1/2}.
\label{eq:uv}
\end{equation}
For a singlet-state pairing, the gap
$\del_{\vect{p}}$ is isotropic, and the quantities
$u_{\vect{p}}$ and $v_{\vect{p}}$ depend only on $p \! = \! |\vect{p}|$.
For a triplet-state pairing, the quantities
$\del_{\vect{p}}$, $u_{\vect{p}}$ and
$v_{\vect{p}}$ depend also on orientation of $\vect{p}$.

Let $q_\nu = (\omega_\nu, {\vect{q}}_\nu)$ and
$q'_\nu = (\omega'_\nu, {\vect{q}}'_\nu)$ be 4-momenta of
neutrino and antineutrino, respectively,
while
$p=(\epsilon,{\vect{p}})$ and $p'=(\epsilon', {\vect{p}}')$ be
4-momenta of annihilating quasi-nucleons.
Using the Golden Rule of quantum mechanics, we can present
the neutrino emissivity due to Cooper pairing
(CP) as:
\begin{eqnarray}
\eqalign{
   Q^{\rec} =  \left( {G_{\rm F} \over 2 \sqrt 2} \right)^2 \,
       {1 \over 2} \, {\cal N}_{\nu} \,
       \int { \dd  {\vect{p}} \over (2 \pi)^3 } \;
       {\dd {\vect{p}}' \over (2 \pi)^3 } \;
       f(\epsilon)f(\epsilon ')
\cr
  \phantom{ Q^{\rec}}
      \times
     \int { \dd {\vect{q}}_\nu \over 2 \omega_\nu (2 \pi)^3 } \;
     { \dd {\vect{q}}'_\nu \over 2 \omega'_\nu (2 \pi)^3 } \;
     \left[c_V^2 I_{00} |l_0|^2 + c_A^2 \, I_{ik} l_i l_k^\ast \right]
\cr
  \phantom{ Q^{\rec}}
     \times
     (2 \pi)^4 \, \del^{(4)} \left( p + p' - q_\nu - q'_\nu \right) \,
     (\omega_\nu + \omega'_\nu),
}
\label{eq:Qgen}
\end{eqnarray}
where ${\cal N}_{\nu}$=3 is the number of neutrino flavors, and
the factor $1/2$ before ${\cal N}_{\nu}$ excludes double counting
of the same quasi-nucleon collisions. The integral is taken over
the range
$(q_\nu + q'_\nu)^2 >0$, where the process is open kinematically;
$f(\epsilon)=1/[\exp(\epsilon/T)+1]$, $\; i,k\, = \, 1,2,3$;
\begin{equation}
   I_{00} = \sum_{\eta\eta'} \,
   |\langle B | \hat{\Psi}^+ \hat{\Psi} | A \rangle |^2,~~~
   I_{ik} = \sum_{\eta\eta'} \,
   \langle B | \hat{\Psi}^+ \sigma_i \hat{\Psi} | A \rangle \,
   \langle B | \hat{\Psi}^+ \sigma_k \hat{\Psi} | A \rangle^\ast.
\label{eq:tensor}
\end{equation}
Here $|A \rangle$ stands for an initial state of the quasi-particle
system in which one-particle states
$({\vect{p}},\eta)$ and $({\vect{p}}',\eta')$ are occupied, and
$|B \rangle$ stands for a final state of the system in which
the indicated one-particle states are empty.

The integral (\ref{eq:Qgen}) is simplified by the standard technique,
as described in Yakovlev et al.\ (1999).
The transformations take into account that nucleons are
non-relativistic and strongly degenerate as well as the fact
that the process is open kinematically in a small domain of phase space
where the quasi-nucleon momenta
$\vect{p}$ and ${\vect{p}}'$ are almost parallel.
The latter circumstance allows one to take smooth functions
$I_{00} ({\vect{p}}, {\vect{p}}')$ and $I_{ik}({\vect{p}}, {\vect{p}}')$
out of the integral over
$\dd {\vect{p}}'$ putting ${\vect{p}}' = - {\vect{p}}$.
After a number of transformations
the final expression for the neutrino emissivity
can be written
(in the standard physical units)
as (Yakovlev et al., 1998, 1999)
\begin{eqnarray}
 Q^{\rec} & = & {4 G_{\rm F}^2
           m_N^\ast \pF \over 15 \pi^5 \hbar^{10}
           c^6} \, (\kB T)^7 \, {\cal N}_{\nu} \; a F =
\nonumber \\
     & = & 1.170 \times 10^{21} \, \left( {m_N^\ast \over m_N }  \right)
      \left( { \pF \over m_N c   } \right) \, T_9^7 \, {\cal N}_{\nu} \;
      aF~~~
      \rate,
\label{eq:Qrec}
\end{eqnarray}
where $T_9= T/(10^9 \; {\rm K})$,
$a$ is a numeric factor (see below), and the function
$F$, in our standard notations (\ref{eq:DimLessVar}), is given
by the integral
\begin{equation}
     F = {1 \over 4 \pi} \, \int {\rm d}\Omega \, y^2
     \int_0^\infty \, {z^4 \, {\rm d}x \over ({\rm e}^z +1)^2}  \, .
\label{eq:F}
\end{equation}
The singlet-state gap $\del_{\vect{p}}$ is isotropic; thus, integration
over $\dd \Omega$ is trivial and gives $4 \pi$.
In the triplet-state case, the function
$F$ contains averaging over positions of a quasi-nucleon on the
Fermi-surface. While using Eq.\ (\ref{eq:Qrec}),
one can take into account that
$p_{\rm F}/(m_N c) \approx 0.353 \, (n_N/n_0)^{1/3}$,
where $n_N$ is the number density of nucleons $N$,
$n_0 = 0.16$~fm$^{-3}$.

The emissivity
 $Q^{\rec}$ depends on superfluidity type through
the factor $a$ and the function $F$.
For the singlet-state neutron pairing,
$a$ is determined by the only vector constant
$c_V$: $a_{nA}=c_V^2=1$.
If we used similar expression for the singlet-state pairing
of protons, we would obtain a very small factor
$a_{pA}=0.0064$, which indicates the weakness of the process.
Under these conditions, one should take into account
the relativistic correction to
$a$, produced by the axial-vector proton current.
Calculating and adding this correction for the singlet-state pairing
of protons, Kaminker et al.\ (1999b) obtained
%
\begin{equation}
a_{pA} = c_V^2+ c_A^2 \left( \frac{\vF_{\!p}}{c}\right)^2
         \left[ \left(\frac{m_p^\ast}{m_p} \right)^2 +\frac{11}{42} \right]
       =
         0.0064 + 1.59 \left(\frac{\vF_{\!p}}{c} \right)^2
   \left[ \left(\frac{m_p^\ast}{m_p} \right)^2 +\frac{11}{42} \right],
\label{eq:AP}
\end{equation}
where $\vF_{\!p}/c = (\pF_{\!p}/m_p c)(m_p/m^\ast_p)$.
The relativistic correction appears to be about 10 -- 50 times
larger than the main non-relativistic term.
This enhances noticeably the neutrino emission
due to the singlet-state proton pairing although
it remains much weaker than the emission due to the neutron pairing.

In the case of the triplet-state pairing
$a$ is determined by the both, vector and axial-vector,
constants of neutral hadron currents:
$a=a_{NB}=a_{NC}=c_V^2+2 c_A^2$
(Yakovlev et al., 1999).
For the neutron pairing, we obtain $a_{nB}=a_{nC}=4.17$.
Notice that in the case of the triplet-state pairing of protons
which is thought to be hardly possible in the NS cores
we would obtain $a_{pB}=a_{pC}=3.18$.
Under such exotic conditions, the neutrino emission
due to proton pairing would be almost as efficient as
the emission due to neutron pairing.

The result for the singlet-state pairing of neutrons,
presented above, coincides with that obtained
in the pioneering article by Flowers et al.\ (1976)
(for two neutrino flavors, ${\cal N}_\nu = 2$).
Similar expressions obtained by
Voskresensky and Senatorov (1986, 1987)
for ${\cal N}_{\nu} = 1$ contain an extra factor $(1+3g_A^2)$.
In addition, the expression for
$Q^{\rec}$
derived by
the latter authors
contains a misprint:
$\pi^2$ in the denominator instead of $\pi^5$
(although numeric estimate of
$Q^{\rec}$ is obtained with the correct factor $\pi^5$).
The cases of the singlet-state proton pairing and
the triplet-state neutron pairing
have been analysed by
Yakovlev et al.\ (1998, 1999) and Kaminker et al.\ (1999b)
for the first time.

Let us mention that the values of
$c_V$, $c_A$ and $a$ can be renormalized in dense
NS matter under in-medium effects. The renormalization
is a difficult task which will be neglected below.

The function $F$, given by Eq.\ (\ref{eq:F}),
depends on the only variable $v$, the gap parameter
(see Eq.\ (\ref{eq:DefGap})). Using Eq.\
(\ref{eq:F}) one can easily obtain the asymptote of this function
and calculate its dependence on
$\tau=T/T_c$ for superfluidity of types
{\bf A}, {\bf B} and {\bf C}
in analogy with calculations presented in
Sects.\ 3--5. This has been done by
Yakovlev et al.\ (1998, 1999).

In a small vicinity of $T \approx T_c$,
in which $v \ll 1$ and $\tau \to 1$, we have:
\begin{eqnarray}
\eqalign{
   F_{\A}(v) = 0.602 \, v^2 = 5.65 \, (1-\tau),
\cr
   F_{\B}(v) = 1.204 \, v^2 = 4.71 \, (1-\tau),
\cr
   F_{\C}(v) = 0.4013 \, v^2 = 4.71 \, (1-\tau).
}
\label{eq:low_v}
\end{eqnarray}
For the low temperatures $T \ll T_c$ the parameter $v \gg 1$,
and the asymptotes of $F(v)$ are
\begin{eqnarray}
\eqalign{
   F_{\A}(v) = {\sqrt{\pi} \over 2} \, v^{13/2} \,
     \exp(-2v) = {35.5 \over \tau^{13/2}} \,
     \exp\left( -\frac{3.528}{\tau} \right) ,
\cr
   F_{\B}(v) = {\pi \over 4 \sqrt{3}} \, v^6 \,
     \exp(-2v) = {1.27 \over \tau^6} \,
     \exp \left( -\frac{2.376}{\tau} \right) ,
\cr
   F_{\C}(v) = {50.03 \over v^2} = 12.1 \, \tau^2.
}
\label{eq:high_v}
\end{eqnarray}

Let us stress that the neutrino emission due to nucleon pairing
differs from other neutrino reactions: first, is has a temperature
threshold (is allowed only for
$T < T_c$); second, its emissivity
is a nonmonotonic function of temperature.
The emissivity grows rapidly with decreasing
$T$ just after superfluidity but then reaches maximum and decreases.
According to Eq.\ (\ref{eq:high_v}), a strong superfluidity
reduces considerably the emissivity
just as it reduces the heat capacity or direct
Urca process: the reduction is exponential
if the gap is nodeless
(cases {\bf A} and {\bf B}) and it is power-law
otherwise
(case {\bf C}); see Sects.\ 3 and 4.

The asymptotes (\ref{eq:low_v}) and (\ref{eq:high_v}),
as well as numeric values of
$F(v)$ for intermediate $v$,
are fitted by the simple expressions
(Yakovlev et al., 1998, 1999)
%
\begin{eqnarray}
\eqalign{
  F_{\A}(v) = (0.602 \, v^2 + 0.5942\, v^4 +
     0.288 \, v^6) \,
     \left( 0.5547 + \sqrt{(0.4453)^2 + 0.0113 \,v^2} \right)^{1/2}
\cr
  \phantom{F_{\A}(v)}
     \times
     \exp \left(- \sqrt{4 \, v^2 + (2.245)^2 } + 2.245 \right),
\cr
  F_{\B}(v) = {1.204 \, v^2 + 3.733 \, v^4 +
     0.3191 \, v^6 \over 1 + 0.3511 \, v^2} \,
     \left( 0.7591 + \sqrt{ (0.2409)^2 + 0.3145 \, v^2 } \right)^2
\cr
 \phantom{F_{\A}(v)}
    \times
    \exp \left( - \sqrt{ 4 \, v^2 + (0.4616)^2} + 0.4616 \right),
\cr
  F_{\C}(v) =
    { 0.4013 \, v^2 - 0.043 \, v^4 + 0.002172 \, v^6 \over
     1 - 0.2018 \, v^2 + 0.02601 \, v^4 - 0.001477 \, v^6
     + 0.0000434 \, v^8}.
}
\label{eq:RecFit}
\end{eqnarray}

\begin{figure}[t]                         
\begin{center}
\leavevmode
\epsfysize=8.5cm
\epsfbox[95 60 440 390]{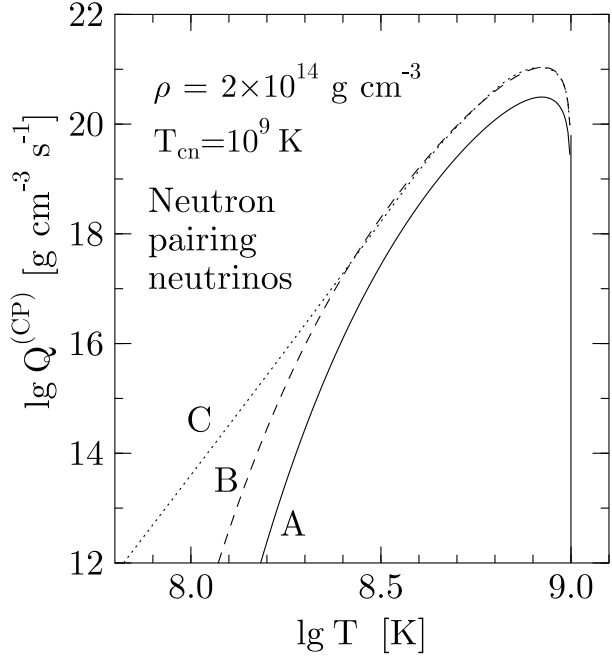}
\end{center}
\caption[]{\footnotesize
        Temperature dependence of the neutrino emissivity
        due to Cooper pairing of neutrons for
        $\rho = 2 \times 10^{14}$ g cm$^{-3}$
        and $T_{cn} = 10^9$ K for superfluidity types {\bf A} (solid line),
        {\bf B} (dashes) and {\bf C} (dots).
}
\label{fig:Qrec}
\end{figure}

Equations (\ref{eq:Qrec}) and (\ref{eq:RecFit}) enable one
to calculate easily the neutrino emissivity
$Q^{\rec}$ due to Cooper pairing of nucleons
for superfluidity of types {\bf A}, {\bf B} and {\bf C}.
Similar expressions describe neutrino emissivity
due to pairing of hyperons. The required values of
$a$ are listed in Yakovlev et al.\ (1999).

Figure \ref{fig:Qrec}
(from
Yakovlev et al., 1999)
shows temperature dependence of the emissivity
$Q^{\rec}$ due to neutron pairing in the NS core for
$\rho = 2 \times 10^{14}$ g cm$^{-3}$.
The adopted equation of state of matter
is described in Sect.\ 7.2. The effective nucleon masses are
set equal to
$m_N^\ast = 0.7 \, m_N$, and the critical temperature is
$\tn = 10^9$ K. The density of study is typical for
transition between the single-state pairing and the triple-state one
(Sect.\ 3.1). Thus, different models of nucleon--nucleon
interaction may lead to different neutron superfluidity type.
We present the curves for the three superfluidity types
considered above.

\begin{figure}[t]                         
\begin{center}
\leavevmode
\epsfysize=8.5cm
\epsfbox[0 20 350 345]{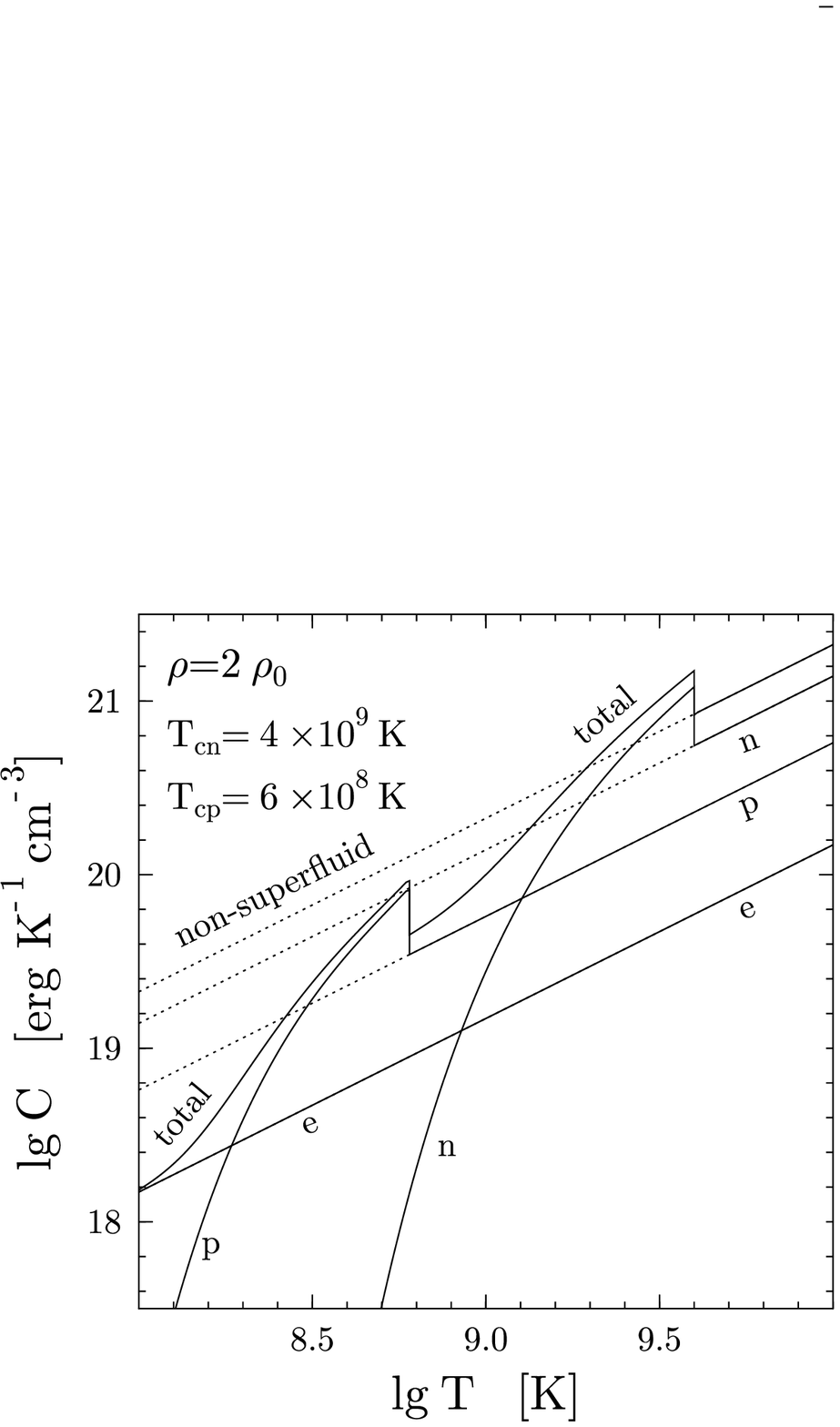}
\end{center}
\caption[]{\footnotesize
        Temperature dependence of the total and
        partial ($n$, $p$, $e$) specific heat capacities
        at $\rho = 2\, \rho_0$
        for neutron superfluidity of type {\bf B} with
        $\tn= 4 \times 10^9$~K
        and proton superfluidity of type {\bf A} with
        $\tp = 6 \times 10^8$~K.
        Dotted lines show corresponding heat capacities in
        non-superfluid matter.
         }
\label{fig:figc1}
\end{figure}

When the temperature falls down below
$T_{cn}$ the neutrino emissivity produced by Cooper
pairing strongly increases. The main neutrino energy release
takes place in the temperature interval
$0.2 \, \tn \! \la \! T \la 0.96 \,\tn$, 
with the maximum at $T \! \approx \! 0.4\, \tn$.
The emissivity may be sufficiently high, compared with
or even larger than the emissivity of the modified Urca process
in non-superfluid matter (Sect.\ 4).
Under certain conditions, neutrino emission
due to pairing of neutrons may be significant
in the inner NS crust
(Yakovlev et al., 1998; Kaminker et al., 1999).
The reaction may be noticeable even in the
presence of the direct Urca process in the inner NS core
if the direct Urca is partly suppressed by the
proton superfluidity (see below).

\subsection{Summary of Sections 3 -- 6}

Let us summarize the results of Sects.\
3 -- 6. For illustration, we use the same equation of state
of matter (neutrons, protons and electrons) in the NS core as
in our cooling simulations
(Sect.\ 7.2), and set
$m_N^\ast = 0.7 \, m_N$.
Let us adopt that the neutron pairing is of type
{\bf B}, while the proton pairing is of type {\bf A}.

\begin{figure}[t]                         
\begin{center}
\leavevmode
\epsfysize=8.5cm
\epsfbox[40 15 540 270]{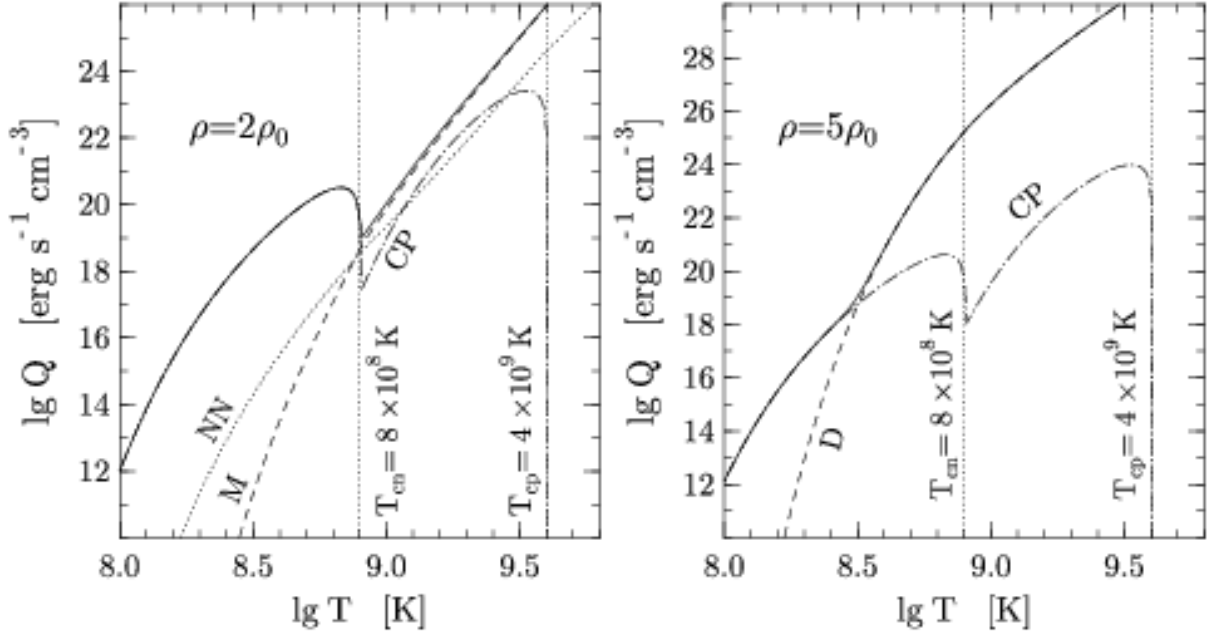}
\end{center}
\caption[]{\footnotesize
        Temperature dependence of the neutrino emissivity
        in different reactions for the neutron
        superfluidity of type {\bf B} with
        $\tn= 8 \times 10^8$~K
        and the proton superfluidity of type {\bf A} with
        $\tp = 4 \times 10^9$~K
        at $\rho = 2\, \rho_0$ (standard neutrino reactions, Fig.\ a)
        and $\rho = 5\, \rho_0$ (direct Urca is allowed, Fig.\ b).
        Dot-and-dash line shows the emissivity due to Cooper
        pairing of neutrons plus protons; solid line presents
        the total emissivity.
        In Fig.\ a the dashed line gives the total emissivity in two
        branches of the modified Urca process;
        the dotted line exhibits the total bremsstrahlung emissivity due to
        $nn$, $np$ and $pp$ scattering.
        Figure b: the dashed line corresponds to the direct Urca process.
         }
\label{fig:QrecMurDur}
\end{figure}

Figure \ref{fig:figc1} illustrates the effect of superfluidity
on the heat capacity for $\rho = 2 \rho_0$,
$\tn = 4 \times 10^9$~K and $\tp = 6 \times 10^8$~K.
In the absence of superfluidity, the main contribution
into the heat capacity comes from neutrons.
The heat capacities of protons and electrons
are lower than that of neutrons by a factor of
2.5 and 9, respectively. After superfluidity appears
with decreasing temperature, the relative contributions
of different particles change.
The jumps of the total and neutron heat capacities
with the fall of temperature below
$T= \tn $ are associated with latent heat release
produced by pairing of neutrons.
However for $T \la 10^9$~K the neutron superfluidity
becomes strong and reduces greatly the partial heat capacity.
The main contribution into the heat capacity
comes now from protons. When the temperature falls below
$\tp$, the proton heat capacity up jumps due to
appearance of the proton superfluidity.
For $T \la 2 \times 10^8$~K the latter superfluidity, in its turn,
becomes strong and reduces exponentially the proton heat capacity.
As a result, at lower temperatures the total heat capacity
is determined by electrons and
does not depend on nucleon superfluidity.
Examining Fig.\ \ref{fig:figc1}, it is easy to predict
relative contributions of different particles into the heat
capacity for any relationships between
$T$, $\tn$ and $\tp$.

The effect of superfluidity on neutrino reactions is more complicated.
For instance, Figs.~\ref{fig:QrecMurDur}
show the neutrino emissivities in different reactions for
$\tn \!=\! 8 \times 10^8$~K and
$\tp \! = \! 4 \times 10^9$~K.
Figure \ref{fig:QrecMurDur}a corresponds to
$\rho \! = \! 2\rho_0$.
The direct Urca process is forbidden at this density
(being allowed at
$\rho_{cr}=4.64 \, \rho_0= 1.30 \times 10^{15}$ g cm$^{-3}$,
for a given equation of state).
In the absence of the neutron superfluidity
($T> \tn$) the dominant mechanism is the modified Urca process.
If, however, the temperature decreases from
$T \! = \! \tn$ to
$T\! \approx \! 10^{8.8}$~K
the total neutrino emissivity increases by about
two orders of magnitude due to Cooper pairing of neutrons.
Therefore, sometimes the appearance of superfluidity
accelerates NS cooling instead of slowing it.

\begin{figure}[t]                         
\begin{center}
\leavevmode
\epsfysize=17.0cm
\epsfbox[45 75 520 740]{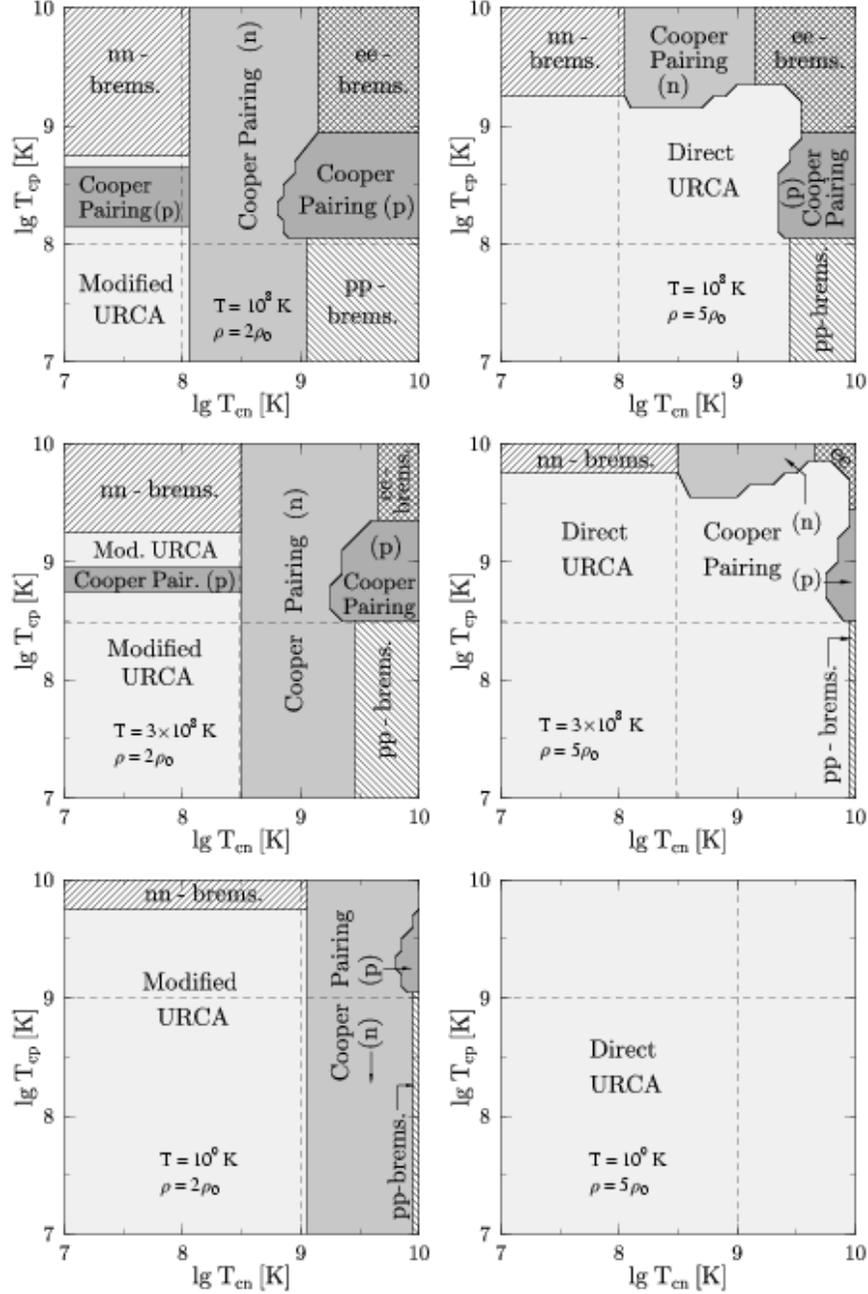}
\end{center}
\caption[]{\footnotesize
        Regions of $\tn$ (superfluidity of type {\bf B})
        and $\tp$ (type {\bf A}), in which different neutrino
        reactions dominate at $T=10^9$,
        $3 \times 10^8$ and $10^8$ K in matter of density
        $\rho = 2\,\rho_0$ (standard cooling)
        and $\rho = 5\, \rho_0$ (enhanced cooling).
        }
\label{fig:qq_murdur}
\end{figure}

Figure \ref{fig:QrecMurDur}b corresponds to a denser matter,
$\rho \! = \! 5 \rho_0$, where the powerful direct Urca process
is switched on. In this case, the neutrino emissivity is
actually determined by two processes, the direct Urca and
Cooper pairing of nucleons. The direct Urca dominates at
$T \! \ga \! 3 \times 10^8$~K. With further decrease of $T$,
the direct Urca and the reactions considered in Sects.\ 4 and 5
are reduced so strongly, that the Cooper pairing becomes dominant.

It is well known that the neutrino emission from a non-superfluid NS
core is mainly determined by a single, most powerful neutrino
emission mechanism: the direct Urca
for the enhanced cooling, or the modified Urca
for the standard cooling.
However, as seen from Fig.\ \ref{fig:QrecMurDur},
this ``simplicity" is violated in the superfluid NS cores.
Different neutrino mechanisms can dominate at different
cooling stages depending on
$T$, $\tn$, $\tp$, and $\rho$.

Figures \ref{fig:qq_murdur} show which neutrino mechanisms
dominate for different
$\tn$ and $\tp$. In addition to the neutrino processes
considered above we have included one more process,
neutrino bremsstrahlung due to electron--electron
collisions
(Kaminker and Haensel, 1999). This mechanism, as a rule,
is sufficiently weak and neglected
in cooling simulations.
Three left Figs.\ \ref{fig:qq_murdur} illustrate the case
of the standard cooling at
$\rho \! = \! 2\, \rho_0$ for
three internal stellar temperatures:
$10^8$, $3 \times 10^8$ and $10^9$~K;
three right figures
correspond to the enhanced cooling
at $ \rho \!= \! 5\, \rho_0$ for the same $T$.
The chosen values of $T$ cover the region
most interesting for practice. Our calculations show
that topology of the figures varies only slightly
with $\rho$ as long as $\rho$ does not cross
the threshold value $\rho=\rho_{cr}$. Therefore, the
presented figures reflect adequately the efficiency
of all neutrino processes in the core of the cooling NS.
One can see that, in the presence of superfluidity,
many different mechanisms can dominate in certain parameter ranges.

Notice that if the direct Urca is open, the modified Urca
is always insignificant and may be neglected.
In the presence of the neutron superfluidity alone,
the neutrino bremsstrahlung due to $pp$ collisions
becomes the main mechanism at
$T \ll \tn$ being independent of the neutron superfluidity.
In the presence of the proton superfluidity alone,
the neutrino bremsstrahlung due to $nn$ collisions dominates at
$T \ll \tp$. The Cooper pairing of neutrons exceeds
the standard neutrino energy losses for
$T \! \la \! 10^9$~K
and for not too strong neutron superfluidity
($0.12 \! \la T/\tn \la \! 0.96$).
This parameter range is very interesting for applications.
The neutrino emission due to Cooper pairing of neutrons
is also significant at early cooling stages, when
$T \ga 10^9$~K, but in a narrower temperature range near
$T \! \approx \! 0.4\, \tn$, or in the presence of the proton superfluidity.
Although the neutrino production due to proton pairing is much weaker,
it can also dominate. The neutrino emission due to
pairing of neutrons and protons can dominate
in the enhanced NS cooling as well provided the nucleons of one species
are strongly superfluid while the other ones are moderately superfluid.
Very strong superfluidities of neutrons and protons
(upper right corners of the figures) switch
off all neutrino processes involving nucleons. As a result,
the neutrino bremsstrahlung due to electron--electron collisions,
which is practically unaffected by superfluidity,
becomes dominant. It is difficult to expect that this
weak process is important under other conditions.

%
%

\section{Cooling of neutron stars}
\subsection{Review of articles on neutron star cooling}
\subsubsection{Overall review}
The theory of NS cooling has been developing over more than 30 years.
The first articles appeared even before the discovery of NSs.
Their authors tried to prove that not too old NSs may
emit sufficiently powerful thermal X-ray radiation
which could serve to discover NSs.
The first estimates of the thermal emission from cooling
NSs were most probably done by R.\ Stabler (1960).
Four years later
Chiu (1964)
repeated these estimates and theoretically proved the possibility
to discover NSs from their thermal emission.
First, simplified calculations of the NS cooling
were done by
Morton (1964),
Chiu and Salpeter (1964) and also
by Bahcall and Wolf (1965b)
after the discovery of X-ray sources in the Crab nebula
and Scorpion constellation
in the balloon experiments by Bowyer et al.\ (1964).

The foundation of the strict cooling theory was laid
in the fundamental paper by
Tsuruta and Cameron (1966)
who explicitly formulated the main elements
of the theory --- the relationship between the internal and
surface NS temperatures,
the neutrino and photon cooling stages, etc.
Later the theory has been developed in many papers.

In 1970s and 1980s, the main attention was paid to
the equation of state and nuclear composition of the NS cores
(including possible appearance of exotic particles),
to neutrino reactions and to the relationship between the
internal and surface temperatures of NSs (particularly in the presence of
strong magnetic fields). The achievements of the theory
in the middle of 1970s were reviewed by
Tsuruta (1979).
A splash of theoretical activity in the beginning of 1980s was concerned
with the launch of the {\it Einstein} space observatory
(Sect.\ 8.1). The articles of the ``Einstein series"
were also reviewed by
Tsuruta (1980, 1981, 1986).
The launch of the {\it ROSAT} observatory in
1990 (Sect.\ 8.1) initiated a new rise of theoretical activity which
is being continued up to now. Recent articles
are mainly focused on the mechanisms of NS reheating
at the late evolutionary stages (see below)
and the effects of superfluidity in the NS cores.
Let us mention recent review by
Tsuruta (1998).
New microscopic theories of dense matter are appearing permanently
complicating cooling models; many problems are still unsolved.

Let us outline some aspects of the theory.

The first articles were devoted to the standard cooling
(produced by the neutrino reactions from the ``standard"
collection, Sect.\ 5). The enhanced cooling has been simulated
since the end of 1970s. It has been assumed in many articles
that the enhanced neutrino luminosity is associated with
exotic composition of the NS cores containing pion condensates
of quark plasma; see, e.g.,
Tsuruta (1979),
Glen and Sutherland (1980),
Van Riper and Lamb (1981),
Yakovlev and Urpin (1981),
Richardson et al.\ (1982),
Umeda et al.\ (1993, 1994a, b),
Schaab et al.\ (1996).
By the end of 1980s a new exotic cooling agent,
kaon condensate, has been introduced
(Page and Baron, 1990;
Schaab et al., 1996;
Page, 1997).
New models of dense matter 
with highly polarized pion degrees of freedom
(Voskresensky and Senatorov, 1984, 1986;
Schaab et al., 1997b) have been proposed.
According to these models the neutrino luminosity is strongly enhanced
due to virtual excitation of pion field even if
density of matter does not exceed the critical density
of actual pion condensation.
Modern cooling theories of stars with the quark cores,
pion or kaon condensates have been described, for instance, by
Pethick (1992)
and Schaab et al.\ (1996).

A new stage of the theory of enhanced cooling has been opened by
Lattimer et al.\ (1991)
who have shown that the direct Urca process can be allowed
in the NS cores with the standard nuclear composition
for many realistic equations of state.
The process initiates rapid NS cooling
(e.g., Page and Applegate, 1992;
Van Riper and Lattimer, 1993;
Gnedin and Yakovlev, 1993;
Page, 1994;
Gnedin et al., 1994;
Page, 1995b;
Levenfish and Yakovlev, 1996;
Schaab et al., 1996)
without invoking ``exotic" hypotheses
(Sect.\ 2.2). Detailed description of the ``non-exotic"
standard and enhanced cooling theories has been given
by
Pethick (1992).

The effect of the NS magnetic field on the relationship between
the internal and surface stellar temperatures has been taken
into account starting from the articles by Tsuruta and
coauthors
(Tsuruta et al., 1972;
Tsuruta 1974, 1975). These articles were the first
where a representative set of neutrino reactions
in the NS crust and core were included
(plasmon decay, annihilation of electron--positron pairs,
photon decay, electron bremsstrahlung due to scattering
off nuclei, neutron branch of the modified Urca process).
A detailed study of the relationship between the
internal and surface temperatures in a NS with the magnetic
field normal to the surface was carried out by
Van Riper (1988).
He also analysed in detail
(Van Riper, 1991a)
the effect of such magnetic fields on the NS cooling.
Page (1995a) as well as
Shibanov and Yakovlev (1996)
considered the cooling of a NS with a dipole magnetic field
and showed that the dipole field affected the cooling much weaker,
and in a qualitatively different way, than the purely radial magnetic field.

Let us mention also a recent series of articles by Heyl and coauthors
(Heyl and Hernquist 1997a, b, 1998a, b, c;
Heyl and Kulkarni, 1998) devoted to the cooling of NSs with
superstrong magnetic fields $10^{14}$--$10^{16}$~G
(the so called ``magnetars'', see Sect.\ 8.1.1).
These fields may strongly reduce thermal insulation
of the NS envelope, making the magnetar's surface
much hotter at the early cooling stage, than the
surface of an ``ordinary" NS. Let us stress that microscopic
properties of matter in superstrong magnetic fields
(equation of state, thermal conductivity) are poorly known,
so that the results by Heyl and coauthors can be regarded as very
preliminary.

The cooling can also be affected noticeably
(Chabrier et al., 1997;
Potekhin et al., 1997;
Page, 1997, 1998a, 1998b)
by the presence of a thin
(of mass $\la 10^{-8} M_\odot$) envelope of light elements
(H, He) at the surface of a non-magnetized or weakly magnetized NS.
Owing to the higher electron thermal conductivity
of plasma composed of light elements, the NS surface
appears to be significantly warmer at the early cooling stages.

An important contribution into the theory
was made in the PhD thesis by
Malone (1974).
He was the first who calculated the standard NS cooling
beyond the approximation of isothermal
internal layers. This allowed him to describe
thermal relaxation of the internal layers in the first
100--1000 years of NS life. Analogous nonisothermal
calculations of the cooling
enhanced by the presence of the pion condensate
were done in the PhD thesis of
Richardson (1980).
The results of both theses were published in one
article
(Richardson et al., 1982).
Later the thermal relaxation of the internal layers
in a young NS was studied in a series of articles by
Nomoto and Tsuruta (1981, 1982, 1983, 1986, 1987)
as well as by
Lattimer et al.\ (1994).
The thermal relaxation is accompanied by propagation of
a cooling wave from the internal layers to the surface.
In principle, the appearance of the cooling wave at the surface
can be observed in young NSs.
The moment of appearance depends on the equation
of state in the central stellar region
(Lattimer et al., 1994).

In 1980
Glen and Sutherland 
and a year later
Van Riper and Lamb (1981)
included the effects of General Relativity
into the equations of NS thermal evolution
(in addition to the equations of hydrostatic
equilibrium as had been done before).
In this connection let us mention a recent article by
Schaab and Weigel (1998)
who carried out the first two-dimensional cooling calculations of
a rotating NS with exact account of the general
relativistic effects produced by rotation.

New direction of study was opened by
Alpar et al.\ (1987)
and
Shibazaki and Lamb (1989).
The authors took into account possible reheating
at late cooling stages
(NS age $t \ga 10^4$ yr)
due to viscous dissipation of rotational
energy inside a NS. The effect is caused by interaction
of superfluid and normal components of matter
in the inner crust of a pulsar which is spinning down
under the action of magnetodipole losses.
The cooling theory with viscous reheating has been developed
further in a number of articles (e.g.,
Van Riper, 1991b;
Van Riper et al., 1991, 1995;
Umeda et al., 1993, 1994b;
Shibazaki and Mochizuki, 1995;
Page, 1997, 1998a, 1998b).

The core of a cooling NS can also be reheated
by the energy release associated with a weak deviation
from beta-equilibrium 
(Reisenegger, 1995).
In addition, the reheating of the star with a non-superfluid core
can be produced by ohmic dissipation
of the core magnetic field 
(Haensel et al., 1990;
Yakovlev, 1993;
Shalybkov, 1994;
Urpin and Shalybkov, 1995)
due to the enhancement of the electric resistance
across the strong magnetic field.
Criticism of this effect by
Goldreich and Reisenegger (1992)
is not convincing because, while calculating
the electric resistance,
the latter authors
neglected motion of neutron component of matter
(inconsistency of such approximation is clearly seen from the
results by
Yakovlev and Shalybkov, 1991).
Finally, the reheating in an old NS
($t \ga 10^7$ yr) can be provided by ohmic decay
of the magnetic field in the NS crust
(Miralles et al., 1998;
Urpin and Konenkov, 1998).
Let us add that the magnetic field dissipation in the crust
represents a very important process which determines evolution of
the surface magnetic field, magnetodipole NS spindown,
NS activity as radio pulsar, etc.
The dissipation process is associated with cooling
since the electric resistance of the curst decreases
in the course of cooling (until it reaches minimum
produced by electron scattering off charged impurities).
Therefore, the dissipation rate depends on
cooling type: it is very weak for the rapid cooling.
On the other hand, the magnetic field decay
weakens magnetodipole spindown of the star.
This means that cooling, magnetic field evolution
and spindown should be considered, in principle, selfconsistently
as {\it magneto-rotational evolution} of a NS.
Corresponding theory has been developed
in a series of papers by Urpin and coauthors
(see, e.g.,
Urpin and Muslimov, 1992;
Urpin and Van Riper, 1993;
Urpin and Konenkov, 1998,
and references therein)
without account for the effect of superfluidity 
on the NS cooling.

\subsubsection{Cooling of superfluid stars}
The effect of nucleon superfluidity in the NS core
on the cooling has been taken into account starting from
the article by Tsuruta et al.\ (1972).
In the first articles, as a rule, it has been included in a simplified
manner; it has been assumed that
the superfluidity completely
switches off corresponding heat capacity or neutrino
emission from the very beginning of the cooling
(which corresponds to infinitely large
critical temperatures). Maxwell (1979)
was the first who studied the dependence of the standard
NS cooling on nucleon critical temperatures.
Friman and Maxwell (1979) considered a simplified NS
model with constant--density
core neglecting the effects of General
Relativity, but implying sufficiently realistic
neutrino luminosity.
For a long time, before publication of the article by
Page and Applegate (1992),
the superfluidity was not considered as a powerful cooling
regulator. It was taken into account because
it appeared inevitably in microscopic theories of
dense matter
(Sect.\ 3.1). However recent studies show that the
superfluidity is one of the most important factors
which affect the standard and enhanced NS cooling.\\[0.5ex]

\begin{figure}[t]                          
\begin{center}
\leavevmode
\epsfysize=8.5cm
\epsfbox[13 90 305 350]{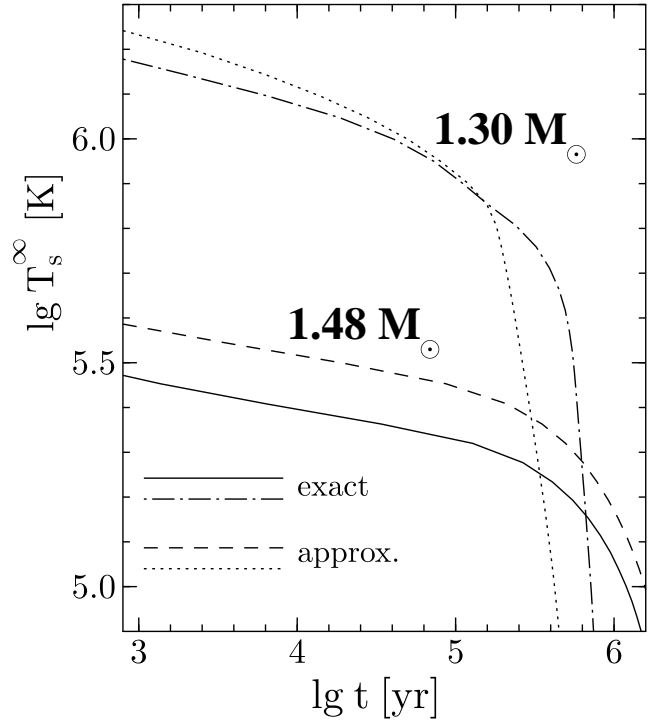}
\end{center}
\caption[]{\footnotesize
Cooling curves calculated using accurate
(solid and dash-dotted lines) and approximate
(dashed and dotted lines) description of the effects
of superfluidity on the neutrino luminosity
and heat capacity (see text).
Solid and dashed lines correspond to
the enhanced cooling of a NS ($M=1.48\, M_\odot$)
in the presence of the proton superfluidity with $\tp = 10^8$~K.
Dash-dotted and dotted lines refer to
the standard cooling of a NS
($1.30\, M_\odot$) in the presence of
neutron and proton superfluidities with
$\tn=10^9$~K and $\tp = 10^8$~K.
The parameters of the models are given in Sect.\ 7.2.2.
Neutrino emission due to nucleon pairing is neglected.
}
\label{fig:cc8}
\end{figure}

{\bf Simulations of the standard cooling} of NSs with
superfluid cores have been performed by many authors,
in particular, by
Tsuruta et al.\ (1972),
Malone (1974),
Tsuruta (1978, 1979, 1980),
Maxwell (1979),
Glen and Sutherland (1980),
Van Riper and Lamb (1981),
Nomoto and Tsuruta (1981, 1986, 1987),
Richardson et al.\ (1982),
Page (1994),
Levenfish and Yakovlev (1996),
Schaab et al.\ (1996),
Levenfish et al.\ (1999),
Yakovlev et al.\ (1999).
While calculating the standard neutrino luminosity of the
NS core, as a rule, one took into account the contribution
of the neutron branch of the modified Urca processes
and the neutrino bremsstrahlung due to
$nn$- and $np$-scattering.
Tsuruta et al.\ (1972) were the first
who studied the NS cooling accompanied by the singlet-state
pairing of neutrons and protons.
The authors
described reduction of the
$n$-branch of the modified Urca-process and
the $np$ bremsstrahlung by the asymptotic expressions
obtained by
Itoh and Tsuneto (1972)
for the modified Urca in the limit of strong
($T \ll  T_c$) singlet-state pairing
of neutrons;
they multiplied
the $nn$-scattering rate by $\exp(-\Del_n/\kB T)$.
Maxwell (1979)
just switched off the modified Urca process at temperatures
below $\tn$ or $\tp$.
In a number of articles various authors used the simplified
reduction factors proposed by
Malone (1974):
$\exp[-(\Del_n + \Del_p)/\kB T]$ for superfluidity
of neutrons and proton at once;
$\exp(-\Del_N/\kB T)$ for singlet-state superfluidity
of neutrons or protons
(for the neutron branch of the modified Urca-process
and $np$-bremsstrahlung).
The reduction of the $nn$-bremsstrahlung by the neutron superfluidity
was described by the expression
$\exp(-2\Del_n/\kB T)$. The accuracy of these approximations
was analysed in Sects.\ 4 and 5. Their effect on the cooling
is demonstrated in Fig.\ \ref{fig:cc8}.

The complete set of neutrino reactions considered in Sects.\
4 and 5 was first used by
Levenfish and Yakovlev (1996).
These authors implied also a more accurate description of the effects
of superfluidity on the heat capacity and neutrino luminosity
(Sects.\ 3 -- 5), but neglected
the neutrino emission due to Cooper pairing of nucleons
(Sect.\ 6). The latter process has been included
only in the recent simulations by
Schaab et al.\ (1997b),
Page (1998a, 1998b),
Levenfish et al.\ (1998, 1999),
and Yakovlev et al.\ (1998, 1999).
Let us mention that
Levenfish et al.\ (1998, 1999) and Yakovlev et al.\ (1998, 1999)
have made use of more accurate expressions for
the neutrino emissivity given in Sect.\ 6
(although neglected the relativistic correction to
the neutrino emissivity due to pairing of protons).

\begin{figure}[t]                          
\begin{center}
\leavevmode
\epsfysize=8.5cm
\epsfbox[40 80 515 340]{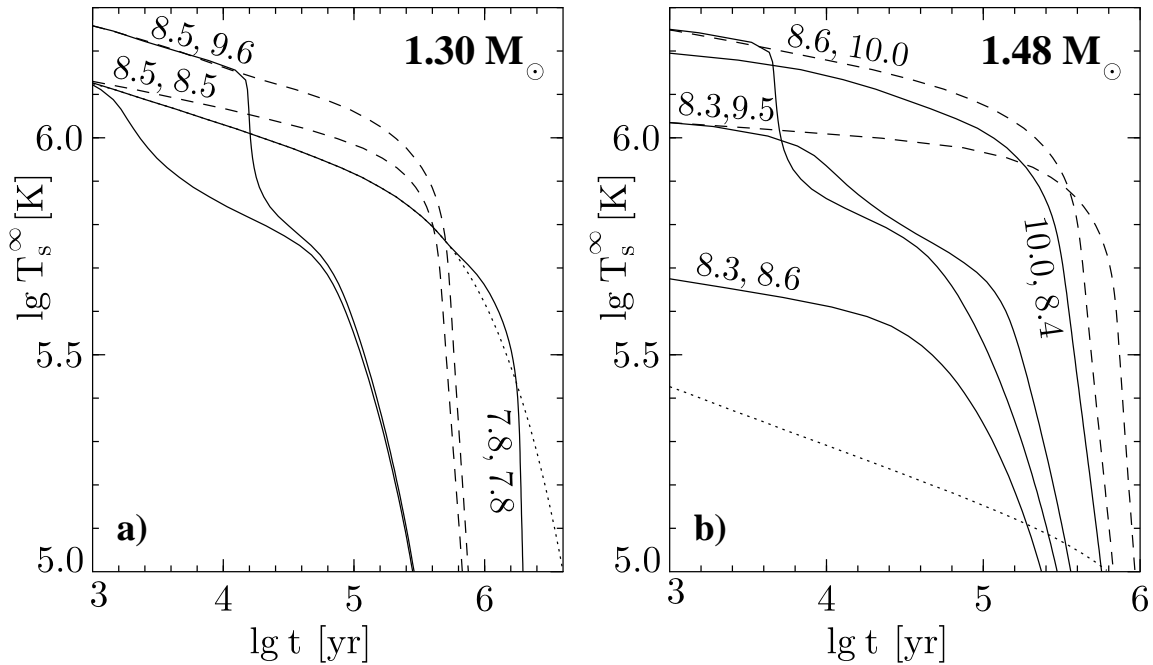}
\end{center}
\caption[]{\footnotesize
Standard cooling of a NS with $M=1.30\, M_\odot$
(Fig.\ a) and enhanced cooling of a NS with $M=1.48 \, M_\odot$
(Fig.\ b) in the presence of superfluidity.
The curves are labeled by the values of
$\lg \tn$ and $\lg \tp$.
Solid lines are obtained including the neutrino emission
due to pairing of nucleons while dashed lines are
obtained neglecting this emission. Dotted lines show cooling
of non-superfluid NSs.
The parameters of the models are given in Sect.~7.2.2.
The solid and dashed lines for
$\lg \tn=7.8$ and $\lg \tp=7.8$ in Fig.\ (a)
coincide with each other and so do the lines
for $\lg \tn=8.3$, $\lg \tp=8.6$ and
$\lg \tn=10.0$, $\lg \tp=8.4$
in Fig.\ (b).
}
\label{fig:rec_es}
\end{figure}

The importance of neutrino emission due to Cooper pairing
of nucleons for the standard NS cooling is illustrated in
Fig.\ \ref{fig:rec_es}a. One can see that the appearance of
superfluidity at the neutrino cooling stage can not only slow down
the NS cooling, as thought before, but, on the contrary,
strongly accelerate it
(see, e.g.,
Yakovlev et al., 1999).
The acceleration can be so fast that
at $t\sim 10^5$--$10^6$ yr
the star can be much cooler that in the case of
 the
enhanced neutrino energy losses.
An exclusion is provided by superfluidity of protons alone
since proton pairing does not produce intense neutrino emission.
In addition, the difference between the cooling curves is small
																	     in the case of normal protons and strongly superfluid neutrons
($\tn \ga 10^9$~K), in which the Cooper pairing of neutrons
is not a dominant process (Sect.\ 6.2). \\[0.5ex]

{\bf Simulation of enhanced cooling}
of NSs with the standard composition of the cores
was started by
Page and Applegate (1992).
These authors were the first who discovered an
interesting feature in the cooling of NSs in which the
direct Urca process was open and neutrons or protons were
superfluid: after thermal relaxation was over
the surface temperature of these stars
fell down rapidly to the value
$T_s=T_s(T_i)$, appropriate to the internal temperature
$T_i = \alpha T_c$, and remained almost constant
during subsequent neutrino cooling stage.
Here $T_c$ is the critical temperature of neutrons or
protons in the core,
and  $\alpha \sim 1$ is a numeric coefficient
(for instance,
Page and Applegate, 1992,
obtained $\alpha \approx 0.2$). Moreover, the temperature
$T_i$ is almost insensitive to other NS parameters
(Page, 1994).
The nature of this phenomenon is very simple ---
the star cools rapidly before the superfluidity onset,
whereas the superfluidity suppresses the cooling and
``freezes" the internal temperature at the level
of $\alpha T_c$. This opens possibility to study superfluidity in those
NSs whose surface temperature is known from observations.

Page (1994) analyzed this possibility
in more detail (neglecting the Cooper--pairing neutrino emission).
He showed that the measured surface temperatures
of PSR 0656+14 
(Finley et al., 1992)
and Geminga
(Halpern and Ruderman, 1993)
could be explained in the standard and enhanced
cooling models. The observations of the Vela pulsar
(\"{O}gelman et al., 1993)
are more easily explained by the enhanced cooling model,
and the observations of
PSR 1055-52
(Brinkmann and \"{O}gelman, 1987) by the standard cooling.
In all these cases one needs (Page, 1994) the presence
of the neutron and proton superfluidities with
high critical temperatures $T_c \sim 10^9$~K
in the entire NS body
where the main neutrino emission occurs.

Let us stress that the results and quantitative conclusions by
Page and Applegate (1992) and
Page (1994)
are considerably modified by including the effects of joint superfluidity
of nucleons and Cooper-pairing neutrinos. However the principal
conclusion by Page and Applegate remains the same:
a NS of age 10$^2$--10$^5$~yr with superfluid nucleons in its core
is a ``thermometer" of this superfluidity ---
one can measure (constrain) the critical temperatures
of neutrons $\tn$ and protons $\tp$ from the values of the surface
temperature.

Aside of
Page and Applegate (1992) and
Page (1994)
,
the enhanced cooling of NSs with the standard nuclear composition
and nucleon superfluidity in the NS cores
has been studied, for instance, by
Gnedin and Yakovlev (1993),
Van Riper and Lattimer (1993),
Gnedin et al.\ (1994),
Levenfish and Yakovlev (1996),
Page (1998a, b),
Levenfish et al.\ (1998, 1999),
Yakovlev et al.\ (1998, 1999).
As in the case of the standard cooling,
the effect of superfluidity on the heat capacity and neutrino luminosity
was described in earlier articles by the approximate factors of the form
$\exp(-\Del_N/\kB T)$, but gradually the theory
presented in Sects.\ 3--5 has become implemented.
The difference of the standard and enhanced cooling curves
calculated using the accurate (Sects.\ 3--5) and approximate
($\exp(-\Del_j/\kB T)$) reduction factors is illustrated in Fig.\
\ref{fig:cc8}. The accurate reduction factors
``freeze" the cooling at the neutrino cooling stage
at a lower internal temperature
$T$ and, accordingly, at a lower surface temperature
(Gnedin and Yakovlev, 1993).

In calculations of the enhanced neutrino luminosity
the contribution from Cooper pairing of nucleons (Sect.\ 6)
has usually been neglected
(excluding the articles by
Schaab et al., 1997b;
Page 1998a, b;
Levenfish et al., 1998, 1999;
Yakovlev et al., 1998, 1999).
According to Yakovlev et al.\ (1999)
this process is especially important in the presence
of the proton superfluidity with $\tp \gg \tn$. Such superfluidity
appears at the early cooling stage and suppresses
the direct Urca process long before the onset of the neutron
pairing. Therefore, a splash of the neutrino emission
associated with the neutron pairing is very pronounced.
In this case the NS cooling does not slow down, as it would be
without neutrino emission due to nucleon 
pairing, but is strongly accelerated
(Fig.\ \ref{fig:rec_es}b). Let us remind that similar
situation takes place in the case of the standard cooling
(also see Sect.\ 7.2). \\[0.5ex]

{\bf As a result, the nucleon superfluidity plays the leading role
for the both, the standard and enhanced, cooling}
(Page 1998a, b;
Levenfish et al., 1998, 1999;
Yakovlev et al., 1998, 1999).
Moreover, the superfluidity mixes the
cooling types, since the enhanced cooling may look
like the standard one while the standard cooling may look
like enhanced (especially if the neutrino emission
due to Cooper pairing is taken into account,
cf.\ Figs.\ \ref{fig:rec_es}a and
\ref{fig:rec_es}b). In these cases, the star, independently
of its cooling type, remains to be a good ``thermometer"
for measuring the nucleon critical temperatures:
even weak variations of $\tn$ or $\tp$  affect strongly the cooling
curves. The only problem is to
``calibrate" properly this thermometer (Sects.\ 3--6).

Finally, let us mention that
Schaab et al.\ (1998b) have considered recently, for the first time,
the standard and enhanced cooling in the
presence of the triplet-state neutron superfluidity of type {\bf C}
(with nodes of the gap at the Fermi surface).
The authors have been based on the model calculations
of neutron pairing in the strong magnetic fields
by Muzikar et al.\ (1980) according to which the field
$B \ga 10^{16}$ G makes the pairing of type
{\bf C} energetically more favourable than that of type
{\bf B}. Let us notice also the first calculations of cooling
(Schaab et al., 1998a)
of a NS with the hyperonic core taken into account
the hyperon superfluidity.
We mention, however, that
Schaab et al.\ (1998a, 1998b)
have made use of approximate reduction factors of neutrino reactions
and approximate expressions for the neutrino emissivity
due to Cooper pairing of particles.

The theoretical conclusion on the leading role of
superfluidity in NS cooling was made at about the same
time when new observational data on the surface temperatures
of some NSs appeared (Sect.\ 8). The latter data brought
the problem of studying superfluidity in the NS cores
to reality.

\subsection{Cooling simulations}
In this section we illustrate the possibility
to explore nucleon superfluidity in the NS cores.
We will mainly follow our recent article
(Levenfish et al., 1999).
In addition, we include into the simulations
the relativistic correction to the neutrino luminosity
produced by proton pairing (Sect.\ 6.1).
It affects slightly the cooling curves at some
values of $\tn$ and $\tp$ but does not change
the main conclusions by
Levenfish et al.\ (1999).

\subsubsection{Simulation scheme}
Below we present the results obtained with the cooling code
(Gnedin and Yakovlev, 1993;
Gnedin et al., 1994;
Haensel and Gnedin, 1994;
Levenfish and Yakovlev, 1996;
Levenfish et al., 1998)
that is based on the approximation of isothermal
stellar interiors.
The code was mainly constructed by O.Y.\ Gnedin.
The isothermal approximation is valid for a star of age
$t>(10$--$10^3)$ yr, inside which the thermal
relaxation is over. 
The superfluidity affects strongly the surface temperature
only after reaching the thermally relaxed stage.
Following
Glen and Sutherland (1980)
we have assumed that the isothermal region is
restricted by the condition
$\rho> \rho_b \! = \! 10^{10}$ g cm$^{-3}$.
The real boundary of the isothermal region is situated
at lower $\rho$ and depends on temperature:
it moves to the surface in the course of cooling.
The chosen value of $\rho_b$ guarantees that the region
$\rho > \rho_b$ is isothermal in a star of age $t \ga 10^3$ yr.
The quantity
$\;T_i(t)=T(r,t) \, \exp[\Phi(r)]$
(which may be called the internal redshifted temperature)
is constant over the isothermal region at any moment of time
$t$. Here $T(r,t)$ is the local temperature,
$\Phi (r)$ is the dimensionless gravitational potential, and
$r$ is radial coordinate.

We take into account explicitly
the effects of General Relativity
on the NS structure and cooling.
Cooling simulation in the isothermal approximation
is reduced to solving the equation of thermal balance
(see, e.g.,
Glen and Sutherland, 1980)
\begin{eqnarray}
   &&C(T_i) \, \frac{\dd T_i}{\dd t} =
    -L_{\nu i} (T_i) - L_{\gamma i} (T_s)\,;
\label{eq:Balance} \\
    &&C = \int c_v \dd V,   \;\;\;\; \dd V = 4 \pi r^2
              \left(1 - \frac{2Gm}{rc^2} \right)^{-1/2} \dd r \, ,
\label{eq:Crel} \\
    &&L_{\nu i} = \int Q \,\exp(2\Phi) \, \dd V,
     \;\;\;\;\;\;
    L_{\gamma i} = 4 \pi {\cal R}^2 \sigma T^{\,4}_s \, \exp(2 \Phi_b).
\label{eq:Lnu}
\end{eqnarray}
In this case, $C$ is the total heat capacity of the star,
$c_v$ is the specific heat capacity, and
$m=m(r)$ is the gravitational mass inside a sphere with
radial coordinate $r$. The quantities
$L_{\nu i}$ and $L_{\gamma i}$ determine the neutrino and
photon NS luminosities, respectively,
$Q$ is the neutrino emissivity,
$\;\Phi_b$ is the value of $\Phi$ at the isothermal region
boundary ($\rho\! = \! \rho_b$), and
$\;\sigma$ is the Stefan--Boltzmann constant.

The photon luminosity of the star depends on the effective
temperature $T_s$ of its surface. The relationship between
$T_s$ and the temperature $T_b=T_i \exp( - \Phi_b)$ at
$\rho \! = \! \rho_b$ is determined by thermal insulation
of the outer envelope ($\rho \! < \! \rho_b$),
where the main temperature gradient is formed.
We have not taken into account the effect of the magnetic field
on the NS cooling and have used the relationship
$T_s=T_s(T_b)$, obtained recently by
Potekhin et al.\ (1997)
for $B=0$. We assume that the NS surface may be
covered by a thin layer (of mass $\la 10^{-13} \, M_\odot$)
of hydrogen or helium. This amount of light elements
is too small to affect the thermal insulation of the envelope
and the NS cooling but it can affect the spectrum of thermal radiation.
Actually, the effect of the dipole surface magnetic fields
$B \la 5 \times 10^{12}$ G on the NS cooling is rather weak
(Shibanov and Yakovlev, 1996). Therefore, our calculations
can be used, at least qualitatively, for the stars with
such fields. In these cases, under
$T_s$ we mean the average effective surface temperature
which determines the total (nonredshifted) photon surface luminosity
of the star as
$L_\gamma= 4 \pi \sigma {\cal R}^2 T^{\,4}_s$.

\subsubsection{Models of cooling neutron stars}
For simplicity, we assume that matter of the NS core
consists of neutrons, protons and electrons (muons and hyperons
are absent). We will adopt
a moderately stiff equation of state
proposed by Prakash et al.\ (1988)
(the version with the compression modulus
$K_0=180$ MeV and the same simplified symmetry factor
$S_V$, that was used by
Page and Applegate, 1992).
The maximum NS mass, for a given equation of state,
is $1.73 \, M_\odot$. In order to study the enhanced and standard cooling
we consider the NS models of two masses. In the first case,
the NS mass is $M=1.48 \, M_\odot$,
radius ${\cal R}=11.44$~km, and the central density
$\rho_c \! = \! 1.376 \times 10^{15}$ g cm$^{-3}$, while in the second case
$M=1.30 \, M_\odot$, ${\cal R}=11.87$~km, and
$\rho_c \! = \! 1.07 \times 10^{15}$ g cm$^{-3}$.
The adopted equation of state opens the direct Urca process
at the density
$\rho \! > \! \rho_{cr}\! = \! 1.30 \times  10^{15}$ g cm$^{-3}$.
Therefore, the $1.48 \, M_\odot$ NS suffers
the {\it enhanced} cooling: the powerful direct Urca process
(\ref{eq:Durca_Nucleon})
is allowed in a small central kernel of radius
2.32~km and mass $0.035 \, M_\odot$
(in addition to the processes (\ref{eq:Murca_N})--(\ref{eq:Brems})
and (\ref{eq:Recomb}) in the entire core).
In the $1.30\, M_\odot$ NS the threshold value $\rho_{cr}$
is not reached, and the star has the {\it standard}
neutrino luminosity, determined by the
processes (\ref{eq:Murca_N})--(\ref{eq:Brems})
and (\ref{eq:Recomb}).
Notice that
Levenfish and Yakovlev (1996),
while calculating the equation of state,
set the parameter
$n_0$ (the standard saturation nuclear--matter density)
equal to 0.165 fm$^{-3}$. In the present calculations we set
$n_0=0.16$ fm$^{-3}$. Owing to this reason the mass
of the rapidly cooling NS model is somewhat different
from that ($1.44\, M_\odot$)
used by
Levenfish and Yakovlev (1996).

In order to calculate the heat capacity of the superfluid 
NS core and its neutrino
luminosity due to the reactions
(\ref{eq:Murca_N})--(\ref{eq:Recomb})
we have used equations of Sects.\ 3--6.
The results obtained in these sections are summarized
in Sect.\ 6.2. The neutrino bremsstrahlung due to $ee$
scattering in the NS core has been neglected: we have shown
that its effect on the cooling is negligible, for our NS models.
We have also taken into account the neutrino luminosity
of the NS crust due to bremsstrahlung of electrons
which scatter off atomic nuclei
(using an approximate formula
proposed by
Maxwell, 1979):
$ L_{br} = 1.65 \times 10^{39}$ $ (M_{cr}/ M_\odot)
           (T_b/10^9 \mbox{K})^6 \exp(2 \Phi_b)$
erg$\;$s$^{-1}$,
where $M_{cr}$ is the crust mass.  In our case,
$ M_{cr}=0.0120 \, M_{\odot}$
for the $1.48 M_{\odot}$ model,
and $ M_{cr}=0.0153 \, M_{\odot}$ for the $1.3 \, M_{\odot}$ model.
The NS heat capacity has been set equal to the sum
of the partial heat capacities of $n$, $p$ and $e$ in the stellar core;
the contribution of the crustal heat capacity
has been neglected due to the smallness
of the crust mass for the chosen NS models.
In the calculations of the neutrino luminosity
and the heat capacity the effective masses
of neutrons and protons in the NS cores have been taken as
$m_N^\ast=0.7 \, m_N$.

Various microscopic theories of Cooper pairing
in the NS cores give very wide scatter of the
critical temperatures of the neutron and proton superfluids
,
$\tn$ and $\tp$ ($\sim 10^7$--$10^{10}$~K), and various density
dependence of these temperatures (Sect.\ 3.1).
Thus we have made a simplified assumption that the critical
temperatures
$\tn$ and $\tp$ are constant throughout the NS core and
may be treated as free parameters. This is the main simplification
of our simulations. We assume that protons suffer
$^1{\rm S}_0$-state pairing (superfluidity of type {\bf A},
Table 1) and neutrons suffer
$^3{\rm P}_2$-state pairing with zero projection of the pair moment
onto quantization axis (superfluidity of type {\bf B}).
We will study the cooling for different values of
$T_{cn}$ and $T_{cp}$,
and we will determine (Sect.\ 8.2.2) those values
which are in better agreement with observations.

As already mentioned in Sect.\ 7.1,
the difference between the enhanced and standard cooling regimes
is smeared out in the presence of superfluidity.
Under these conditions, the cooling regime formally indicates
if the direct Urca is allowed or forbidden in the NS core.

\subsubsection{Results}
We have calculated about 2000 NS cooling curves.
The curves describe the dependence of the effective surface
stellar temperature
$T_s^\infty = T_s \sqrt{1- {\cal R}_g/{\cal R}}$,
as detected by a distant observer, on age $t$; here,
${\cal R}_g$ is the gravitational NS radius.
For  $M=1.30 \, M_\odot$ and $M=1.48 \, M_\odot$ we have
$T^\infty_s/T_s= 0.822$ and $0.786$, respectively.
The critical temperatures of neutrons and protons,
$\tn$ and $\tp$, in the NS core have been varied in a wide
interval from $10^6$ to $10^{10}$~K.

\begin{figure}[t]                          
\begin{center}
\leavevmode
\epsfysize=8.5cm
\epsfbox[75 250 500 650]{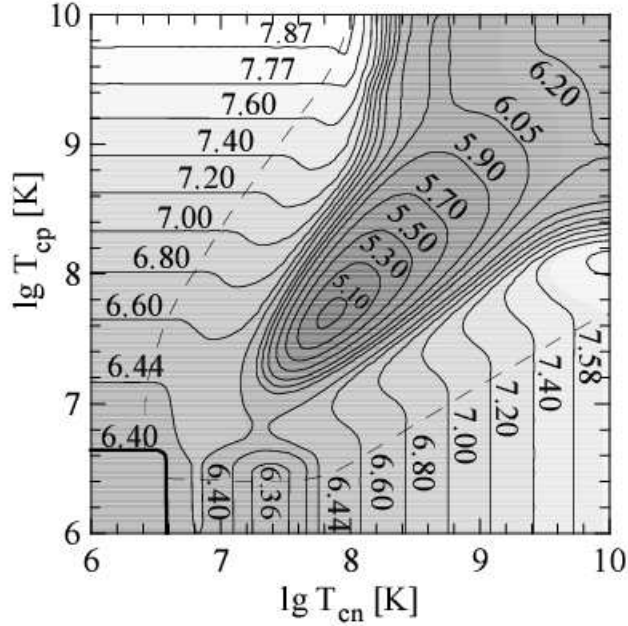}
\end{center}
\caption[]{\footnotesize
Lines of values of $T_{cn}$ and $T_{cp}$ which
correspond to certain internal temperatures
$T_i$ (the values of $\lg T_i$ are given near the curves)
or surface temperatures
$T_s^\infty$ (given in Table \protect{\ref{etab:tstotc}})
of a NS with the enhanced neutrino luminosity
($1.48 \, M_\odot$)
and the Geminga's age ($3.4 \cdot 10^5$ yr). The darker the
background the cooler is the star. The region of joint
neutron and proton superfluidity (in the center and the upper
right corner) is enclosed by the dashed lines.
A small region where the superfluidity does not
appear by the given age (the left lower corner) is separated
by the thick solid line.
}
\label{fig:Gem_2}
\end{figure}

\begin{figure}[t]                          
\begin{center}
\leavevmode
\epsfysize=8.5cm
\epsfbox[75 260 500 650]{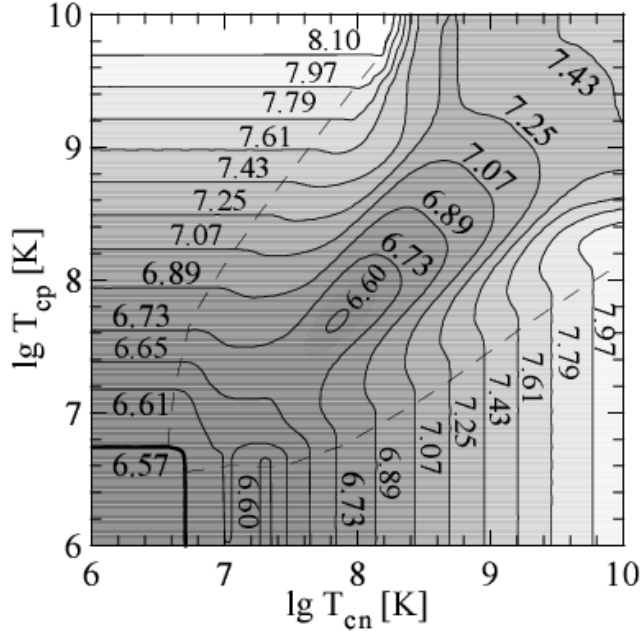}
\end{center}
\caption[]{\footnotesize
Same as in Fig.\ \protect{\ref{fig:Gem_2}}, but for a NS
of the PSR$\, 0656+14$ age ($10^5$ yr).}
\label{fig:Gem_3}
\end{figure}

\begin{figure}[t]                          
\begin{center}
\leavevmode
\epsfysize=8.5cm
\epsfbox[100 270 500 660]{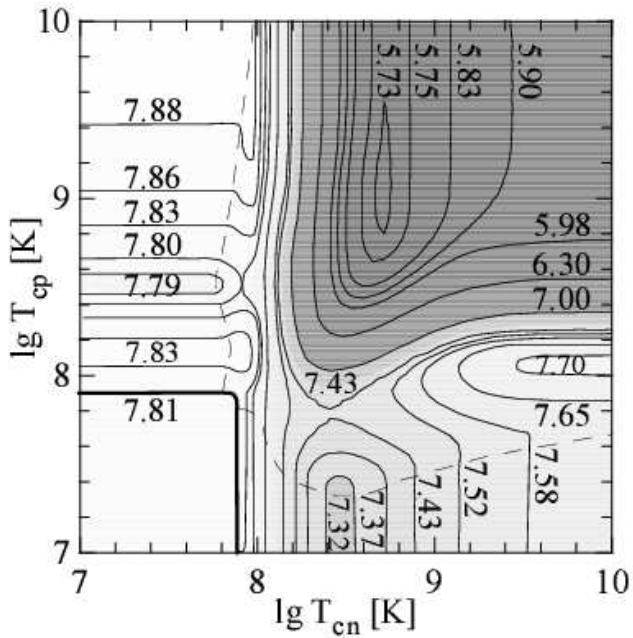}
\end{center}
\caption[]{\footnotesize
Isotherms of the internal temperature $T_i$ (or
the surface temperature $T_s^\infty$,
Table \ref{etab:tstotc}) for a NS of the Geminga's age,
as in Fig.\ \protect{\ref{fig:Gem_2}},
but for the standard cooling ($1.30 \, M_\odot$). }
\label{fig:Gem_4}
\end{figure}

\begin{figure}[t]                          
\begin{center}
\leavevmode
\epsfysize=8.5cm
\epsfbox[80 280 490 660]{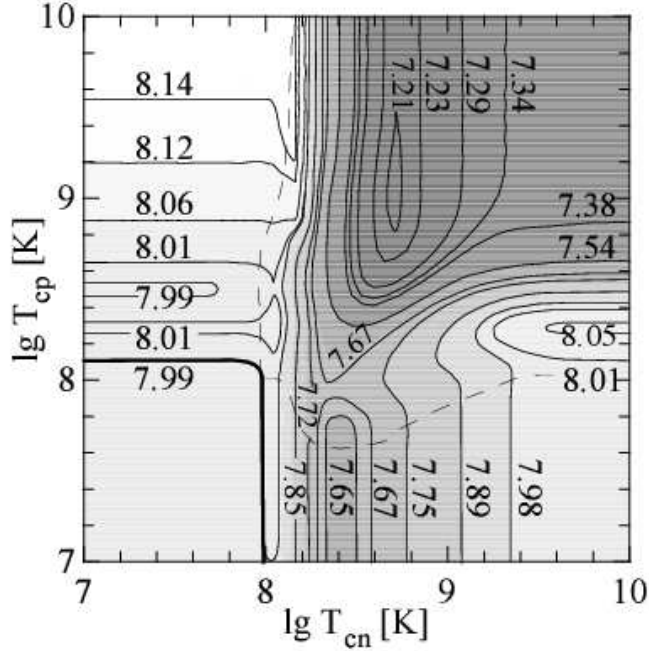}
\end{center}
\caption[]{\footnotesize
Same as in Fig.\ \protect{\ref{fig:Gem_4}},
but for a NS of the PSR$\, 0656+14$ age.}
\label{fig:Gem_5}
\end{figure}

The examples of the cooling curves for selected values of
$\tn$ and $\tp$ were already discussed above
(Figs.\ \ref{fig:rec_es}a and \ref{fig:rec_es}b).
However it is rather inconvenient to analyse the results
in the form of the cooling curves. It is better
to plot the values of ($\tn,\,\tp $), which lead to certain
surface temperatures $T^\infty_s$ of the NS of given age $t$.
Figures \ref{fig:Gem_2}--\ref{fig:Gem_5} present the results
in this way. For example, we have chosen the values of
$t$, which correspond to the ages of the Geminga pulsar
($3.4 \times 10^5$~yr, Figs.\ \ref{fig:Gem_2} and \ref{fig:Gem_4})
and PSR$\,0656+14$ ($10^5$~yr, Figs.\ \ref{fig:Gem_3} and \ref{fig:Gem_5}).
Observational data on these and other cooling NSs are
given in Sect.\ 8.1.

Before discussing the results of calculations let us remind
some relationships which determine the main features of the
NS cooling. We will use these relationships for describing
Figs.\ \ref{fig:Gem_2}--\ref{fig:Gem_5}.
		
The effects of the neutron and proton superfluidities on the
heat capacity are different. In a non-superfluid NS,
the heat capacities of $p$ and $n$
constitute $\sim 1/4$ and $\sim 3/4$ of the total heat
capacity $C_{tot}$, respectively. Therefore, a strong
superfluidity of $p$ reduces $C_{tot}$ by $\sim 25\%$,
while a strong superfluidity of
$n$ reduces it by about 4 times.
When the ratio of the internal stellar temperature
$T_i$ to a critical nucleon temperature
$T_{cN}$ decreases from 0.3 to 0.1 (for instance, in the course
of NS cooling), the heat capacity of nucleons
$N=n$ or $p$ decreases by more than three orders of magnitude
and becomes much lower than the heat capacity of electrons.
Further reduction of the nucleon heat capacity does not
change the total heat capacity. The appearance of the weak
superfluidity of nucleons $N$ almost doubles their heat
capacity due to the latent heat release at the phase transition.
The heat capacity of nucleons remains higher than the
heat capacity of non-superfluid nucleons as long as
$T_i/T_{cN}$ decreases from 1 to 0.3, i.e., as long as
the difference of temperature logarithms is
$\lg T_{cN} - \lg T_i \leq 0.5$.
					
The effects of superfluidities of $n$ and $p$ on the NS neutrino
luminosity are also different. For the direct and modified
Urca processes this difference is weak. If Cooper--pairing
neutrino emission were neglected, the asymmetry of Figs.\
\ref{fig:Gem_2}--\ref{fig:Gem_5} with respect to inversion
of the axes $\tn \lr \tp$, would be mainly explained by different
contributions of $n$ and $p$ into the heat capacity.
This can be proved by comparing Figs.\ \ref{fig:Gem_2}--\ref{fig:Gem_5}
with analogous figures in Levenfish and Yakovlev (1996),
where the neutrino emission due to Cooper pairing was neglected.
Inclusion of the Cooper--pairing neutrinos greatly amplifies
the asymmetry: the neutrino emission due to proton
pairing is weak, while the emission due to neutron pairing
dominates over the standard neutrino losses for
$T_i \la 10^9$~K $ \la \tn$  and over the direct
Urca process for $T_i \la \tn \ll \tp$.
The main neutrino energy release in the process takes place at
$0.2 \la T_i/T_{cn} \la 0.96$
(as long as the difference of the temperature logarithms is
$\lg T_{cn} - \lg T_i \leq 0.7$). \\[0.2ex]
		
\noindent 
{\bf Enhanced cooling.}
%
%
Figures \ref{fig:Gem_2} and \ref{fig:Gem_3} illustrate the enhanced cooling
of NSs of the ages of the Geminga and PSR$\,0656+14$ pulsars.

Figure \ref{fig:Gem_2} shows isotherms of the internal
temperature $T_i$ of the NS of the Geminga's age versus
$\tn$ and $\tp$. Since the temperature is related to
the surface temperature, the same lines are isotherms
of the surface temperature
$T_s^\infty$ (Table \ref{etab:tstotc}).
Dashes exhibit the auxiliary lines which
enclose the region of the joint superfluidity of nucleons.
The region to the left of the upper dashed line corresponds
to the superfluidity of protons and normal neutrons, while
the region to the right of the lower line corresponds to
the superfluidity of neutrons and normal protons.
The dashed lines intersect at the isotherm of the temperature
$T_i=10^{6.4}$~K, to which a non-superfluid star would cool down
by the moment $t$. This isotherm, plotted by the thick line,
encloses the region where the nucleon superfluidity
has not appeared by the moment $t$ and does not affect the cooling.
Notice, that owing to the effects of General Relativity,
the isotherms $T_i$ correspond to somewhat higher local
temperatures of matter (see above). Thus, the isotherm
$T_i = 10^{6.4}$~K is associated with the values of
$T_{cN}$ which are slightly higher than $T_i$.

First we discuss behaviour of
iso\-therms to the right of the lower auxiliary line.
The vertical segments of the isotherms reveal that
the NS cooling is governed by the only superfluidity of $n$.
With increasing $\tn$ at $\lg T_{cn} \ga 7.7$,
the core temperature $T_i$ grows up which can be explained like this.
The higher $\tn$, the earlier the neutron superfluidity appears.
Accordingly, the powerful direct Urca process is suppressed
earlier, cooling delay is longer, and the NS is
hotter at given $t$. For low $\tn$, the NS cools down
in a non-superfluid state almost all its life
and is sufficiently cold by
the age $t$. The neutrino luminosity of such a star is rather
weak and becomes comparable to the photon luminosity.
Additional neutrino energy losses due to neutron pairing at
$\lg \tn \ga 6.6$ slightly accelerate the cooling
at $6.6 \la \lg \tn \la 6.6 + 0.7$.
However, for $\lg  \tn \la 6.6 + 0.5$ the acceleration
is compensated by the latent heat release produced by
the jump of the neutron heat capacity. The values
0.7 and 0.5 have been explained above.

Vertical segments of the isotherms intersect
the auxiliary line at temperatures
$\tp$, at which the proton superfluidity appears.
With further increase of $\tp$,
the NS heat capacity (determined by protons for $\lg \tn \ga 7.4$)
jumps up by about twice.
The latent heat release increases
$T_i$: in a strip of width 
$\sim$0.5 in $\lg \tp$ above the lower dashed line
the isotherms shift to the left. With further growth of
$\tp$ the heat capacity is strongly reduced, the NS becomes
cooler and the isotherms shift to the right.

In a similar fashion, exchanging the proton and neutron superfluids,
we can explain horizontal segments of
 the
isotherms to the left of the upper auxiliary line.
In this case acceleration of the NS cooling in the range
$\lg \tp \ga  6.6$ does not appear because of
the weakness of the neutrino emission produced by
the proton pairing. A strong proton superfluidity suppresses
the heat capacity weaker than a neutron superfluidity
with the same critical temperature. Therefore,
at high $\tp$ the star is warmer than for the same
$\tn$ at vertical segments of isotherms
in the lower right part of the figure.

\newcommand{\s}{$\;\;$}
\begin{table}[t]
\caption[]{Relationship between $T_i$ and $T_s^{\infty}$ [K] for Fig.\
\protect{\ref{fig:Gem_2}--\ref{fig:Gem_5}}. }
\label{etab:tstotc}
\begin{center}
\begin{tabular}{|p{0.8cm}|p{14.7cm}|}
\hline
\multicolumn{2}{|c|}{$ M= 1.48\, M_\odot^{\rrr}$}\\[1ex]
\hline
\parbox[t]{1.2cm}{lg$T_{i}^{\rrr}$\\[-0.3ex]
lg$T_{s}^{\infty}$\\[-1ex]
}
&
\parbox[t]{15.8cm}{5.10\s 5.30\s 5.50\s 5.70\s 5.90\s 6.05\s 6.20\s 6.36\s
6.40\s 6.44\s 6.57\s 6.60\s 6.61\s
6.65\s 6.73  \\[-0.3ex]
4.53\s 4.61\s 4.68\s 4.76\s 4.84\s 4.91\s 4.98\s 5.05\s
5.07\s 5.09\s 5.16\s 5.17\s 5.18\s
5.20\s 5.24  \\[-1ex]
} \\
\hline
\parbox[t]{1.2cm}{lg$T_{i}^{\rrr}$\\[-0.3ex]
lg$T_{s}^{\infty}$\\[-1ex]
}
&
\parbox[t]{15.8cm}{6.80\s 6.89\s 7.00\s 7.07\s 7.20\s 7.25\s 7.40\s 7.43\s
7.60\s 7.61\s 7.77\s 7.79\s 7.87\s
7.97\s 8.10  \\[-0.3ex]
5.28\s 5.32\s 5.38\s 5.42\s 5.49\s 5.52\s 5.60\s 5.62\s
5.71\s 5.72\s 5.81\s 5.82\s 5.86\s
5.92\s 5.99 \\[-1ex]
} \\
\hline
\end{tabular}\\[2ex]
\begin{tabular}[c]{|p{0.8cm}|p{11.7cm}|}
\hline
\multicolumn{2}{|c|}{$ M= 1.30\, M_\odot^{\rrr}$}\\[1ex]
\hline
\parbox[t]{0.8cm}{lg$T_{i}^{\rrr}$\\[-0.3ex]
                  lg$T_{s}^{\infty}$\\[-1ex]
}
&
\parbox[t]{15.3cm}{5.73\s 5.75\s 5.83\s 5.90\s 5.98\s 6.30\s 7.00\s 7.21\s
7.23\s 7.29\s 7.32\s 7.34 \\[-0.3ex]
\s 4.78\s 4.79\s 4.82\s 4.85\s 4.88\s 5.02\s 5.37\s 5.48\s
5.49\s 5.53\s 5.54\s 5.55 \\[-1ex]
}\\
\hline
\parbox[t]{1.2cm}{lg$T_{i}^{\rrr}$\\[-0.3ex]
lg$T_{s}^{\infty}$\\[-1ex]
}
&
\parbox[t]{15.8cm}{7.38\s 7.43\s 7.52\s 7.54\s 7.58\s 7.65\s 7.67\s 7.70\s
7.75\s 7.79\s 7.80\s 7.81\\[-0.3ex]
\s 5.57\s 5.60\s 5.65\s 5.66\s 5.69\s 5.72\s 5.74\s 5.75\s
5.78\s 5.80\s 5.81\s 5.81 \\[-1ex]
}\\
\hline
\parbox[t]{1.2cm}{lg$T_{i}^{\rrr}$\\[-0.3ex]
lg$T_{s}^{\infty}$\\[-1ex]
}
&
\parbox[t]{15.8cm}{7.83\s 7.85\s 7.86\s 7.88\s 7.89\s 7.98\s 7.99\s 8.01\s
8.05\s 8.06\s 8.12\s 8.14 \\[-0.3ex]
\s 5.82\s 5.84\s 5.84\s 5.85\s 5.86\s 5.91\s 5.91\s 5.93\s
5.95\s 5.95\s 5.99\s 6.00 \\[-1ex]
} \\
\hline
\end{tabular}
\end{center}
\end{table}

The horizontal segments of the isotherms intersect
the upper auxiliary line at temperatures
$\tn$ which switch on the neutron superfluidity.
This superfluidity induces the latent heat release
and associated slight NS heating (dips on the isotherms
to the right of the auxiliary line). With increasing
$\tp$, the NS heating produced by the neutron superfluidity
becomes weaker and disappears (the dips vanish).
The effect is mainly caused by the neutrino emission
due to neutron pairing which is more pronounced
if the direct Urca process is strongly suppressed.
At high $\tp$ the neutron pairing becomes the main neutrino
emission mechanism. Since the neutrino emission due to
the neutron pairing is more efficient than
the latent heat release the appearance of the neutron superfluidity
with the growth of $\tn$ does not delay the cooling but, on the contrary,
accelerates it. A noticeable NS cooling
via Cooper-pairing neutrinos takes place
in a strip of the width 0.7 (in $\lg \tn$) to the right of
the upper dashed line. Lower and to the right of this strip
the cooling is associated with reduction of
the neutron heat capacity.

Now consider the region of
$\tn$, $\tp$ between the auxiliary lines in more detail.
Some increase of $T_i$ at
$\lg \tn \la 6.6 +0.5$ and $\lg \tp \la 6.6 + 0.5$
is caused by the latent heat release at onset
of superfluidity of $n$ and $p$. Further growth of $\tn  \sim  \tp$
induces initially rapid decrease and then weak increase
of $T_i$. The decrease is explained by exponential reduction
of the heat capacity by the joint superfluidity of $n$ and $p$
while the weak increase is associated with reduction of the
direct Urca process at the early cooling stages.

Figure \ref{fig:Gem_3} is analogous to Fig.\ \ref{fig:Gem_2},
but corresponds to a younger NS of age $t=10^5$~yr.
Figures \ref{fig:Gem_2} and \ref{fig:Gem_3} show
that one needs $\tp \ll \tn$ or $\tn \ll \tp$
to support high surface temperature $T_s^\infty$ (or $T_i$)
for a longer time.\\[0.2ex]

\noindent 
{\bf Standard cooling.}
%
%
Standard cooling of a $1.30\, M_\odot$ NS is illustrated in
Figs.~\ref{fig:Gem_4} and \ref{fig:Gem_5} for the stars of
Geminga and PSR $\, 0656+14$ ages, respectively.
Isotherms are qualitatively different from
those for the enhanced cooling
(cf.\ with Figs.\ \ref{fig:Gem_2} and \ref{fig:Gem_3}):
even an approximate symmetry of the neutron and proton
superfluidities is absent. The asymmetry can be attributed to
the weakness of the standard neutrino energy losses.
First, in the absence of such a powerful cooling regulator as
the direct Urca process the difference
of the heat capacities of $n$ and $p$ (see above) is more pronounced.
Second, the Cooper-pairing neutrino emission becomes
more important on the background of weaker neutrino emission
produced by other neutrino reactions; the Cooper-pairing emission
is asymmetric itself, being more efficient for neutrons
than for protons.

If the superfluidity is absent, a NS
($1.30 \, M_\odot$) enters the photon cooling stage
at $t_\nu \sim 1.6 \times 10^5$~yr.
Thus the PSR$\, 0656+14$ pulsar appears at the transition stage
and Geminga is at the photon stage.
The neutrino luminosity is already weak and 
the superfluidity affects the cooling mainly either through
the heat capacity or through the neutrino emission
due to Cooper pairing of neutrons and, to a less extent, of protons.

Consider, for instance,  Fig.\ \ref{fig:Gem_4}.
Two dashed auxiliary lines enclose the domain of
joint superfluidity of nucleons.
To the left of the upper line protons are superfluid and
neutrons not, while below the lower line only neutrons are
superfluid. The lines intersect at the isotherm of the temperature
$ \lg T_i \approx 7.81$, which a non-superfluid NS
would have by the age $t$. The superfluidity with
$\lg \tn \la 7.9$ and $\lg \tp \la 7.9$ does not appear
by the moment $t$ and does not affect the cooling.

Horizontal segments of isotherms to the left of the upper dashed line
show that the cooling is regulated by the proton superfluidity alone.
For $ \lg \tp \ga 7.9$~K, the superfluidity appears just before
the given moment $t$ and is weak.
In the range $7.9 \la \lg \tp \la 7.9 + 0.5$
it initiates the latent heat release and a weak increase of
the total heat capacity (determined mainly by normal neutrons).
In the range $7.9+0.1 \la \tp \la 7.9+0.7$ the weak raise of the
heat capacity is compensated by the (also weak) increase
of the neutrino emission due to pairing of protons.
As a result, $T_i$ initially increases, and then decreases.
For $\lg \tp \ga 8.4$, the proton heat capacity is
reduced, and the total NS heat capacity decreases by $\sim$ 25\%.
On the other hand, for high $\tp$, the proton superfluidity onset
is shifted to the neutrino cooling stage. The cooling delay produced by
the suppression of the neutrino luminosity in the neutrino era
is somewhat stronger than the cooling
acceleration produced by the effect of superfluidity on the
heat capacity. Thus $T_i$ continues its growth with
increasing $\tp$.

Horizontal segments of the isotherms end at temperatures
$\tn \sim 10^8$~K, at which the neutron superfluidity is switched on.
The neutron pairing induces a splash of neutrino emission and 
cooling acceleration in a strip of the width 
$\sim$0.7 (in $\lg \tn$) to the right of the auxiliary line.
The minimum of $T_i$ takes place in the interval
$8.5 \la \lg \tn \la 8.7$,
because at $8.0 \la \lg \tn \la 8.0 + 0.5 $ the neutrino energy losses
are partly compensated by the latent heat release. The lowest temperatures
$T_i$ are realized in the case $\tp \gg 10^8$~K, in which
the nucleon heat capacity suffers the strongest suppression.

On vertical segments of the isotherms below the lower dashed line
the cooling is regulated by the neutron superfluidity alone.
With growing $\tn$ in this domain, the temperature
$T_i$ varies in the same manner as with growing
$\tp$ in the domain of the purely proton superfluidity.
This happens because at $\lg \tn \ga 7.9$ the neutrino emission
due to the neutron pairing is important.
It speeds up the NS cooling in the range
$7.9 \la \lg \tn \la 7.9 + 0.7 $ (see above).
If $7.9 \la \lg \tn \la$ 7.9 + 0.5, the neutrino cooling is
partly compensated by the latent heat release. Therefore
the minimum of $T_i$ takes place in the range
$7.9+0.5 \la \lg \tn \la 7.9+0.7$.
It is not so deep as in the upper part of Fig.\
\ref{fig:Gem_4} since the nucleon heat capacity is
now suppressed only partly.

A strong neutron superfluidity reduces the heat capacity stronger
and the neutrino luminosity weaker than a strong proton superfluidity.
Owing to the weakness of the neutrino energy losses
this difference is sufficient for a NS with high $\tn$ and
normal $p$ to cool in a different way than at equally
high $\tp$ and normal $n$. The strong neutron superfluidity
delays the cooling of those NSs
which would be at the neutrino cooling stage
or at the neutrino-photon transition
stage if they were non-superfluid.
This is demonstrated in Fig.\ \ref{fig:Gem_5} 
(for PSR$\, 0656+14$);
in the absence of superfluidity at $t=10^5$~yr the pulsar
would be at the transition stage. The strong neutron superfluidity
accelerates the cooling of older NSs, e.g.,
of the Geminga's age (Fig.\ \ref{fig:Gem_4}).

Vertical segments of isotherms in Fig.\
\ref{fig:Gem_4} intersect the auxiliary line at temperatures
$\tp$, at which the proton superfluidity appears.
This superfluidity leads to the latent heat release
and to the growth of $T_i$ in a strip of the width
$\sim$0.5 in $\lg \tp$ above the lower dashed line.
With further growth of $\tp$ the heat capacity is strongly
reduced and the cooling is accelerated. For a very strong joint
superfluidity of $n$ and $p$
($ \tn \gg 10^{7.9}$~K, $\tp \gg 10^{7.9}$~K, the very right upper corner
of Fig.~\ref{fig:Gem_4}), the nucleon heat capacity and
the neutrino luminosity of the NS core are fully suppressed,
and the cooling is governed by the electron heat capacity.

Figure \ref{fig:Gem_5} is analogous to Fig.~\ref{fig:Gem_4},
but corresponds to a younger NS. Its neutrino luminosity
is somewhat higher and the relative contribution
of nucleon--pairing neutrinos is smaller.
The neutrino emission produced by pairing
affects the cooling weaker.

Comparison of the present calculations with observations
will be made in the next section.


\section{Thermal emission from neutron stars and
         superfluidity in their cores}

\subsection{Thermal emission from neutron stars}

\subsubsection{Observations of X-ray emission from neutron stars}

In the very first article devoted to NSs
Baade and Zwicky (1934) predicted that
NSs should be born hot and cool gradually emitting thermal radiation.
Since NSs are compact, their radiation is weak.
Modern detectors are able to register it only from the
closest ($D \la 1 $--$ 2 $ kpc) and sufficiently hot
($T_s \ga 10^5$--$ 10^6$~K) isolated NSs. The main radiation
flux for the surface temperatures
$T_s$ mentioned above is emitted in the soft X-ray and
hard ultraviolet spectral ranges (0.01--1~keV),
which are thus most favourable for observations.
This radiation, however, is strongly absorbed by
the Earth atmosphere and can be detected only
from balloons, rockets and space observatories.

The attempts to discover NSs from their thermal surface
emission undertaken in the middle of 1960s in the
first balloon experiments with X-ray detectors were not successful.
The NSs were discovered later, in 1967, as radio pulsars
(Hewish et al., 1968).
Soon afterwards they were also discovered in X-rays as X-ray pulsars,
X-ray bursters and transients; however, radiation from
these objects is determined by accretion of matter
on NSs in binary systems, but not by outflow of thermal
energy from the interiors of cooling NSs. A search for intrinsic
thermal radiation from NSs was being continued.

Wolff et al.\ (1975) as well as
Toor and Seward (1977)
attempted to detect thermal radiation of the Crab pulsar
in a balloon experiment during lunar occultation of the pulsar.
The thermal radiation was not detected but the upper limit
of the surface temperature of the pulsar was established which
does not differ considerably from the present upper limit.

Subsequent observations were mainly carried out
in the soft X-rays with the X-ray telescopes on board of the
orbital observatories
{\it Einstein} (1978--1981),
{\it EXOSAT} (The European
X-Ray Observatory Satellite, 1983--1986),
{\it ROSAT} (The R\"{o}ntgen Satellite,
1990-1998),
{\it ASCA} (The Advanced Satellite for
Cosmology and Astrophysics, operates since 1993)
and {\it RXTE} (The Rossi X-ray Timing Explorer,
launched in December, 1995).
In these studies, various types of detectors have been used:
gaseous (scintillation) photon proportional counters
which possess moderate (from $\sim 8$ to $ 40\%$, depending
on detector scheme and photon energy) spectral resolution
and rather large angular resolution ($\sim 12''$--$3' $);
solid-state microchannel photodetectors with high
angular resolution (down to $3''$) but giving almost no spectral
information; solid-state CCD matrices with higher spectral resolution
(to $ 3\%$), than the gaseous detectors, but worse angular resolution
($1'$), than the microchannel photodetectors.
Temporal resolution
(0.001--0.5 ms), in most cases, enables one to study variations
of radiation fluxes concerned, for instance, with NS rotation.
By the moment of completing this review the best angular,
spectral and temporal resolutions possessed, respectively:
the microchannel X-ray detector
HRI (High Resolution Imager) on board of {\it ROSAT}
($\sim 5''$ in the energy range 0.1--2.5~keV), the detector
SIS (Solid-state Imaging Spectrometer)
on board of {\it ASCA} (0.4--10~keV,
$\Delta E/E \sim 2\% $ for photon energy $E \sim 6$~keV) and
PCA (Proportional Counter Array)
on {\it RXTE} ($\sim 1$~$\mu$s in 2--60~keV range).
X-ray detectors have high sensitivity. For instance, the
{\it ROSAT} observatory was able to detect the sources
with a flux
$\sim 5 \times 10^{-14}$~erg$\,$cm$^{-2}\,$s$^{-1}$.
More detailed information on active and future
detectors can be found, for instance, through the
Internet site
{\it http://heasarc.gsfc.nasa.gov/docs/heasarc/missions.html}.

The orbital observatories have led to a great progress in search
for the thermal NS radiation. Soft X-ray radiation has
been detected from about thirty radio pulsars and from several
isolated radio silent NSs. Thermal component has been identified
sufficiently reliably in $\sim 20\%$ of cases.
   
A search for the thermal radiation from the NS surface
was also carried out with the ultraviolet observatory
{\it EUVE} (The Extreme Ultra-Violet Explorer, operates since 1992)
using the DS (Deep Survey) and ST (Scanning Telescope)
telescopes, sensitive in the photon energy range
70--760~eV, with the angular resolution
$\sim 30$--$45''$ (using wideband filters
40--190~\AA\/ and 160--238~\AA). About 20 known NSs were observed
(Korpela and Bowyer, 1996).
Five sources were detected in the 40--190~\AA\/ band.
Two of them, PSR~0656+14 and Geminga, are the cooling pulsars, and
one, RX~J1856--37, is an isolated (probably cooling) radio silent
NS. For other objects, the upper limits to the ultraviolet fluxes
were obtained.
   
Since 1978 wideband photometric observations
of isolated NSs have been carried out
in optics using the sensitive ground-based telescopes
such as ESO-NTT (The New Technology Telescope
of the European South Observatory), the Keck Telescope,
the 6-meter Telescope, and the orbital HST
(The Hubble Space Telescope, launched in 1990).
About ten radio pulsars have been identified as
optical sources
(Mignani, 1998).
As a rule, the luminosity of isolated NSs in optics
is several orders of magnitude weaker than in X-rays.
This serves as an additional argument in identifying point-like
X-ray sources as isolated NSs, especially if they are not pulsars.
The point-like X-ray sources associated with accreting NSs
in close binaries in pairs with normal stars
(X-ray pulsars, bursters, transients)
show usually higher optical fluxes due to contribution
of companions or accretion disks.
Multicolour photometric observations have been carried out
for several isolated NSs
(PSR~0540--69, PSR~1509--58, PSR~0656+14, Crab, Vela, Geminga),
while for others
(PSR~1929+10, PSR~0950+08, PSR~1055--52, RX~J185635--3854, RX~J0720.4--3125)
the fluxes have been measured in one or two filters only.
Recently Martin et al.\ (1998),
using the 10-meter Keck Telescope,
obtained the first optical spectra (with $\sim 2$~\AA\/
resolution), of Geminga, the isolated middle-aged pulsar.
Earlier the spectroscopic observations were carried out
only of two young pulsars, Crab and Vela, whose optical
emission is non-thermal.
Recently Shearer et al.\ (1998), using the MAMA
(Multiple Anode Microchannel-Plate Array) detector
with high temporal resolution
and the 6-meter Telescope,
discovered temporal variations of the
optical emission (in the B-filter) from
Geminga and PSR~0656+14, cooling middle-aged NSs;
radiation pulsates with the NS spin periods.
Earlier, pulsating optical emission has been observed
only from several young pulsars
(including Crab, Vela, PSR 0540-69)
with the spectra of clearly non-thermal origin.
The energy of the non-thermal emission is taken
mainly from the NS spin energy.

The optical and ultraviolet observations complement the X-ray spectra
with the data in Rayleigh--Jeans part of the spectrum.
This enables one to get more definite conclusions
on the spectrum type (thermal or non-thermal) and on parameters
of the NS atmospheres. Interpretation of the thermal
radiation has to be consistent with
observations of the non-thermal emission from pulsars and
nearby space in radio, X-rays and gamma-rays.
Finally, one should use modern atmospheric models of NS radiation
for a correct interpretation of observations.
      
\subsubsection{Interpretation of observations}

Discovery and interpretation of the thermal radiation from cooling
NSs is a complicated problem. As mentioned above, the radiation is weak
since the NSs are compact. The radiation has to be separated
from the surrounding background which is especially
difficult for young NSs in supernova remnants.
Strong background is created by non-thermal emission
of synchrotron nebulae (plerions) which are formed around young
and active pulsars. The background is produced also
in non-thermal processes of emission of X-ray and optical
quanta in the magnetospheres of radio pulsars.
The distances to NSs are known poorly (uncertainty is often
about some ten percent and higher). The next difficulty
is provided by interstellar gas which absorbs soft
part of the spectrum (from a fraction of keV down to
optics), especially for the objects which are more distant than
some
hundred pc or are obscured by local clumps of interstellar
gas. Finally, the thermal radiation from the main part of the
pulsar's surface is superimposed by the thermal radiation
from the hotter
($\sim (2$--$3) \times 10^6$~K) polar caps heated by fluxes of energetic
particles from the magnetosphere. The spectra of the both
thermal components and the power-law spectrum of the possible
non-thermal component strongly overlap which complicates interpretation
even more.

It is clear that the middle-aged radio pulsars
($\sim 10^4$--$10^6$ yr) are
most favourable for studies of the thermal
radiation. Their surface is still rather hot,
$T_s \sim 10^5$--$10^6$~K, to be observable
but the objects are already insufficiently active to support
synchrotron nebulae around them. Supernova remnants are
dissolved by given age, and the non-thermal magnetospheric
processes weaken due to pulsar spin down.
All these factors reduce the background and increase
the chances to discover radiation from the pulsar surface.
The isolated NSs which, due to some reason, do not show
pulsar activity from the early stage, for instance, due to
the weak magnetic field or rapid spin down
are of special interest for the cooling theory.
These stars should possess lower background of
the non-thermal emission which simplifies detection
of the thermal component (see below).

An observer which studies X-ray radiation from
cooling NSs has usually a poor set of observational data
in the form of photon count rates in various spectral
channels of a detector; the count rates do not exceed, for instance,
1--10 counts per second for the brightest objects
detected with
PSPC (Position Sensitive Proportional Counter)
on board of {\it ROSAT}. For the majority of isolated NSs observed with
{\it ROSAT}, the count rate is
$\sim 10^{-3}$--$10^{-2}$ counts per second;
therefore, the exposure
time should exceed ten hours to accumulate statistically
significant number of counts (e.g., 100--1000).
Data processing and reconstructing the real spectrum
of the NS radiation is an example of a classical incorrect
inverse problem which can be solved only under
additional assumptions on the spectrum of the source.
The spectral analysis is usually carried out by calculating
the count rates in different channels by ``transmitting"
a model spectral flux of the object through
a response matrix of a detector; calculated values
are compared then with observations.
A model flux depends on NS parameters such as distance,
column density of interstellar hydrogen, etc.
These parameters are varied
(for instance, by $\chi^2$ method)
to obtain the best
agreement with observations. However the response matrix
is usually known with some uncertainty.
Its parameters vary in time and require permanent checking
using calibration sources on board of a space observatory
or cosmic standard sources. The model spectrum may be multicomponent;
for instance, it may contain contributions of thermal emission
from the main part of the pulsar surface and from the hot polar caps
as well as non-thermal magnetospheric emission. 
While fitting the spectrum one can,
in principle, determine all the varying quantities:
parameters of the NS thermal radiation (effective temperature,
magnetic field and chemical composition of the surface),
parameters of the hard radiation component 
(temperature and sizes of polar caps), spectral
index and intensity of the non-thermal component, spin period,
NS radius and mass, distance, column density of interstellar gas, etc.
Additional constraints on the variable quantities can be obtained
from similar fitting of the X-ray light curves.

A complete realization of the above scheme is not possible, to date,
mainly because of poor statistics of photon counts and
poor spectral resolution of the detectors.
Another obstacle comes from incompleteness of the theory of
magnetospheric pulsar emission and the theory
of atmospheres of isolated NSs with strong magnetic fields.
The non-thermal magnetospheric radiation is usually
described by a power-law spectrum with an index determined
from fitting. In the absence of the atmosphere models,
the thermal NS radiation is often described by the black body
spectrum. However, properties of radiation
emerging from a NS are actually determined by a thin
atmosphere with the temperature growing inside the star.
The atmospheric spectrum
may noticeably differ from the black body.
This is proved by the well elaborated models of
NS atmospheres with
``weak'' magnetic fields $B \la 10^9$--$10^{10}$ G
(Zavlin et al., 1996;
Rajagopal and Romani, 1996;
Pavlov and Zavlin, 1998)
and hot atmospheres ($T_s > 10^6$~K)
with strong fields $B \ga 10^{11}$~G
(Pavlov et al., 1995;
Pavlov and Zavlin, 1998).
In particular, the radiation spectra from the weakly magnetized NSs
covered with the atmospheres of light elements (hydrogen, helium)
appear to be harder for the same effective temperature $T_s$.
For typical temperatures $T_s \sim 10^5$--$10^6$~K,
the light elements are almost completely ionized; the main contribution
into opacity comes from free-free transitions
which is well described by the Kramers formula
(the spectral opacity decreases  as $E^{-3}$ with the growth of photon energy
$E$). This makes the atmosphere more transparent
for higher-energy photons, which emerge thus from deeper and hotter
layers, leading to hardening of the spectrum.
In the atmospheres of heavier elements (iron), which are
more difficult to ionize, the photoeffect and bound--bound
transitions are most important. This leads to weaker
energy dependence of the opacity on photon energy.
The radiation spectrum becomes closer to the black body, with
the temperature corresponding to some mean depth from which
the quanta emerge. In this case almost Planckian continuum
contains deep and wide photoionization jumps and 
absorption lines of atoms and ions of heavy elements.
In contrast to the black body emission, the
atmospheric radiation is anisotropic even for
the NSs with weak magnetic fields.
The anisotropy depends on photon energy and chemical composition;
the well known ``limb darkening" effect is quite pronounced.
The anisotropy is greatly amplified by a strong magnetic field.
In this case, the anisotropy axis is aligned along
the magnetic field and radiation beaming can be
strongly asymmetric. An allowance for the anisotropy
and gravitational bending of photon trajectories near the NS surface
(Zavlin et al., 1995)
is especially important for interpretation of the light curves.

To date, among the models of NS atmospheres with
strong magnetic fields, $B \ga 10^{10}$--$10^{11}$~G,
only the models
of hot atmospheres ($T_s > 10^6$~K)
composed of fully ionized plasma are reliable.
According to these models,
the radiation spectrum becomes softer, than for weaker magnetic fields,
but remains noticeably harder than the Planck spectrum.
For lower temperatures, the magnetized atmospheres
become partly ionized. The strong magnetic fields drastically distort
structure of atoms and ions and affect the spectral opacity.
Detailed calculation of the opacity is complicated
and has not been done yet even for the simplest case
of hydrogen plasma. The magnetic field increases
the binding energies of electrons in atoms. For instance, the field
$B \sim 10^{12}$~G amplifies the binding energy of the hydrogen atom
to $\sim 150$~eV.
The number density of neutral atoms becomes larger increasing
the role of photoeffect and bound--bound transitions; these processes
must be calculated with allowance for thermal motion of atoms.
Such calculations are complicated since the electric
fields, induced in an atom-comoving reference frame due to atomic motion
across the magnetic field, are so strong that they change
the structure of atomic energy levels and probabilities
of radiative transitions (the dynamic Stark--effect; see, e.g.,
Potekhin, 1994). While calculating
the opacity of this, partly ionized plasma
it is important to take into account the effects of
plasma non-ideality and pressure ionization.
These calculations are being performed at present.
Only the models of hydrogen and iron atmospheres 
neglecting the dynamic Stark effect have been constructed so far
(Shibanov et al, 1993; Rajagopal et al., 1997).
The dependence of the radiation spectrum
on chemical composition has the same tendencies as for the
atmospheres with weak field. The spectra of the iron atmospheres are
softer and closer to the black body, than the spectra of the hydrogen
atmospheres. An account for the induced electric fields,
which strongly broaden the photoionization jumps and other
spectral features,
will enable one to check the above statement.

Owing to the effects mentioned above an ``atmospheric"
temperature $T_s$ appears to be noticeably
(typically, by a factor of 1.5 -- 3) lower than a `` black body"
temperature in interpretation of the same observational data.
If the validity conditions of the available
atmosphere models are fulfilled,
then the spectra of the thermal radiation and the NS light curves
are better fitted by the atmosphere models than by the black body.
Parameters inferred from atmospheric interpretation
(emitting surface area, distance to a NS, column density
of the interstellar gas, etc.) are, as a rule, in better
agreement with the data provided by independent observations
in other spectral bands (see, e.g.,
Page et al., 1996; Zavlin et al., 1998).
The black body interpretation often gives less realistic
parameter values and meets some difficulties in explaining all
set of observational data.

Since statistics of count rates is commonly poor
and the number of adjustable parameters is large,
the confidence ranges of these parameters are often 
too wide. To make them narrower some parameters
are fixed (NS mass and radius, distance, etc.).
The parameters become then more constrained. In a sense, this
is an illusion if not confirmed by the results of independent
observations (e.g., in other spectral bands). For instance,
radio observations of non-thermal pulsar emission are very
useful for interpretation of X-ray observations. They give
additional constraints on distance from parallax measurements,
on interstellar extinction from dispersion measure determination,
on orientation of NS spin and magnetic axes from polarimetric
measurements, on pulsar age from measurements of spin down
rate, on stellar mass from measurements of orbital
parameters of binary systems containing NSs, etc.
Additional constraints can also be obtained from optical
and gamma-ray observations.

\subsubsection{Observational results}

As a result of more than 30-year search for thermal radiation,
X-ray radiation has been discovered or upper limits
of the surface temperature have been obtained for some
tens of isolated NSs
(Becker and Tr\"{u}mper, 1997, 1999).
However, so far the discovery of the thermal component can be regarded
as definite only in several cases. These cases include four
closest middle--aged radio pulsars (Vela, Geminga,
PSR~0656+14 and PSR~1055--52) and three sufficiently young
radio silent NSs
in supernova remnants (1E~1207--52 in the remnant PKS~1209--51/52,
RX~J0002+62 near CTB-1 and RX~J0822--4300 in Puppis A).
All seven sources are reliably identified; their nature as
isolated NSs is confirmed by observations of periodic
pulsations (excluding 1E~1207--52), produced by stellar rotation,
and/or large ratio of the X-ray to optic luminosities.
These NSs are listed
in Table \ref{etab:NS_data} in accordance with their age. For each star,
we give its characteristic age (determined either from
NS spin down rate or from morphology of a supernova remnant),
spin period,
method of interpretation of thermal spectrum, the effective
surface NS temperature
$T_s^\infty=T_s \sqrt{1-({\cal R}_g/{\cal R}) }$ (Sect.\ 7.2.3)
determined by this method, and distance to the source.

\begin{table*}[t]   
\caption[]{Observational data}
\label{etab:NS_data}
\begin{center}
\begin{tabular}{|p{2.7cm}|p{1cm}|c|@{}p{10.05cm}|}
\hline
\hline
 Source       & \begin{tabular}{@{}c@{}}
 $\lg t\rrr\;$ \\[-0.3ex] $\!\![{\rm yr}]$
                  \end{tabular}
                & \begin{tabular}{@{}c@{}}
 $P$ \\[-0.3ex] $[{\rm ms}]$
                  \end{tabular}
                & \begin{tabular}[c]{p{1.4cm}|p{1.6cm}|p{1.1cm}|p{2.4cm}|p{1.9cm}@{}}
$\!$Model{\raisebox{0.5ex}{$^{a)}$}}
                &                     \begin{tabular}{@{}c@{}}
 $\;\;\lg T_{\!s}^{\infty}\rrr$ \\[-0.3ex] $[{\rm K}]\rrr$
                                      \end{tabular}
				                      &
               $\;\;$b)
                &                      \begin{tabular}{c}
$\!\!\!$Distance \\[-0.3ex] $[{\rm kpc}]$
                                       \end{tabular}
                &
 Refs.
{\raisebox{0.5ex}{$^{\!\!f)}$}}
                  \end{tabular} 
		  \\
\hline
\hline
RX$\,$J0822-43 & 3.57 & 75 &
        \begin{tabular}[c]{p{1.4cm}|p{1.6cm}|p{1.1cm}|p{2.4cm}|p{1.9cm}@{}}
      $\;$\hh H & $ 6.23^{+0.02\rrr}_{-0.02} $
                                             & 95.5\%
                                             & \hh 1.9 -- 2.5
                                             &
                                   Z 1999b \\[0.5ex]
           \hline
      $\;$\hb bb & $ 6.61^{+0.05\rrr}_{-0.05}  $
                                             & 95.5\%
                                             & \hh 2.5 -- 3.5
                                             &
                                   Z 1999b \\[0.5ex]
    \end{tabular}    \\
\hline
   1E$\,$1207-52  & 3.85 & --- &
        \begin{tabular}[c]{p{1.4cm}|p{1.6cm}|p{1.1cm}|p{2.4cm}|p{1.9cm}@{}}
        $\;$\hh H &  $ 6.10^{+0.05\rrr}_{-0.06} $
                                             & $\,$90\%
                                             & \hh 1.6 -- 3.3
                                             &
                                   Z 1998 \\[0.5ex]
           \hline
         $\;$\hb bb & $ 6.49^{+0.02\rrr}_{-0.01}  $
                                             & $\,$90\%
                                             & \hh$\;$11 -- 13
                                             & Z 1998 \\[0.5ex]					     
    \end{tabular}    \\
\hline
RX$\,$J0002+62 & $3.95^{c)}$ & $242^{d)}$ &
        \begin{tabular}[c]{p{1.4cm}|p{1.6cm}|p{1.1cm}|p{2.4cm}|p{1.9cm}@{}}
        $\;$\hh H &  $ 6.03^{+0.03\rrr}_{-0.03} $
                                             & 95.5\%
                                             & \hh 2.7 -- 3.5
                                             &
                                     Z 1999a \\[0.5ex]
           \hline
         $\;$\hb bb & $ 6.18^{+0.18\rrr}_{-0.18} $
                                             & 95.5\%
                                             & \hh$3.1\pm 0.4$
                                             & Z 1999a \\[0.5ex]
    \end{tabular}    \\
\hline
 \begin{tabular}{@{}l}
   PSR~0833-45 \\
     (Vela)
 \end{tabular}
             & $4.4^{e)}$ & 89 &
        \begin{tabular}[c]{p{1.4cm}|p{1.6cm}|p{1.1cm}|p{2.4cm}|p{1.9cm}@{}}
         $\;$\hh H &  $ 5.90^{+0.04\rrr}_{-0.01}$
                                             & $\,$90\%
                                             & \hh$0.4\pm 0.12$
                                             &
                                             Pa 1996\\[0.5ex]
           \hline
         $\;$\hb bb & $ 6.24^{+0.03\rrr}_{-0.03} $
                                             & $\;\;\;$---
                                             & \hh\hh 0.5
                                             &
                                              O 1995 \\[0.5ex]					      
    \end{tabular}    \\
\hline
 PSR~0656+14 & 5.00 & 385 &
        \begin{tabular}[c]{p{1.4cm}|p{1.6cm}|p{1.1cm}|p{2.4cm}|p{1.9cm}@{}}
          $\;$\hh H & $ 5.72^{+0.04\rrr}_{-0.02} $
                                             & $\;\;\;$---
                                             & \hh$0.28^{+0.06\rrr}_{-0.05}$
                                             &
                                             A 1993 \\[0.5ex]
           \hline
          $\;$\hb bb & $ 5.96^{+0.02\rrr}_{-0.03} $
                                             & $\,$90\%
                                             & \hh $\;\;\;$ 0.76
                                             &
                                            Po 1996 \\[0.5ex]
    \end{tabular}    \\
\hline
 \begin{tabular}{@{}l}
     PSR~0630+178\\
     (Geminga)
 \end{tabular}
             & 5.53 & 237 &
        \begin{tabular}[c]{p{1.4cm}|p{1.6cm}|p{1.1cm}|p{2.4cm}|p{1.9cm}@{}}
          $\;$\hh H &  $ 5.25^{+0.08\rrr}_{-0.01} $
                                             & $\,$90\%
                                             &$\,$0.008$\,$--$\,$0.022
                                             &
                                       M 1994 \\[0.5ex]
                  \hline
         $\;$\hb bb & $ 5.75^{+0.05\rrr}_{-0.08} $
                                             & $\,$90\%
                                             & \hh $\;\;\;$ 0.16
                                             &
                                    HW 1997 \\[0.5ex]
				    
   \end{tabular}    \\
\hline
 PSR~1055-52 & 5.73 &  197 &
        \begin{tabular}[c]{p{1.4cm}|p{1.6cm}|p{1.1cm}|p{2.4cm}|p{1.9cm}@{}}
          $\;$\hb bb & $ 5.88^{+0.03\rrr}_{-0.04} $
                                             & $\;\;\;$---
                                             & \hh\hh 0.9
                                             &
                                        O 1995 \\[0.5ex]					
     \end{tabular} \\
\hline
\multicolumn{4}{@{}l@{}}{
 \begin{tabular}{@{}l@{}}
  $^{a)}\,\rrr${\footnotesize  ``H'' observations are interpreted
  with a hydrogen atmosphere model,
  }\\[-0.7ex]
  $\phantom{^{a)}\,}$\rrr{\footnotesize ``bb'' with black body spectrum.}\\[-0.7ex]
  $^{b)}\,$\rrr{\footnotesize Confidence$\:$level$\:$of temperature$\:T_s^\infty$
              (95.5\%$\:$corresponds to$\:2\sigma$ level$,\:$a$\:$90\% corresponds
              to 1.64$\sigma$);}\\[-0.7ex]
  $\phantom{^{b)}\,}$\rrr{\footnotesize   dash means that
               the level is not indicated
               in cited references.}\\[-0.7ex]
  $^{c)}\,$\rrr{\footnotesize The mean age taken according
             to Craig et al.\ (1997).
                     }\\[-0.7ex]
  $^{d)}\,$\rrr{\footnotesize According to
                  Hailey and Craig (1995).}\\[-0.7ex]
  $^{e)}\,$\rrr{\footnotesize According to
        Lyne et al.\ (1996).}\\[-0.7ex]
  $^{f)}$ \rrr{\footnotesize Z -- Zavlin et al., Pa -- Page et al., 
               A -- Anderson et al.,
               Po -- Possenti et al,}\\[-0.7ex]
$\phantom{^{f)}\,}$\rrr{\footnotesize 
	        O -- \"Ogelman, M -- Meyer et al.,
	       HW -- Halpern \& Wang.}\\[-0.7ex]
%
%
%
\end{tabular}
 } \\
\end{tabular}
\end{center}
\end{table*}

Among the ``atmospheric" interpretations presented in Table
\ref{etab:NS_data}, those for 1E~1207--52,
RX~J0002+62, RX~J0822--43 and the Vela pulsar seem to be most
reliable. The interpretation of their spectra with the hydrogen--helium
atmosphere models has an additional advantage: the fits of observational
data with iron atmosphere models or with the black body
spectrum are of lower statistical significance.
Geminga and PSR~0656+14 are older and cooler. An important contribution
into opacity of their atmospheres comes from the effects
of motion of neutral and partly ionized atoms across strong magnetic
field
(see
Potekhin and Pavlov, 1997;
Bezchastnov et al., 1998; and also Sect.\ 8.1.2).
The presented interpretations have been made using the simplified
atmosphere models in which these effects have been neglected.
The hydrogen composition of the Geminga's atmosphere
is additionally confirmed by a possible discovery
of the proton cyclotron line in its optical spectrum
(Bignami et al., 1996; Martin et al., 1998).
The ``atmospheric" temperatures of Geminga and PSR 0656+14
are likely to be closer to reality than the ``black body" temperatures
but less reliable than for younger NSs.

Comparison of the surface temperatures of the objects from Table
\ref{etab:NS_data} with the simulations of NS cooling will be
discussed in Sect.\ 8.2. In this discussion we will assume that
the ``atmospheric" and ``black body" interpretations of observations
are equally acceptable and obtain constraints on 
critical temperatures of nucleon superfluidity
in the NS cores. Here we will outline some other
observations.

The upper limits of $T_s$ for the majority of other objects are
too high to be of interest for the NS cooling theory.
Let us mention only the upper limit of the surface temperature
of the young Crab pulsar. Numerous attempts to
detect the thermal radiation from the pulsar have failed
due to the powerful non-thermal radiation from the magnetosphere
and supernova remnant in the wide spectral range.
The most stringent upper limit
$T_s^\infty =1.55 \times 10^6$ K was obtained by
Becker and Aschenbach (1995)
from the {\it ROSAT} observations. Taking into account the uncertainty
in distance and interstellar absorption, this upper
limit does not contradict to the standard cooling of the Crab pulsar
(with the non-superfluid core) although for stiffer equations
of state the agreement is rather questionable.
Let us mention that the surface temperature of Crab
can be estimated independently from observations of after-glitch
relaxation. According to the theory, this relaxation
depends of the internal NS temperature, and the internal temperature
is related to the surface one. Such estimates are not very definite
but give $T_s^\infty \approx 1.6 \times 10^6$ K
(Alpar et al., 1985), in excellent agreement with the above upper limit.
  
There are about ten more potential candidates to the
cooling isolated NSs among radio silent compact
X-ray sources in addition to those listed in Table \ref{etab:NS_data}.
New observations are required to confirm that these sources
are the cooling NSs. Two most probable candidates are
the point-like X-ray objects in the young supernova remnants:
1E~1841--045 in the center of Kes~73 and
1E~161348--5055 in the center of RCW~103. The both objects
have close black body temperatures,
$T_s \sim 0.7$ and $\sim 0.6$~keV. Such temperatures
are too high for the ordinary cooling NSs of supernova
remnant ages, $\sim 2000$--4000~yr
(Gotthelf and Vasisht 1997;
Gotthelf et al., 1997).
However, for 1E~1841--045, the above age agrees with its dynamic age
determined from spin down rate (from observations of coherent
pulsations of X-ray emission discovered recently),
$P= 11.8$~s and
$\dot P=4.73 \times 10^{-11}$~s s$^{-1}$
(Vasisht and Gotthelf, 1997). The standard (for radio pulsars)
estimate of the magnetic field from $P$ and $\dot P$
reveals the enormous field strength $B\sim 10^{15}$~G.
The optical and radio emission from the objects has not been discovered.
If confirmed in future observations
(in accord with preliminary estimates
by Heyl and Hernquist, 1997a),
these data can be explained by the thermal radiation from
young cooling NSs with superstrong magnetic fields and
with outer shells composed of light elements.
The presence of the superstrong field and the shell increases the thermal
conductivity of the outer layers
and, as a consequence, rises the surface temperature
at the early cooling stage. For the second source,
1E~161348--5055, this hypothesis should be treated with great care
due to recently discovered mysterious raise of
the luminosity for several months accompanied by the increase
of the X-ray flux by one order of magnitude with the
unchanged spectrum
(Gotthelf et al., 1999).
A search for pulsations with NS spin period 
has not been successful so far for this object,
which has been observed regularly in X-rays since 1979.
 
The sources of repeating soft gamma-ray emission
(SGRs -- {\it soft gamma repeaters})
belong possibly to the same family of
cooling NSs with the superstrong magnetic field
(called ``magnetars",
Thompson and Duncan, 1996).
However, they probably undergo different evolutionary stage.
All of them are situated near the supernova remnants which
indicates they are young. For two of them,
SGR~1900+14 and SGR~1806-20, coherent pulsations of X-ray radiation
with the periods 5.16 and 7.47~s, respectively, 
have been discovered recently
(Kouveliotou et al., 1998a, b; 
Hurley et al., 1999).
Both sources have about the same large braking index
as the source in Kes~73. Estimated surface temperatures
of the objects ($T_s \la 1$~keV) exceed typical
temperatures of ordinary cooling NSs of their ages
($t \la 1000$--10000~yr), if the ages are the same as for
parental supernova remnants.

Let us mention one more type of the sources. They are the
so called ``anomalous X-ray pulsars", the X-ray sources with
the soft thermal spectrum
($T\la 1$~keV; e.g.,
Mereghetti et al., 1998)
and almost sinusoidal light curve. Typical periods of
their brightness variation are about ten seconds.
The magnitude of pulsating component of their radiation
is anomalously small and more typical for
the thermal radiation from the surface of a cooling NS than
for an X-ray pulsar. One cannot exclude that at least some
of these objects also belong to the family of
cooling magnetars. No binary companions were found for these objects,
and no accretion disks which would provide accreted matter for
X-ray pulsars. A powerful magnetodipole radiation
of a rotating NS with superstrong magnetic field
would explain naturally rapid spin down;
an increase of the thermal conductivity of the outer layers
would produce sufficiently high surface temperature.
Young ages of some anomalous pulsars are also confirmed
by coincidence of their positions with the centers
of supernova remnants. They are also radio silent.
 
Finally, let us mention two more radio silent isolated
NSs. They have been discovered quite recently; first,
as bright point--like sources of the soft X-ray radiation
in the {\it ROSAT} surveys, and later as weak optical
objects in special observations with the ground-based
Keck Telescope and the Hubble Space Telescope.
These stars are not associated with the supernova remnants.

One of them, RX~J1856.5-3754
(Walter et al., 1996;
Walter and Matthews, 1997),
has the black body surface temperature
$(8.5 \pm 1.4) \times 10^5$~K
(Pavlov et al., 1996),
for the adopted values of
${\cal R}=10$~km and $M=1.4M_{\odot}$, and is situated at a distance
$D \la 130$~pc.
Radiation fluxes in the optical filters
F606W (6000 \AA) and  F300W (3000 \AA),
detected with the Hubble Space Telescope,
are 25.9 and 23.48 stellar magnitudes, respectively
(Walter and Matthews, 1997). They agree with the hypothesis
on the thermal origin of the spectrum in the spectral range
from optics to X-rays. The upper limit on the pulsating
component is rather high (about $15\%$). Further accumulation
of photon statistics will enable one to continue search for
periodic pulsations. On the other hand, the observed thermal radiation
can be explained by accretion of interstellar gas on
a cold, weakly magnetized NS which moves slowly
through interstellar matter
(Walter and Matthews, 1997;
Zane et al., 1996). In order to choose between these two
hypotheses one needs to measure parallax and proper motion of the
object.
 
The other NS, RX~J0720.4-3125, is somewhat hotter
(the black body surface temperature is $79 \pm 4$~eV).
Its X-ray flux is pulsating with the period
8.39~s  and modulation $\sim 12\%$
(Haberl et al., 1997). There are not very reliable
indications of the spin down,
$\dot P\sim 10^{-11}$--$10^{-12}$~s s$^{-1}$. Optical observations with
the Keck Telescope indicate the presence of an object
identified with this source. The detected fluxes are
$26.6 \pm 0.2$ and $26.9 \pm 0.3$ stellar magnitudes in
the B and R filters, respectively
(Kulkarni and van Kerkwijk, 1998).
For a distance to the source
$D\la 100$~pc, the weakness of the optical emission is in favour
of the hypothesis
that the object is an isolated NS. One can expect to deal either
with a cooling NS--magnetar or with an old, weakly magnetized NS
accreting matter from interstellar medium.
A choice of correct hypothesis can be done only on the basis of
future observations.

Let us mention also two more radio silent NSs,
RX~J0806.4-4123
(Haberl et al., 1998)
and
1RX~J130848.6 +212708
(Schwope et al., 1999),
discovered most recently in the {\it ROSAT} surveys;
they are very similar to the two latter NSs.
They are possibly the cooling NSs
(Neuh\"{a}user and Tr\"{u}mper, 1999) 
but the observations are still insufficient
to drew final conclusions.

\subsection{Theory and observations of cooling neutron stars}

\subsubsection{Necessity of superfluidity for interpretation
               of observations}

\begin{figure}[t]                         
\begin{center}
\leavevmode
\epsfysize=8.5cm
\epsfbox[120 410 470 792]{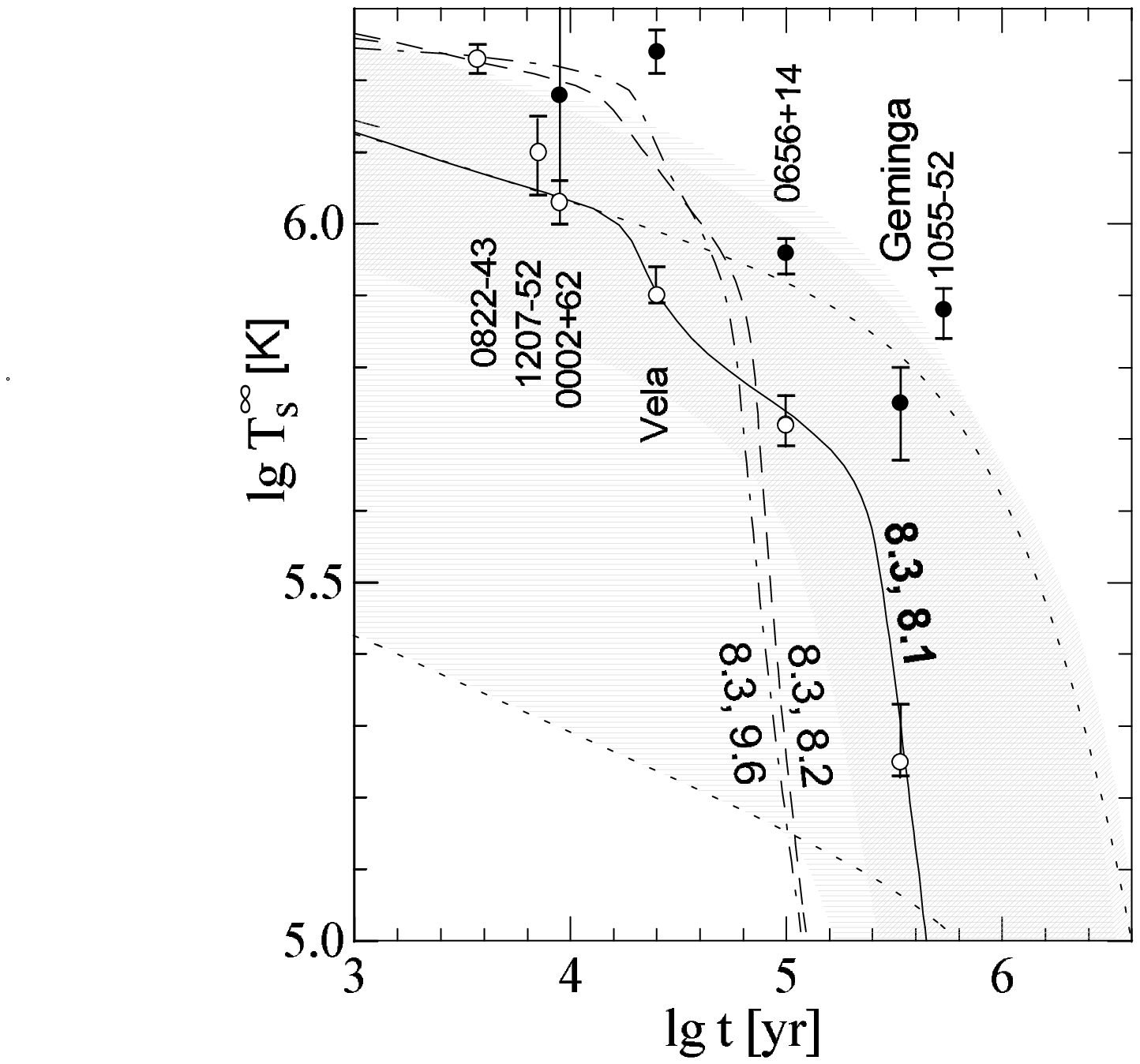}
\end{center}
\caption[]{\footnotesize
  Observational data on surface temperatures of NSs obtained
  (Table \ref{etab:NS_data}) in the models of
  black body radiation (filled circles) and hydrogen atmosphere
  (open circles). Shaded regions show the ranges of $T_s^\infty$
  filled by the standard (diagonal shading)
  and the enhanced (horizontal shading) cooling curves
  of NSs with different $\tn$ and $\tp$ (from $10^6$ to $10^{10}$~K).
  The solid line shows the standard cooling ($1.30 \, M_\odot$)
  for specified values $\lg \tn$ and $\, \lg \tp$ (given near the curve).
  The dotted lines exhibit the standard (upper line)
  and enhanced (lower line) cooling of a non-superfluid NS.
  The dashed and dot-and-dashed lines show, respectively,
  the standard and enhanced cooling of a NS which possesses
  an envelope of mass
  $7 \times 10^{-10} M_\odot$ composed of light elements
  (see Sect.\ 8.2.1).
}
\label{fig:NS_data}
\end{figure}

The aim of this section is to confront the cooling theory
of superfluid NSs (Sect.\ 7.2) with observations (Sect.\ 8.1).
Table \ref{etab:NS_data} presents the data on
the surface temperatures of seven isolated NSs.
In Fig.\ \ref{fig:NS_data} these data (circles)
are compared with the cooling curves (Sect.\ 7.2).
Diagonal and horizontal
shadings show the regions of the surface temperature
$T_s^\infty(t)$ filled, respectively, by
different standard and enhanced cooling
curves calculated by varying
$\tn$ and $\tp$ from $10^6$ to $10^{10}$ K.
The dotted curves show cooling of non-superfluid stars.

The ``non-superfluid'' curves are seen to be in poor
agreement with observations. On the other hand, the observations
can be explained by assuming superfluidity in the NS cores.
This is illustrated by the
standard cooling curve (the solid line).
The values of $\tn$ and $\tp$ are chosen in such a way
to hit the maximum number of observational points at once.

According to Fig.\ \ref{fig:NS_data},
all the ``atmospheric'' temperatures as well as the
``black body'' temperatures of RX~J0002+62, PSR~0656+14 and Geminga
are located in
the allowed regions of the standard and enhanced cooling of
superfluid NSs. Thus our cooling calculations can be
compared with the ``atmospheric" and ``black body" temperatures
$T_s$ of these sources.

High black body surface temperatures of
RX~J0822-43, 1E~1207-52 (not presented in
Fig.\ \ref{fig:NS_data}, but given in Table
\ref{etab:NS_data}), Vela, and PSR~1055-52 are not explained by
our models but can be explained by other models of cooling NSs,
particularly, the models with superfluid cores. For instance,
the observations of the Vela pulsar agree with the standard
cooling of a superfluid NS
($\tn=10^7$~K, $\tp=10^{10}$~K) possessing an outer envelope
of mass $\sim 10^{-9}\, M_\odot$ composed of light elements
(Potekhin et al., 1997). High black body temperatures of
RX~J0822-43, 1E~1207-52 and PSR~1055-52 may indicate either the
presence of some additional reheating mechanism inside these
sources (Sect.\ 8.1) or the presence of
superstrong ($B \ga 10^{14}$~G) magnetic fields (Heyl and Hernquist, 1997a).
Finally, one cannot exclude that the black body
interpretation of their spectra is incorrect (Sect.\ 8.1.2).

\subsubsection{Confronting calculations and observations}
Let us analyse which critical temperatures
of nucleon superfluidity, $\tn$ and $\tp$, in the NS cores
used in the cooling simulations (Sect.\ 7.2) agree
with the observations of the NS thermal radiation
(Table \ref{etab:NS_data}).
As seen from Fig.\ \ref{fig:NS_data}, we can explain
the majority of NS observations either by the
standard or by the enhanced NS cooling
assuming the presence of nucleon superfluidity.
These observations include six ``atmospheric" interpretations
(RX~J0822-43, 1E~1207-52, RX~J0002+62, Vela, PSR~0656+14, Geminga)
and three ``black body'' ones (RX J0002+62,
PSR~0656+14, Geminga).

Let us assume that the NS atmospheres may contain light elements
whose surface density is insufficient
(Sect.\ 7.2) to affect the cooling. In this case, we can use
the both, ``black body" and ``atmospheric", interpretations
of the spectra of thermal radiation. Although the values of
$M$ and ${\cal R}$ and, accordingly, the gravitational redshift
for our models (Sect.\ 7.2.2) are somewhat different from those
which have been obtained (or adopted) in interpretation of
the observations, the temperatures
$T_s^\infty$ from Table \ref{etab:NS_data}
and Fig.\ \ref{fig:NS_data} can be compared with our cooling curves
due to several reasons.
First, our calculations are not very sensitive to variation
of a NS mass or radius (see Sect.\ 8.3 below).
Second, the confidence regions of $M$ and ${\cal R}$,
inferred from interpretation of observations, include, as a rule,
the values we use in modelling.

For simplicity, we assume that
the internal structure of the indicated NSs is the same, and, particularly,
the NSs have the same mass. Then the critical temperatures
of nucleons in their cores should be the same.
Let us analyse the confidence regions of
$\tn$ and $\tp$ constrained by the observations.
Including the observations of several NSs enables us to reduce
these regions.

\begin{figure}[t]                         
\begin{center}
\leavevmode
\epsfysize=8.5cm
\epsfbox[20 310 565 590]{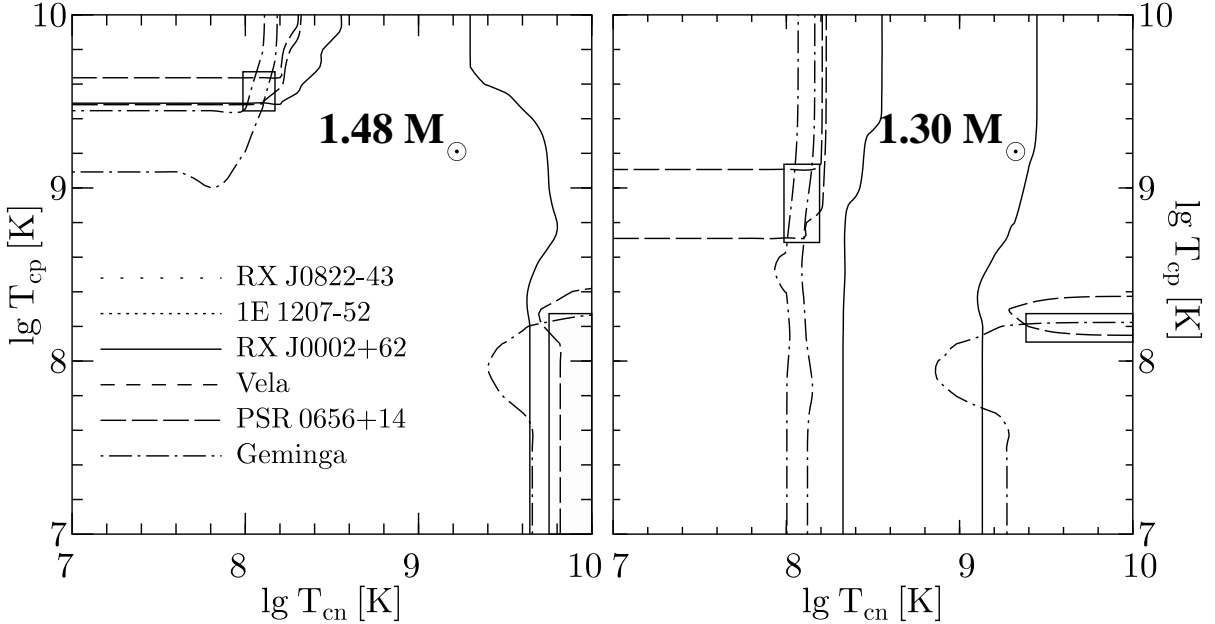}
\end{center}
\caption[]{\footnotesize
  Confidence regions of $T_{cn}$ and $T_{cp}$ which
  correspond to the ``black body" surface temperatures
  (Table \ref{etab:NS_data},  Fig.\ \protect{\ref{fig:NS_data}})
  of RX J0002+62, PSR~0656+14 and Geminga
  in the models of
  enhanced (left) and standard (right) NS cooling.
         }
\label{fig:all_bb}
\end{figure}

\begin{figure}[t]                         
\begin{center}
\leavevmode
\epsfysize=8.5cm
\epsfbox[20 310 570 590]{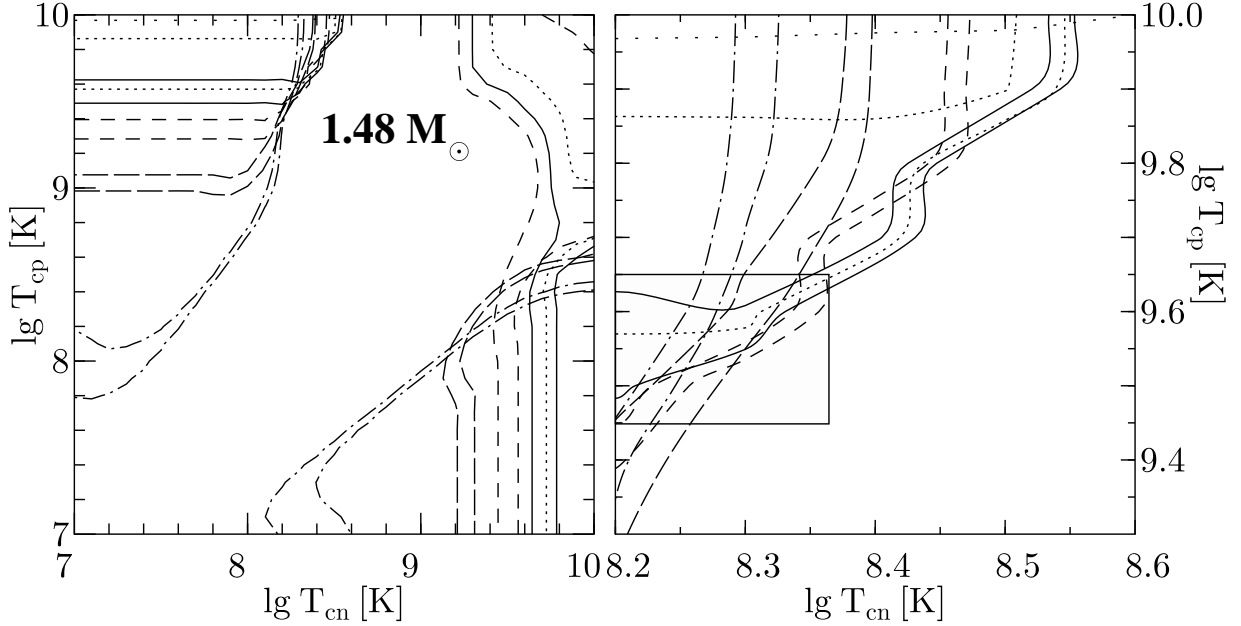}
\end{center}
\caption[]{\footnotesize
  Same as in Fig.\ \protect{\ref{fig:all_bb}}, but for the enhanced
  cooling and ``atmospheric" interpretation
  (Table \ref{etab:NS_data}) of spectra of thermal
  radiation from RX~J0822-43, 1E~1207-52, RX~J0002+62, Vela,
  PSR~0656+14 and Geminga. On the right panel,
  we show in more detail the region (shaded rectangle)
  in which the allowed values of $T_{cn}$ and $T_{cp}$
  of five latter objects  are either close or intersect.
        }
\label{fig:fast_atm}
\end{figure}

\begin{figure}[t]                         
\begin{center}
\leavevmode
\epsfysize=8.5cm
\epsfbox[20 345 560 620]{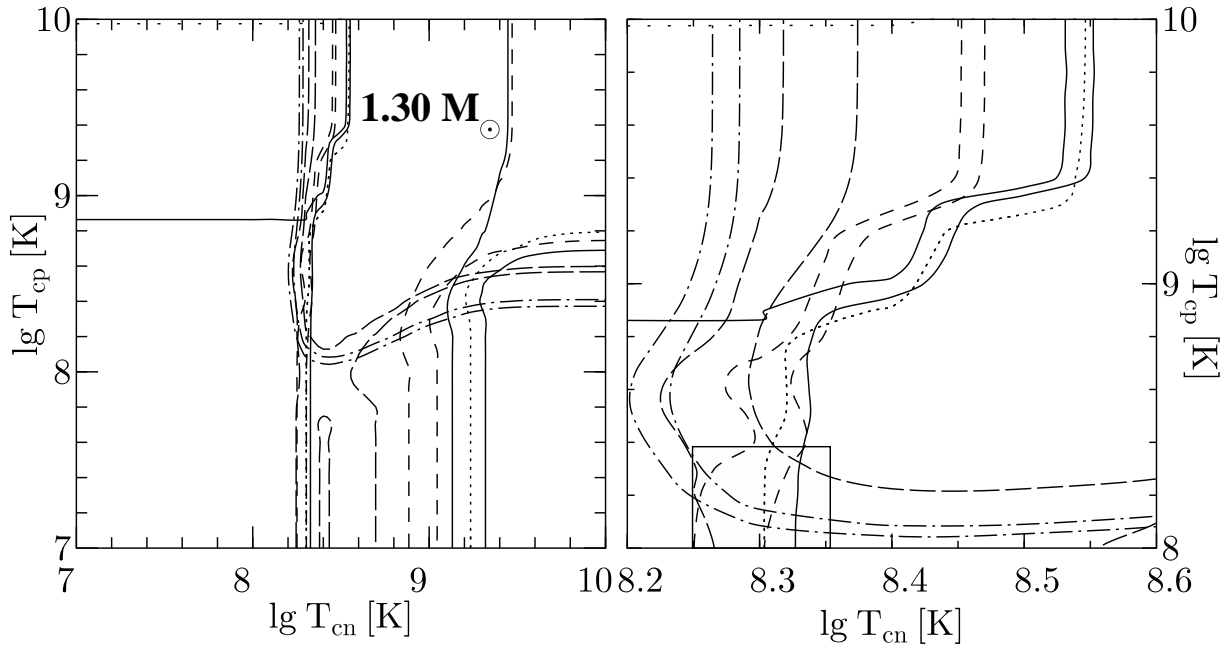}
\end{center}
\caption[]{\footnotesize
  Same as in Fig.\ \protect{\ref{fig:fast_atm}},
  but for the standard cooling.
         }
\label{fig:std_atm}
\end{figure}

The regions in question are plotted in Figs.\
\ref{fig:all_bb}--\ref{fig:std_atm}.
Figure \ref{fig:all_bb} corresponds to the standard and
enhanced cooling of NSs
(with masses $1.30\, M_\odot$ and $1.48 \, M_\odot$, respectively)
with the black body spectrum.  Figure
\ref{fig:fast_atm} corresponds to the enhanced cooling of the stars
($1.48 \, M_\odot$) possessing hydrogen atmospheres.
Finally, the standard cooling of NSs ($1.30 M_\odot$) with
the hydrogen atmospheres is shown in Fig.\ \ref{fig:std_atm}.
In each figure, the lines of different types
enclose the confidence regions of $T_{cn}$ and $T_{cp}$ 
associated with the error bars of the
observed NS surface temperatures
$T_s^\infty$ (Table \ref{etab:NS_data}). Correspondence of the
lines to the selected NSs is displayed in Fig.\ \ref{fig:all_bb}.
For the  PSR~0656+14 and Geminga pulsars, the isotherms are taken
from Figs.\ \ref{fig:Gem_2}--\ref{fig:Gem_5}. In each figure
the actual overall confidence region of $\tn$ and $\tp$ lies at the overlap
of the confidence regions of all objects.

Figures \ref{fig:all_bb}--\ref{fig:std_atm} show that the
observations of several NSs at once can be explained
either by the standard or by the enhanced cooling,
adopting either black body or atmospheric interpretations of the
spectra. In all the cases, there are the
ranges of $\tn$ and $\tp$ close or joint for all NSs; they do not
contradict to microscopic theories of nucleon
superfluidity in NSs (Sect.\ 3.1).

According to the left panel of Fig.\ \ref{fig:all_bb},
by adopting the enhanced cooling and the
black body interpretation of the observations we obtain two
confidence regions of $\tn$ and $\tp$;
each explains cooling of three
objects at once. The first region
corresponds to a moderate neutron superfluidity
($\lg \tn \approx 8.1$) and a strong proton superfluidity
($9.6 \la \lg \tp \approx 9.8$); the second, wider region
corresponds to a strong neutron superfluidity
($9.75 \la \lg \tn \approx 10.0$)
and a moderately weak proton superfluidity
($\lg \tp < 8.3$). For the standard cooling of a NS with the
black body spectrum (the right panel of Fig.\ \ref{fig:all_bb}),
there are also two regions of $\tn$ and $\tp$ for the three NSs,
but they are somewhat different than for the enhanced cooling.
The first region corresponds to a moderate neutron superfluidity
($8.0 \la \lg \tn \la 8.2$) and a moderately strong
proton superfluidity ($8.7 \la \lg \tp \la 9.1$), while the
second one is associated with a strong neutron superfluidity
($\lg \tn > 9.3$) and a moderately weak proton one
($8.1 < \lg \tp < 8.3$).

In the case of the enhanced cooling and the ``atmospheric"
surface temperatures (Fig.\ \ref{fig:fast_atm})
there is the only region of joint or very close values of
$\tn$ and $\tp$ for five NSs
(the shaded rectangle on the right panel). It refers
to a moderate neutron 
($8.2 \la \lg \tn \la 8.35$) and strong proton 
($9.54 \la \lg \tp \la 9.65$) superfluidity. Such superfluidity
($\lg \tp =8.3$, $\lg \tp = 9.6$) enables us to describe also the sixth
object from the ``atmospheric'' set (RX~J0822-43),
if we assume that it possesses a thermally insulating envelope
of light elements with the mass of $ 7 \times 10^{-10} M_\odot$
(see Fig.\ \ref{fig:NS_data}). However, even in the absence of
the envelope, the confidence region of
$\tn$ and $\tp$ for this object is sufficiently close
to the joint confidence region indicated above.

Finally, for the standard cooling and the ``atmospheric" spectrum
(Fig.\ \ref{fig:std_atm}), there is again the only 
confidence region of $\tn$ and $\tp$,
where the critical temperatures are
nearly the same for the five NSs;
it corresponds to a moderately strong superfluidity of $n$ and $p$
($8.25 \la \lg \tn \la 8.35$ and $8.0 \la \lg \tp \approx 8.4$,
shaded rectangle). The observations of
RX~J0822-43 can also be explained by the presence of the
same superfluidity
($\lg \tn = 8.3$, $\lg \tp = 8.2$) assuming that the object
possesses an envelope of light elements of the 
$7\times 10^{-10} M_\odot$ mass (see Fig.\ \ref{fig:NS_data}).
In the absence of the envelope, the confidence region of
$\tn$ and $\tp$ is outside 
of the joint confidence region 
since the error bar of $T_s^\infty$ for this object
lies at the boundary of the shaded region (filled by 
the cooling curves of superfluid NSs).

\subsection{Discussion}

As follows from the results of Sect.\ 8.2,
the majority of NS observations can be explained using
the adopted NS models by the standard and enhanced cooling
for the ``black body" and ``atmospheric" interpretations of the
spectra observed. In all these cases there are
the values of $\tn$ and $\tp$ close or common for
all NSs which do not contradict to the microscopic theories
(Sect.\ 3.1), but depend on cooling type and interpretation
of the thermal NS radiation.

The existence of the same critical temperatures
for several NSs at once is quite unexpected.
Initially we have expected different joint confidence regions of
$T_{cn}$ and $T_{cp}$ for different pairs of NSs.
The result is even more surprising taken into account
simplicity of our cooling models. It would be interesting
to confirm (or reject) these results using more advanced
cooling models and a larger number of objects.

We have checked that our results are rather insensitive
to variations of the NS mass $M$ as long as $M$
does not pass through the threshold value
$M_{cr}=1.442\, M_\odot$ (of switching on the direct Urca).
This fact is a consequence of the main simplification of our
models, constancy of critical temperatures
throughout the NS core. In particular, the standard cooling curves
for the NSs with masses lower than
$M_{cr}$ practically coincide (see, e.g., Fig.\ \ref{fig:PA}).
The same takes place for the enhanced cooling curves
of NSs (with masses $M \ga M_{cr}$), if protons or neutrons in their cores
possess moderate or strong superfluidity with $T_c \ga 10^8$~K
(just as for the
joint confidence regions  of $\tn$ and $\tp$ obtained above).

We have assumed in our analysis that the NS ages $t$ are known
(Sect.\ 8.1.3). Then the confidence regions of
$\tn$ and $\tp$ for each star have been determined from an
error bar of the surface temperature
$T_s$ obtained from the observations.
Introduction of the error bars of $\Delta t$
instead of the fixed values of $t$ would lead to
additional broadening of the confidence ranges of
$\tn$ and $\tp$ for each NS and to slight broadening of the
joint confidence regions of
$\tn$ and $\tp$. Let us mention that the characteristic ages
of PSR~0656+14, Geminga and PSR1055-52 are determined within a factor
of $\sim 3$. The uncertainty in ages of young NSs is higher
which, however, should not cause stronger broadening of the
$\tn$ and $\tp$ regions due to weak slope of the cooling curves
at $t=10^3$--$10^4$~yr (see Fig.\ \ref{fig:NS_data}).

For the black body
interpretation of the thermal radiation from NSs,
our simulations predict
a strong neutron superfluidity,
$\tn \approx 10^{9.7}$ -- $10^{10.0}$~K,
and weak proton superfluidity in the NS cores.
This conclusion is in a qualitative agreement with
the result by
Page (1994).
A strong proton superfluidity is also possible
in the presence of a moderate neutron superfluidity.
In such cases one superfluidity ($n$ or $p$) is noticeably stronger
than the other.
In the both cases of ``black body" and ``atmospheric" interpretations
of the observations the largest contrast of
$\tn$ and $\tp$ takes place for the NS cooling
with the allowed direct Urca process.
Notice that such a superfluidity is predicted by
Takatsuka and Tamagaki (1997b).

In neither case
we need simultaneously strong superfluidities of $n$ and $p$
to explain the observations. This indicates that the equation
of state of the NS cores cannot be too soft
(the softness would mean weak nucleon--nucleon repulsion at small separations
which would induce especially strong pairing, Sect.\ 3.1).

Finally let us mention that we can satisfy
the observations by varying the only parameter, $\tp$,
if we assume the presence of the neutron superfluidity
with $\tn \approx 10^{8.1}$--$10^{8.3}$~K. For the standard
cooling, this parameter should lie in the moderate range
$10^{8.0}$--$10^{9.0}$~K, while for the enhanced cooling
it should be larger, $10^{9.45}$--$10^{9.65}$~K.

\section{Conclusions}

We have described one of a few methods to study fundamental
properties of matter of density around the nuclear matter density
or several times higher. The method is based on theoretical
simulations of cooling of neutron stars and comparison
of the results with the observations of the thermal radiation from
isolated cooling neutron stars.

We have described all the basic elements of the theory of cooling of
neutron stars with the standard nuclear composition:
stellar heat capacity (Sect.\ 3), neutrino cooling (Sects.\ 4--6),
relationship between the internal and surface stellar temperatures
(Sect.\ 7.2.1). However we have not intended to discuss
all the cooling problems in detail but focused on a relatively
new direction of the theory which had been initiated by the remarkable
paper by
Page and Applegate (1992).
These authors were the first who noticed that the neutron star
cooling could be very sensitive to the superfluidity of neutrons
and protons in their cores. The temperatures of the
superfluidity onset, in their turn, are sensitive to the equation of
state of superdense matter. This opens the principal possibility
to explore the nucleon superfluidity and, accordingly,
the equation of state in the neutron star cores.

The effects of superfluidity had been included into
cooling simulations long before
Page and Applegate (1992).
However they had not been the objects of special attention;
systematic analysis of the effects of superfluidity on
the neutrino luminosity and the heat capacity had been absent,
and the effects had been described by the oversimplified
expressions (Sect.\ 7.1).
We have tried to fill in this gap and presented the results
of detailed calculations of the heat capacity and neutrino
emissivities in superfluid matter in the frame of
the Bardeen--Cooper--Schrieffer theory (Sects.\ 3 --6).
Our consideration is done in a unified manner. The results
are mainly presented in the form of simple fit expressions
convenient for practical implications. Thus we have
produced a rather elaborated method
to include the effects of
superfluidity into simulations of neutron--star cooling.
This method can be extended easily to the neutron star models with
superfluid hyperons.

For illustration, we have presented
(Sect.\ 7.2) the results of cooling simulations of neutron stars
whose cores contain superfluid neutrons and protons.
For simplicity, the critical temperatures
$\tn$ and $\tp$ have been assumed constant
over the neutron star core. We have analyzed the effect of superfluidity
on the enhanced and standard cooling and showed that this effect
is really crucial. If the standard cooling of middle--aged
($10^2$--$10^5$~yr) neutron stars with non-superfluid cores
goes much slower than the cooling enhanced by the
direct Urca process, this is not so in the presence of superfluidity.
As a whole, the nucleon superfluidity strongly reduces the difference
between the standard and enhanced cooling.
This enables one, in principle, to explain the majority of
observational data by the standard cooling of the stars with
superfluid cores. On the other hand, this circumstance
disproves a widely accepted point of view
that the direct Urca process necessarily induces rapid
cooling.

We have compared (Sect.\ 8.2) the results of cooling simulations
with the observational data on the thermal radiation
of neutron stars. Almost in all the cases
(for different methods of inferring the surface temperatures from
observational data) the observations can be explained
in the models of the standard and enhanced cooling of a neutron star
with the superfluid core. It is important that the required
values of the critical temperatures
$\tn$ and $\tp$ do not contradict to the values
predicted by various microscopic theories of
nucleon superfluidity in superdense matter (Sect.\ 3.1).
However our results should be regarded as preliminary; they
cannot allow us to identify reliably
superfluidity and equation of state in the neutron star cores.
To impose more stringent constraints one needs to carry out
more elaborated simulations of neutron star cooling
taking into account density dependence of
$\tn$ and $\tp$ in the stellar core. It would be very desirable
to make an equation of state of matter, used in cooling simulations,
consistent with the critical temperatures. For this purpose,
one needs a representative set of equations of state
of superdense matter (from soft to stiff ones),
supplemented by the values of $T_{cn}(\rho)$ and
$T_{cp}(\rho)$, calculated from the same microscopic models.
This would enable one to construct a set of more realistic
cooling curves and improve thus ``calibration" of neutron stars
as natural thermometers of superfluidity in their cores.
Comparing theory with observations one would be able to
constrain strongly the set of allowable equations of state.
Unfortunately, simultaneous microscopic calculations
of the equations of state and superfluid gaps are almost absent.
On the other hand, the surface temperatures of the neutron stars
cannot be determined uniquely from observations so far.
 
We hope that after appearance of self-consistent
calculations of the equations of state and superfluid gaps
and after launch of orbital observatories of new generation
({\it AXAF, XMM, ASTRO-E, SXG}$\,$),
which would produce more reliable data on
the surface temperatures of neutron stars, this method
will enable one to reach deeper understanding of the nature
of superdense matter.

{\bf Acknowledgments.}
We are grateful to D.A.\ Baiko, V.G.\ Bezchastnov, P.\ Haensel,
A.B.\ Koptsevich, G.G.\ Pavlov,  M.G.\ Urin,
D.N.\ Voskresensky, and V.E.\ Zavlin
for discussions of the problems included in this review.
Our special thanks are to A.D.\ Kaminker and A.Y.\ Potekhin
for reading the manuscript of the review and making
useful comments. The review is partly based on the course
of lectures ``White dwarfs and neutron stars""
given by one of the authors (D.G.\ Yakovlev)
to 6-year students specializing in ``Cosmic studies"
at the St.-Petersburg State Technical University.
The work was supported in part by
RFBR (grant No.\ 99-02-18099) and INTAS (96-0542).

%
%



\begin{thebibliography}{222}

\bibitem{awp89}
    Ainsworth~T~L,  Wambach~J,  Pines~D
    {\it Phys.\ Lett.\ } {\bf B222} 173 (1989)




\bibitem{anp85}
      Alpar~M~A, Nandkumar~R, Pines~D
      {\it Astrophys.\ J.\ }  {\bf 288} 191 (1985)


\bibitem{abko87}
     Alpar~M~A, Brinkmann~W, \"{O}gelman~H,  Kizilo\u{g}lu~\"{U}
     {\it Astron.\ Astrophys.\ } {\bf 177} 101 (1987)


\bibitem{ao85a}
      Amundsen~L, {\O}stgaard~E
      {\it Nucl.\ Phys.\ } {\bf A437}  487  (1985a)

\bibitem{ao85b}
      Amundsen~L,  {\O}stgaard~E
      {\it Nucl.\ Phys.\ } {\bf A442} 163 (1985b)

\bibitem{am61}
      Anderson~P~W, Morel~P
      {\it Phys.\ Rev.\ }  {\bf 123} 1911 (1961)

\bibitem{acprt93}
     Anderson~S~B, C\'{o}rdova~F~A, Pavlov~G~G, Robinson~C~R, Thompson~R~J
     {\it Astrophys.\  J.\ }  {\bf 414} 867 (1993)


\bibitem{bz34}
    Baade~W, Zwicky~F
    {\it Phys.\ Rev.\ } {\bf 45} 138 (1934)


\bibitem{bw65a}
    Bahcall~J~N, Wolf~R~A
    {\it Phys.\ Rev.\ } {\bf 140} 1445 (1965a)

\bibitem{bw65b}
     Bahcall~J~N, Wolf~R~A
     {\it Phys.\ Rev.\ } {\bf 140} 1452 (1965b)

\bibitem{by99}
    Baiko~D~A, Yakovlev~D~G
    {\it Astron.\ Astrophys.\ } {\bf 342} 192 (1999)


\bibitem{bl84}
    Bailin~D, Love~A
    {\it Phys.\ Rep.\ } {\bf 107} 325
    (1984)

\bibitem{bb98}
   Balberg~S, Barnea~N
   {\it Phys.\ Rev.\ } {\bf C57} 409 (1998)

\bibitem{bcll90}
   Baldo~M, Cugnon~J, Lejeune~A, Lombardo~U
   {\it Nucl.\ Phys.\ } {\bf A515} 409 (1990)

\bibitem{bcll92}
   Baldo~M, Cugnon~J, Lejeune~A, Lombardo~U
   {\it Nucl.\ Phys.\ } {\bf A536} 349 (1992)

\bibitem{bw63}
   Balian~R, Werthamer~N~R
   {\it Phys.\ Rev.\ } {\bf 131} 1553 (1963)

\bibitem{bcs57}
   Bardeen~J, Cooper~L~N, Schrieffer~J~R
   {\it Phys.\ Rev.\ } {\bf 108} 1175 (1957)

\bibitem{bbev98}
   Barranco~F, Broglia~R~A, Esbensen~H~E, Viggezi~E
   {\it Phys.\ Rev.\ } {\bf C58} 1257 
   (1998)

\bibitem{bp91}
   Baym~G, Pethick~C~J
   {\it Landau Fermi-Liquid Theory}
   (Wiley, New York, 1991)

\bibitem{bpp69}
   Baym~G, Pethick~C~J, Pines~D
   {\it Nature } {\bf 224} 673 (1969)

\bibitem{ba95}
   Becker~W, Aschenbach~B
   {\it The Lives of Neutron Stars} (Eds Alpar~M~A et al)
   (Kluwer, Dordrecht, 1995) p.~47

\bibitem{bt97}
   Becker~W, Tr\"{u}mper~J
   {\it Astron.\ Astrophys.\ } {\bf 326} 682 (1997)

\bibitem{bt99}
   Becker~W, Tr\"{u}mper~J.
   {\it Astron.\ Astrophys.\ } {\bf 341} 803 (1999) 


\bibitem{bpv98}
   Bezchastnov~V~G, Pavlolv~G~G, Ventura~J
   {\it Phys.\ Rev.\ } {\bf A58} 180 (1998)



\bibitem{bcmeb96}
   Bignami~G~F et al.
  {\it Astrophys.\ J.\ Letters} {\bf 456} L111 (1996)




\bibitem{boguta81}
    Boguta~J
    {\it Phys.\ Lett.\ } {\bf B106} 255
    (1981)

\bibitem{bmp58}
    Bohr~A, Mottelson~B~R, Pines~D
    {\it Phys.\ Rev.\ } {\bf 110} 936 (1958)

\bibitem{bm90}
    Botermans~W, Malfliet~R
    {\it Phys.\ Rep.\ } {\bf 198} 115 (1990)

\bibitem{bbcf64}
    Bowyer~S, Byram~E~T, Chubb~T~A, Friedman~H
    {\it Nature} {\bf 201} 1307
    (1964)

\bibitem{bo87}
    Brinkmann~W, \"{O}gelman~H
    {\it Astron.\ Astrophys.\ } {\bf 182} 71 (1987)

\bibitem{bbllp94}
    Broglia~R~A, De Blasio~F, Lazzari~G, Lazzari~M, Pizzochero~P~M
    {\it Phys.\ Rev.\ } {\bf D50} 4781 (1994)

\bibitem{bkpp88}
    Brown~G~E, Kudobera~K, Page~D, Pizzochero~P~M
    {\it Phys.\ Rev.\ } {\bf D37} 2042 (1988) 

\bibitem{burrows80}
    Burrows~A
    {\it Phys.\ Rev.\ Lett.\ } {\bf 44} 1640 (1980)




\bibitem{ccy72}
   Chao~N-C, Clark~J~W, Yang~C-H
   {\it Nucl.\ Phys.\ } {\bf A179} 320 (1972)

\bibitem{cpy97}
   Chabrier~G, Potekhin~A~Y, Yakovlev~D~G
   {\it Astrophys.\ J.\ Lett.\ } {\bf 477} L99 (1997)

\bibitem{ccdk93}
   Chen~J~M, Clark~J~W, Dav\'{e}~R~D, Khodel~V~V
   {\it Nucl.\ Phys.\ } {\bf A555} 59 (1993)

\bibitem{ccks86}
   Chen~J~M, Clark~J~W, Krotscheck~E, Smith~R~A
   {\it Nucl.\ Phys.\ }{\bf A451} 509 (1986)



\bibitem{cherepashchuk96}
      Cherepashchuk~A~M
      Uspekhi Fiz.\ Nauk {\bf 166} 809 (1966)


\bibitem{chiu64}
       Chiu~H-Y
       Ann.\ Phys.\ {\bf 26} 364 
       (1964) 

\bibitem{cs64}
       Chiu~H-Y, Salpeter~E~E
       {\it Phys.\ Rev.\ Lett.\ } {\bf 12} 413 (1964)  

\bibitem{crv97}
  Civitareze~O, Reboiro~M, Vogel~P
  {\it Phys.\ Rev.\ } {\bf C56} 1840 
  (1997)

\bibitem{ckyc76}
   Clark~J~W, K\"{a}llman~C-G, Yang~C-H, Chakkalakal~D~A
   {\it Phys.\ Lett.\ }{\bf B61} 331 (1976)

\bibitem{cms59}
      Cooper~L~N, Mills~R~L, Sessler~A~M
      {\it Phys.\ Rev.\ } {\bf 114} 1377 (1959)


\bibitem{chp97}
     Craig~W~W, Hailey~Ch~J, Pisarski~R~L
     {\it Astrophys.\ J.\ } {\bf 488} 307 (1997)

\bibitem{dl95}
      De Blaiso~F~V, Lazzari~G
      {\it Phys.\ Rev.\ }{\bf C52} 418 
      (1995)  


\bibitem{eeho96a}
     Elgar{\o}y~{\O}, Engvik~L, Hjorth-Jensen~M, Osnes~E
     {\it Phys.\ Rev.\ Lett.\ } {\bf 77} 1428
     (1996a)

\bibitem{eeho96b}
     Elgar{\o}y~{\O}, Engvik~L, Hjorth-Jensen~M, Osnes~E
     {\it Nucl.\ Phys.\ } {\bf A604} 466 (1996b)

\bibitem{eeho96c}
     Elgar{\o}y~{\O}, Engvik~L, Hjorth-Jensen~M, Osnes~E
     {\it Nucl.\ Phys.\ } {\bf A607} 425 (1996c)

\bibitem{eeodhl96}
     Elgar{\o}y~{\O}, Engvik~L, Osnes~E, De Blasio~F~V,
     Hjorth-Jensen~M, Lazzari~G
     {\it Phys.\ Rev.\ } {\bf D54} 1848 (1996d)

\bibitem{eh98}
     Elgar{\o}y~{\O}, Hjorth-Jensen~M
     {\it Phys.\ Rev.\ } {\bf C57} 1174
     (1998)


\bibitem{fok92}
     Finley~J~P, \"{O}gelman~H, Kizilo\u{g}lu~\"{U}
     {\it Astrophys.\ J.\ Lett.\ } {\bf 394} L21 (1992)


\bibitem{frs76}
     Flowers~E~G, Ruderman~M, Sutherland~P~G
     {\it Ap.\ J.\ } {\bf 205} 541 (1976)

\bibitem{fsb75}
     Flowers~E~G, Sutherland~P~G, Bond~J~R
     {\it Phys.\ Rev.\ }{\bf D12} 315 (1975)

\bibitem{fm79}
     Friman~B~L, Maxwell~O~V
     {\it Astrophys.\ J.\ } {\bf 232} 541 (1979)


\bibitem{ginzburg69}
    Ginzburg~V~L
    {\it Uspekhi Fiz.\ Nauk} {\bf 97} 601 (1969)

\bibitem{gk64}
     Ginzburg~V~L, Kirzhnits~D~A
     {\it Zh.\ Teor.\ Eksper.\ Fiz.\ } {\bf 47} 2006 (1964)


\bibitem{gs80}
    Glen~G, Sutherland~P
    {\it Astrophys.\ J.\ } {\bf 239} 671 (1980)

\bibitem{gy93}
      Gnedin~O~Y, Yakovlev~D~G
      {\it Astron. Lett.\ } {\bf 19} 104 (1993)

\bibitem{gys94}
     Gnedin~O~Y, Yakovlev~D~G, Shibanov~Yu~A
     {Astron.\ Lett.\ } {\bf 20} 409 (1994)

\bibitem{gr92}
     Goldreich~P, Reisenegger~A
     {\it Astrophys.\ J.\ } {\bf 395} 250 (1992)

\bibitem{gv97}
    Gotthelf~E~V, Vasisht~G
    {\it Astrophys.\ J.\ Lett.\ } {\bf 486} L133 (1997)

\bibitem{gph97}
    Gotthelf~E~V, Petre~R, Hwang~U
    {\it Astrophys.\ J.\ Lett.\ } {\bf 487} L175 (1997)

\bibitem{gpv99}
    Gotthelf~E~V, Petre~R, Vasisht~G
    {\it Astrophys.\ J.\ Lett.\ }
    accepted (1999) (astro-ph/9901371)


\bibitem{gpe83}
    Gudmundsson~E~H, Pethick~C~J, Epstein~R~I
    {\it Astrophys.\ J.\ } {\bf 272} 286 (1983)


\bibitem{hmp98}
Haberl~F, Motch~C, Pietsch~W {\it Astron.\ Nachr.\ }
{\bf 319} 97 
(1998)

\bibitem{hmbzp97} 
   Haberl~F et al. 
   {\it Astron.\ Astrophys.\ } {\bf 326} 662 (1997)

\bibitem{haensel87}
   Haensel~P
  {\it Progr.\ Theor.\ Phys.\ Suppl.\ }
  {\bf 91} 268 (1987)

\bibitem{haensel94}
   Haensel~P
   {\it Acta Physica Polonica } {\bf B25} 373 (1994)

\bibitem{hg94}
   Haensel~P, Gnedin~O~Y
  {\it Astron.\ Astrophys.\ } {\bf 290} 458 (1994)

\bibitem{hp94}
   Haensel~P, Pichon~B
   {\it Astron.\ Astrophys.\ } {\bf 283} 313 (1994)

\bibitem{huy90}
   Haensel~P, Urpin~V~A, Yakovlev~D~G
   {\it Astron.\ Astrophys.\ } {\bf 229} 133 (1990)


\bibitem{hc95}
        Hailey~Ch~J, Craig~W~W
        {\it Astrophys.\ J.\ Lett.\ } {\bf 455} L151 (1995)



\bibitem{hr93}
    Halpern~J~P, Ruderman~M
    {\it Astrophys.\ J.\ } {\bf 415} 286 (1993)


\bibitem{hw97}
   Halpern~J~P, Wang~F~Y-H
   {\it Astrophys.\ J.\ } {\bf 477} 905 (1997)



\bibitem{hw64}
   Henley~E~M, Wilets~L
   {\it Phys.\ Rev.\ } {\bf B133} 1118 (1964)

\bibitem{hbpsc68} 
   Hewish~A, Bell~S~J, Pilkington~J~D~H, Scott~P~F, Collins~R~A
   {\it Nature } {\bf 217} 709 (1968)

\bibitem{hh97a}
   Heyl~J~S, Hernquist~L
   {\it Astrophys.\ J.\ Lett.\ } {\bf 491} L95 
   (1997a)

\bibitem{hh97b}
     Heyl~J~S, Hernquist~L
    {\it Astrophys.\ J.\ Lett.\ } {\bf 489} L67
    (1997b)

\bibitem{hh98a}
    Heyl~J~S, Hernquist~L
    {\it MNRAS} {\bf 298} 17 (1998a)

\bibitem{hh98b}
    Heyl~J~S, Hernquist~L
    {\it MNRAS} {\bf 300} 599 (1998b)

\bibitem{hh98c}
    Heyl~J~S, Hernquist~L
    {\it MNRAS} (1998c) (in press)

\bibitem{hk98}
    Heyl~J~S, Kulkarni~S~R
    {\it Astrophys.\ J.\ Lett.\ } {\bf 506} L61 
    (1998)


\bibitem{hgrr70}
   Hoffberg~M, Glassgold~A~E, Richardson~R~W, Ruderman~M
   {\it Phys.\ Rev.\ Lett.\ } {\bf 24} 775 (1970)

\bibitem{hurleyetal99}
  Hurley K et al
  A giant, periodic flare from the soft gamma-repeater SGR 1900+14
  {\it Nature } {\bf 397} 41 (1999) 


\bibitem{in82}
    Imshennik~V~S, Nadyozhin~D~K
    {\it ``Soviet Scientific Reviews", Section E: Astrophysics
    and Space Physics Reviews} (R~A~Sunyaev ed.) {\bf 2} 75
    (1982) Harwood Academic Publishers

\bibitem{in88}
    Imshennik~V~S, Nadyozhin~D~K
    {\it Uspekhi Fiz.\ Nauk.\ } {\bf 156} 561
    (1988)

\bibitem{itty63}
   Ishihara~T, Tamagaki~R, Tanaka~H, Yasuno~M
   {\it Prog.\ Theor.\ Phys.\ } {\bf 30} 601 (1963)


\bibitem{ikms84}
   Itoh~N, Kohyama~Y, Matsumoto~N, Seki~M
   {\it Astrophys.\ J.\ } {\bf 285} 304 (1984)


\bibitem{ihnk96}
    Itoh~N, Hayashi~H, Nishikawa~A, Kohyama~Y
    {\it Astrophys.\ J.\ Suppl.\ } {\bf 102} 411 (1996)

\bibitem{it72}
    Itoh~N, Tsuneto~T
    {\it Prog.\ Theor.\ Phys.\ }{\bf 48} 1849 (1972)

\bibitem{iwamoto80}
    Iwamoto~N
    {\it Phys.\ Rev.\ Lett.\ }  {\bf 44} 1637 (1980)

\bibitem{iwasaki95}
   Iwasaki~M
   {\it Prog.\ Theor.\ Phys.\ } {\bf 120} 187
   (1995)

\bibitem{kppty99}
   Kaminker~A~D, Pethick~C~J, Potekhin~A~Y, Thorsson~V, Yakovlev~D~G
   {\it Astron.\ Astrophys.\ } {\bf 343} 1009
   (1999a)

\bibitem{kh99}
   Kaminker~A~D, Haensel~P
   {\it Acta Phys.\ Polonica} {\bf 30} 1125 (1999) 

\bibitem{kyh97}
    Kaminker~A~D, Yakovlev~D~G, Haensel~P
    {\it Astron.\ Astrophys.\ } {\bf 325} 391 (1997)

\bibitem{khy99}
   Kaminker~A~D, Haensel~P, Yakovlev~D~G
   {\it Astron.\ Astrophys.\ } {\bf 345} L14 (1999b)

\bibitem{kn86}
   Kaplan~D~B, Nelson~A~E
   {\it Phys.\ Lett.\ } {\bf B175} 57
   (1986); 
   erratum: {\bf 179} 409 (1986)

\bibitem{kennedy68}
   Kennedy~R~C
   {\it Nucl.\ Phys.\ } {\bf A118} 189 (1968)

\bibitem{kwh64}
   Kennedy~R, Wilets~L, Henley~E~M
   {\it Phys.\ Rev.\ } {\bf B133} 1131 (1964)


\bibitem{kb96}
  Korpela~E~J, Bowyer~S.
  {\it Astron.\ Astrophys. Suppl.\ } {\bf 188} 4301 (1996)

\bibitem{kouveliotouetal98}
  Kouveliotou~C et al. 
 {\it Nature} {\bf 393} 235 (1998a)

\bibitem{kfwk98}
   Kouveliotou~C, Fishman G~J, Woods P, Kippen M
   {\it IAU Circular} 7003 (1998b)

\bibitem{krotscheck72}
   Krotscheck~E
   {\it Zeit\. Phys.\ }  {\bf 251} 135 (1972)

\bibitem{kk98}
   Kulkarni~S~R, van Kerkwijk~M~H
   {\it Astrophys.\ J.\ Lett.\ } {\bf 507} L49 (1998) 

 


\bibitem{lpph91}
     Lattimer~J~M, Pethick~C~J, Prakash M, Haensel~P
     {\it Phys.\ Rev.\ Lett.\ } {\bf 66} 2701 (1991)

\bibitem{lvpp94}
     Lattimer~J~M, Van Riper~K, Prakash~M, Prakash~M
     {\it Astrophys.\ J.\ } {\bf 425} 802 (1994)

\bibitem{ly93}
  Levenfish~K~P, Yakovlev~D~G
  {\it Strongly Coupled Plasma Physics}
  (Eds Van Horn~H~M, Ichimaru~S) (Univ.\ of Rochester Press,
  Rochester, 1993) p.~167

\bibitem{ly94a}
   Levenfish~K~P, Yakovlev~D~G
   {\it Astron.\ Reports\ } {\bf 38}  247 (1994a)

\bibitem{ly94b}
   Levenfish~K~P, Yakovlev~D~G
   {\it Astron.\ Lett.\ } {\bf 20} 43 (1994b)

\bibitem{ly96}
   Levenfish~K~P, Yakovlev~D~G
   {\it Astron.\ Lett.\ } {\bf 22} 47 (1996)

\bibitem{lsy98}
   Levenfish~K~P, Shibanov~Yu~A, Yakovlev~D~G
   {\it Physica Scripta} {\bf T77} 79 (1998)

\bibitem{lsy99}
  Levenfish~K~P, Shibanov~Yu~A, Yakovlev~D~G
  {\it Astron.\ Lett.\ } {\bf 25} 
   (1999)

\bibitem{lvv95}
Lewin~W~H~G, Van Paradijs~J, Van den Huevel E~P~G (eds),
{\it X-ray Binaries} (Cambridge: Cambridge University Press, 1995)

\bibitem{lp73}
   Lifshitz~E~M, Pitaevskii~L~P
   {\it Statistical Physics, Part~2}
   (Pergamon, Oxford 1980)

\bibitem{lrp93}
    Lorenz~C~P, Ravenhall~D~G, Pethick~C~J
    {\it Phys.\ Rev.\ Lett.\ }
    {\bf 70} 379 (1993)

\bibitem{lpgc96}
   Lyne~A~G, Pritchard~R~S, Graham-Smith~F, Camilo~F
   {\it Nature } {\bf 381} 497 (1996)

\bibitem{malone74}
    Malone~R~C
    {\it Ph.D.\ thesis} (Cornell University, 1974)



\bibitem{mhs98}
   Martin~C, Halpern~J~P, Schiminovich~D
   {\it Astrophys.\ J.\ Lett.\ } {\bf 494} L211 (1998)

\bibitem{maxwell79}
    Maxwell~O~V {\it Astrophys.\ J.\ } {\bf 231} 201 (1979)

\bibitem{maxwell87}
    Maxwell~O~V
    {\it Astrophys.\ J.\ } {\bf 316} 691 (1987)

\bibitem{mbcdm77}
    Maxwell~O~V, Brown~G~E, Campbell~D~K, Dashen~R~F, Manassah~J~T
    {\it Astrophys.\ J.\ } {\bf 216} 77 (1977)

\bibitem{mis98}
    Mereghetti S, Israel G~L, Stella L
    {\it MNRAS} {\bf 296} 689 (1998)

\bibitem{mpm94}
    Meyer~R~D, Pavlov~G~G, M\'{e}sz\`{a}ros~P
    {\it Astrophys.\ J.\ } {\bf 433} 265 (1994)

\bibitem{migdal59}
   Migdal~A~B
   {\it Nucl.\ Phys.\ } {\bf 13} 655 (1959)

\bibitem{migdal71}
    Migdal~A~B
    {\it Uspekhi Fiz.\ Nauk} {\bf 105} 781 (1971)


\bibitem{mignani98}
  Mignani~R~P {\it Neutron Stars and Pulsars}
  (Eds Shibazaki~N et al.) 
  (Universal Academy Press, Tokyo, 1998) p.~335 



   (1998)

\bibitem{muk98}
    Miralles~J~A, Urpin~V~A, Konenkov~D~Yu
    {\it Astrophys.\ J.\ } {\bf 503} 368 
    (1998)

\bibitem{morton64}
    Morton~D~C
   {\it Nature } {\bf 201} 1308 (1964)


\bibitem{muhlschlegel59}
   M\"{u}hlschlegel~B
   {\it Zeit.\ Phys.\ } {\bf 155} 313 (1959)

\bibitem{mt88}
   Muto~T, Tatsumi~T
   {\it Progr.\ Theor.\ Phys.\ } {\bf 79} 461 
   (1988) 

\bibitem{mss80}
   Muzikar~P, Sauls~J~A, Serene~J~W
   {\it Phys.\ Rev.\ } {\bf D21} 1494 (1980)

\bibitem{nv73}
   Negele~J~W, Vautherin~D
   {\it Nucl.\ Phys.\ } {\bf A207} 298 (1973)

\bibitem{nk87}
   Nelson~A~E, Kaplan~D~B
   {\it Phys.\ Lett.\ } {\bf B192} 193 
   (1987)

\bibitem{nt99}
Neuh\"{a}user R, Tr\"{u}mper J~E
{\it Astron.\ Astrophys.\ } {\bf 343} 151 (1999)

\bibitem{nt81}
   Nomoto~K, Tsuruta~S
   {\it Astrophys.\ J.\ Lett.\ } {\bf 250} L19 (1981)

\bibitem{nt82}
    Nomoto~K, Tsuruta~S
    {\it Accreting Neutron Stars}
    (Eds Tr\"{u}mper~J, Brinkmann~W)
    (Max Plank Institute, Garching, 1982) p.~275

\bibitem{nt83}
    Nomoto~K, Tsuruta~S
   {\it Supernova Remnants and Their X-ray Emission }
   (Eds Danziger~J, Gorenstein~P) (Reidel,
   Dordrecht, 1983)  p.~509

\bibitem{nt86}
   Nomoto~K, Tsuruta~S
   {\it Astrophys.\ J.\ Lett.\ } {\bf 305} L19 (1986)

\bibitem{nt87}
   Nomoto~K, Tsuruta~S
   {\it Astrophys.\ J.\ } {\bf 312} 711 (1987)

\bibitem{ogelman95}
   \"{O}gelman~H
  {\it Lives of Neutron Stars} 
  (Eds Alpar~M~A, Kizilo\u{g}lu \"{U}, van Paradjis J.)
  (Kluwer, Dordrecht, 1995)
  p.~101


\bibitem{ofz93}
   \"{O}gelman~H, Finley~J~P, Zimmermann~H~U
   {\it Nature } {\bf 361} 136 (1993)


\bibitem{okun90}
    Okun'~L~B {\it Leptons and Quarks} (Nauka, Moscow, 1990)



\bibitem{page94}
   Page~D
   {\it Astrophys.\ J.\ } {\bf 428}  250 (1994)

\bibitem{page95a}
     Page~D
     {\it Astrophys.\ J.\ } {\bf 442} 273 (1995a) %

\bibitem{page95b}
      Page~D
      {\it Revista Mexicana de Fisica } {\bf 41} Supl.\ 1 178 (1995b)

\bibitem{page97}
    Page~D
    {\it Astrophys.\ J.\ Lett.\ } {\bf 479} L43 (1997)

\bibitem{page98a}
    Page~D
    {\it The Many Faces of Neutron Stars}
    (Eds Buccheri R, van Paradijs J, Alpar M~A)
    (Kluwer, Dordrecht, 1998a) p.\ 538

\bibitem{page98b}
   Page~D
   {\it Neutron Stars and Pulsars} (Eds Shibazaki~N et al)
   (Universal Academy Press, Tokyo, 1998b) p.~183

\bibitem{pa92}
   Page~D,  Applegate~J~H
   {\it Astrophys.\ J.\ Lett.\ } {\bf 394} L17 (1992)

\bibitem{pb90}
   Page~D, Baron~E
   {\it Astrophys.\ J.\ Lett.\ } {\bf 354} L17 (1990)
   erratum: {\it Astrophys.\ J.\ } {\bf 382} L11


\bibitem{psz96}
   Page~D, Shibanov~Yu~A, Zavlin~V~E
   {\it R\"{o}ntgenstrahlung from the Universe}
   (Eds Zim\-mer\-mann H~U, Tr\"{u}mper~J~E, Yorke~H)
   (Max-Planck Institute f\"{u}r Extraterrestrische Physik, Garching, 1996)
   p.~173

\bibitem{pandharipande71}
    Pandharipande~V~R
    {\it Nucl.\ Phys.\ }  {\bf A174} 641 
    (1971)

\bibitem{ppt95}
    Pandharipande~V~R, Pethick~C, Thorsson~V
    {\it Phys.\ Rev.\ Lett.\ } {\bf 75} 4567 (1995) 

\bibitem{ps75}
     Pandharipande~V~R, Smith~R~A
     {\it Phys.\ Lett.\ } {\bf B59} 15 (1975)


\bibitem{pszm95}
      Pavlov~G~G, Shibanov~Yu~A, Zavlin~V~E, Meyer~R~D
      {\it The Lives of the Neutron Stars}
      (Eds Alpar~M~A, Kizilo\u{g}lu~\"{U}, van Paradijs~J)
      (Kluwer, Dordrecht, 1995) p.~71



\bibitem{pz98}
      Pavlov~G~G, Zavlin~V~E
      {\it  Neutron Stars and Pulsars}
      (Eds  Shibazaki~N et al)  
      (Universal Academy Press, Tokyo, 1998) p.~327 

\bibitem{pztn96}
      Pavlov~G~G, Zavlin~V~E, Tr\"{u}mper~J, Neuh\"{a}user~R
      {\it Astrophys.\ J.\ Lett.\ } {\bf 472} L33 (1996)


\bibitem{pethick92}
      Pethick~C~J
      {\it  Rev.\ Mod.\ Phys.\ } {\bf 64} 1133 (1992)  

\bibitem{pr95}
      Pethick~C~J, Ravenhall~D~G
      {\it Ann.\ Rev.\ Nucl.\ Particle Sci.\ } {\bf 45} 429 (1995)

\bibitem{pr98}
      Pethick~C~J, Ravenhall~D~G
      {\it The Many Faces of Neutron Stars}
      (Eds  Buccheri~R, van Paradijs~J, Alpar~M~A)
      (Kluwer, Dordrecht, 1998) p.~49

\bibitem{pt94}
      Pethick~C~J, Thorsson~V
      Phys.\ Rev.\ Lett.\ {\bf 72} 1964 
      (1994)

\bibitem{pines91}
      Pines~D
      {\it Neutron Stars: Theory and observation}
      (Eds Ventura~J, Pines~D) (Kluwer, Dordrecht, 1991)
      p.~57


\bibitem{pizzochero98}
     Pizzochero~P~M 
{\it Astrophys.\ J.\ Lett.\ } {\bf 502} L153 (1998)

\bibitem{pmc96}
     Possenti~A, Mereghetti~S, Colpi~M
     {\it Astron.\ Astrophys.\ } {\bf 313} 565 (1996)

\bibitem{potekhin94}
     Potekhin~A~Y
     {\it J.\ Phys.\ B.: At.\ Mol.\ Opt.\ Phys.\ } {\bf 27} 1073 (1994)

\bibitem{pp97}
    Potekhin~A~Y, Pavlov~G~G
    {\it Astrophys.\ J.\ } {\bf 483} 414 (1997)

\bibitem{pcy97}
   Potekhin~A~Y, Chabrier~G, Yakovlev~D~G
   {\it Astron.\ Astrophys.\ } {\bf 323} 415 (1997)

\bibitem{pal88}
   Prakash~M, Ainsworth~T~L, Lattimer~J~M
   {\it Phys.\ Rev.\ Lett.\ } {\bf 61} 2518 (1988)

\bibitem{pplp92}
   Prakash~M, Prakash~M, Lattimer~J~M, Pethick~C~J
   {\it Astrophys.\ J.\ Lett.\ } {\bf 390} L77 (1992)

\bibitem{rr96}
    Rajagopal~M, Romani~R~W
    {\it Astrophys.\ J.\ } {\bf 461} 327 (1996)

\bibitem{rrm97}
    Rajagopal~M, Romani~R~W, Miller~M~C
   {\it Astron.\ J.\ } {\bf 479} 347 (1997)

\bibitem{reid68}
   Reid~R~V
   {\it Ann.\ Phys.\ } {\bf 50} 411 (1968)

\bibitem{reisenegger95}
   Reisenegger~A
    {\it Astrophys.\ J.\ } {\bf 442} 749 
    (1995)


\bibitem{richardson80}
     Richardson~M~B
     {\it Ph.D.\ thesis} (State University of Alabama, 1980)

\bibitem{rvrm82}
    Richardson~M~B, Van Horn~H~M, Ratcliff~K~F, Malone~R~C
    {\it Astrophys.\ J.\ } {\bf 255} 624 (1982)


\bibitem{ruderman91}
    Ruderman~M
    {\it Astrophys.\ J.\ } {\bf 366} 261 (1991)

\bibitem{sawyer72}
    Sawyer~R~F
    {\it Phys.\ Rev.\ Lett.\ } {\bf 29} 382 (1972)

\bibitem{scalapino72}
    Scalapino~D~J
    {\it Phys.\ Rev.\ Lett.\ } {\bf 29} 386 (1972)

\bibitem{sbs98}
    Schaab~C, Balberg~S, Schaffner-Bielich~J
    {\it Astrophys.\ J.\ Lett.\ } {\bf 504} L99 (1998a)

\bibitem{shww97}
     Schaab~Ch, Hermann~B, Weber~F, Weigel~M~K
     {\it Astrophys.\ J.\ Lett.\ } {\bf 480} L111 (1997a)

\bibitem{svsww97}
    Schaab~Ch, Voskresensky~D, Sedrakian~A~D, Weber~F, Weigel~M~K
    {Astron.\ Astrophys.\ } {\bf 321} 591 (1997b) 

\bibitem{swwg96}
    Schaab~Ch, Weber~F, Weigel~M~K, Glendenning~N~K
    {\it Nucl.\ Phys.\ A} {\bf A605} 531 (1996)

\bibitem{sww98}
   Schaab~Ch, Weber~F, Weigel~M~K
   {\it Astron.\ Astrophys.\ } {\bf 335} 596 
   (1998b)

\bibitem{sw98}
   Schaab~Ch, Weigel~M~K
   {\it Astron.\ Astrophys.\ } {\bf 336} L13 
   (1998)

\bibitem{shshs99}
Schwope~A~D, Hasinger~G, Schwarz~R, Haberl~F, Schmidt~M
{\it Astron.\ Astrophys.\  (Lett.)} (1999) in press

\bibitem{sal97}
    Sedrakian~A, Alm~T, Lombardo~U
   {\it Phys.\ Rev.\ } {\bf C55} 582 
   (1997)

\bibitem{shalybkov94}
    Shalybkov~D~A
    {\it Astron.\ Lett.\ } {\it 20} 182 (1994)

\bibitem{st85}
     Shapiro~S~L, Teukolsky~S~A
     {\it Black Holes, White Dwarfs, and Neutron Stars}
     (Wiley-Interscience, New York, 1983)


\bibitem{sheareretal98}
  Shearer~A et al.
   {\it Astron.\ Astrophys.\ } {\bf 335} L21 
   (1998)

\bibitem{spzv93}
   Shibanov~Yu~A, Pavlov~G~G, Zavlin~V~E, Ventura~J
   {\it Isolated Pulsars} (Eds Van Riper~K~A, Epstein~R~I, Ho~C)
   (Cambridge Univ.\ Press, Cambridge, 1993) p.~174

\bibitem{sy96}
    Shibanov~Yu~A, Yakovlev~D~G
    {Astron.\ Astrophys.\ } {\bf 309} 171 (1996)


\bibitem{sl89}
    Shibazaki~N, Lamb~F~K
    {\it Astrophys.\ J.\ } {\bf 346} 808 (1989)

\bibitem{sm95}
  Shibazaki~N, Mochizuki~Y
 {\it Astrophys.\ J.\ } {\bf 438} 288 
 (1995)

\bibitem{sb79}
   Soyeur~M, Brown~G~E
   {\it Nucl.\ Phys.\ } {\bf A324} 464 (1979) 

\bibitem{stabler60}
    Stabler~R  {\it Ph.D.\ thesis}  (Cornell University, 1960)


\bibitem{tabakin64}
   Tabakin~F
   {\it Ann. Phys.\ } {\bf 30} 51 (1964)

\bibitem{takatsuka72a}
   Takatsuka~T
    {\it Prog.\ Theor.\ Phys.\ } {\bf 47} 1062 (1972a)

\bibitem{takatsuka72b}
   Takatsuka~T
   {\it Prog.\ Theor.\ Phys.\ } {\bf 48} 1517 (1972b)

\bibitem{takatsuka73}
   Takatsuka~T
   {\it Prog.\ Theor.\ Phys.\ } {\bf 50} 1754 (1973)

\bibitem{takatsuka84}
   Takatsuka~T
  {\it Prog.\ Theor.\ Phys.\ } {\bf 71} 1432 (1984)

\bibitem{tt71}
   Takatsuka~T, Tamagaki~R
   {\it Prog.\ Theor.\ Phys.\ } {\bf 46} 114 (1971)

\bibitem{tt93}
   Takatsuka~T, Tamagaki~R
   {\it Progr.\ Theor.\ Phys.\ Suppl.\ } {\bf 112} 27 (1993)

\bibitem{tt95}
   Takatsuka~T, Tamagaki~R
   {\it Progr.\ Theor.\ Phys.\ } {\bf 94} 457 (1995)

\bibitem{tt97a} 
   Takatsuka~T, Tamagaki~R
   {\it Progr.\ Theor.\ Phys.\ } {\bf 97} 263 (1997a)

\bibitem{tt97b} 
     Takatsuka~T, Tamagaki~R
     {\it Progr.\ Theor.\ Phys.\ } {\bf 97} 345 
     (1997b)

\bibitem{tt97c} 
   Takatsuka~T, Tamagaki~R
   {\it Progr.\ Theor.\ Phys.\ } {\bf 98} 393 
   (1997c)

\bibitem{tamagaki68}
    Tamagaki~R
    {\it Prog.\ Theor.\ Phys.\ } {\bf 39} 91 (1968)

\bibitem{tamagaki70}
   Tamagaki~R
   {\it Prog.\ Theor.\ Phys.\ } {\bf 44} 905 (1970)


\bibitem{td96}
  Thompson~C, Duncan~R~C
  {\it Astrophys.\ J.\ } {\bf 473} 322 (1996)

\bibitem{thorne77}
  Thorne K~S
  {\it Astrophys.\ J.\ } {\bf 212} 825 (1977)




\bibitem{ts77}
   Toor~A, Seward~F~D
  {\it Astrophys.\ J.\ } {\bf 216} 560 (1977)

\bibitem{tsuruta74}
   Tsuruta~S
   {\it Physics of Dense Matter}
   (Ed Hansen~C~J) (Reidel, Dordrecht, 1974) p.~209

\bibitem{tsuruta75}
   Tsuruta~S
   {\it Astrophys.\ Space Sci.\ } {\bf 34} 199 (1975)

\bibitem{tsuruta78}
   Tsuruta~S
   {\it Physics and Astrophysics of Neutron Stars and Black Holes}
   (Societa Italiana di Fisica, North Holland, Amsterdam, 1978)
    p.~635

\bibitem{tsuruta79}
   Tsuruta~S
   {\it Phys.\ Rep.\ } {\bf 56} 237 (1979)

\bibitem{tsuruta80}
   Tsuruta~S
   {\it  X-ray astronomy} 
   (Eds Giacconi~R, Setti~G)
   (Reidel, Dordrecht, 1980)  p.~73

\bibitem{tsuruta81}
   Tsuruta~S
   {\it Pulsars} 
   (Eds Sieber~W, Wielebinski~R)  (Reidel, Dordrecht, 1981)
   p.~331


\bibitem{tsuruta86}
      Tsuruta~S
      {\it Comments on Astrophysics} {\bf 11} 151 (1986)

\bibitem{tsuruta98}
      Tsuruta~S
      {\it Phys.\ Rep.\ } {\bf 292} 1 
      (1998)

\bibitem{tc66}
       Tsuruta~S, Cameron~A~G~W
       {\it Canad.\ J.\ Phys.\ } {\bf 44} 1863 (1966)

\bibitem{tclr72}
        Tsuruta~S, Canuto~V, Lodenquai~J, Ruderman~M
        {\it Astrophys.\ J.\ } {\bf 176} 739 (1972)


\bibitem{untmt94}
    Umeda~H, Nomoto~K, Tsuruta~S, Muto~T, Tatsumi~T
    {\it Astrophys.\ J.\ } {\bf 431} 309 (1994a)

\bibitem{usnt93}
        Umeda~H, Shibazaki~N, Nomoto~K, Tsuruta~S
        {\it Astrophys.\ J.\ } {\bf 408} 186 (1993)

\bibitem{utn94}
        Umeda~H, Tsuruta~S, Nomoto~K
        {\it Astrophys.\ J.\ } {\bf 433} 256 (1994b)


\bibitem{uk98}
   Urpin~V~A, Konenkov~D~Yu
  {\it  Neutron Stars and Pulsars} (Eds Shibazaki~N et al)
  (Universal Academy Press, Tokyo, 1998)~p.~171.

\bibitem{um92}
   Urpin~V~A, Muslimov~A~G
   {\it MNRAS} {\bf 256} 261 (1992)

\bibitem{uv93}
   Urpin~V~A, Van Riper~K~A
   {\it Astrophys.\ J.\ } {\bf 411} L87 (1993)

\bibitem{us95}
   Urpin~V~A, Shalybkov~D~A
   {\it Astron.\ Rep.\ } {\bf 39} 332 (1995)


\bibitem{vanriper88}
    Van Riper~K~A
    {\it Astrophys.\ J.\ } {\bf 329} 339 (1988)

\bibitem{vanriper91a}
    Van Riper~K~A
    {\it Astrophys.\ J.\ Suppl.\ } {\bf 75} 449 (1991a)

\bibitem{vanriper91b}
    Van Riper~K~A
    {\it Atrophys.\ J.\ } {\bf 372} 251 (1991b)

\bibitem{vem91}
    Van Riper~K~A, Epstein~R~I, Miller~G~S
    {\it Astrophys.\ J.\ Lett.\ } {\bf 381} L47 (1991)

\bibitem{vl81}
    Van Riper~K~A, Lamb~F
    {\it Astrophys.\ J.\ Lett.\ } {\bf 244} L13 (1981)

\bibitem{vl93}
    Van Riper~K~A, Lattimer~J~M
    {\it Isolated Pulsars} (Eds Van Riper~K~A, Epstein~R, Ho~C)
    (Cambridge Univ.\ Press, Cambridge, 1993) p.~122

\bibitem{vle95}
    Van Riper~K~A, Link~B, Epstein~R~I
    {\it Astrophys.\ J.\ } {\bf 448} 294 (1995)

\bibitem{vg97}
    Vasisht~G, Gotthelf~E~V.
    {\it Astrophys.\ J.\ Lett.\ } {\bf 486} L129 (1997)



\bibitem{vs84}
     Voskresensky~D~N, Senatorov~A~V
     {\it Pisma Zh.\ Eksper.\ Teor.\ Fiz.\ } {\bf 40} 395 (1984)


\bibitem{vs86}
     Voskresensky~D~N, Senatorov~A~V
     {\it Pisma Zh.\ Eksper.\ Teor.\ Fiz.\ } {\bf 90} 1505 (1986)

\bibitem{vs87}
     Voskresensky~D~N, Senatorov~A~V
     {\it  Sov.\ J.\ Nucl.\ Phys.\ }
     {\bf 45} 411 (1987)

\bibitem{wwn96}
  Walter~F~M, Wolk~S~J, Neuh\"{a}user~R
  {\it Nature} {\bf 379} 233 (1996)

\bibitem{wm97}
    Walter~F~M, Matthews~L~D
    {\it Nature} {\bf 389} 358 (1997)

\bibitem{wap91}
   Wambach~J, Ainsworth~T~L, Pines~T~L
   {\it Neutron Stars: Theory and Observation} (Eds Ventura~J, Pines~D)
   (Kluwer, Dordrecht, 1991) p.~37

\bibitem{wap93}
   Wambach~J, Ainsworth~T~L, Pines~D
   {\it Nucl.\ Phys.\ } {\bf A555} 128 (1993)


\bibitem{wolf66}
   Wolf~R~A
   {\it Astrophys.\ J.\ } {\bf 145} 834 (1966)

\bibitem{wkkn75}
  Wolff~R~S,  Kestenbaum H~L, Ku~W, Novick R
  {\it Astrophys.\ J.\ } {\bf 202} L15 (1975)

\bibitem{yakovlev93}
   Yakovlev~D~G
  {\it Strongly Coupled Plasma Physics} (Eds Van Horn~H~M, Ichimaru~S)
  (Univ.\ of Rochester Press, Rochester, 1993) p.~157



\bibitem{ykl98}
   Yakovlev~D~G, Kaminker~A~D, Levenfish~K~P
   {\it Neutron Stars and Pulsars} (Ed Shibazaki~N et al.)
   (Universal Academy Press, Tokyo, 1998) p.~195 

\bibitem{ykl99}
     Yakovlev~D~G, Kaminker~A~D, Levenfish~K~P
    {\it Astron.\ Astrophys.} {\bf 343} 650 (1999)

\bibitem{yl95}
    Yakovlev~D~G, Levenfish~K~P
    {\it Astron.\ Astrophys.\ } {\bf 297} 717 (1995)

\bibitem{yu81}
    Yakovlev~D~G, Urpin~V~A
    {\it Pisma Astron.\ Zh.\ } {\bf 7} 157 (1981)

\bibitem{ys91}
    Yakovlev~D~G, Shalybkov~D~A
    {\it Astrophys.\ Space Sci.\ } {\bf 176} 171 (1991);
    {\bf 176} 191 (1991)


\bibitem{ztt96}
  Zane~S, Turolla~R, Treves~A
  {\it Astrophys.\ J.\ } {\bf 471} 248 (1996)


\bibitem{zps96}
  Zavlin~V~E, Pavlov~G~G, Shibanov~Yu~A
   {\it Astron.\ Astrophys.\ } {\bf 315} 141 (1996)

\bibitem{zpt98}
  Zavlin~V~E, Pavlov~G~G, Tr\"{u}mper~J
  {\it Astron.\ Astrophys.\ } {\bf 331} 821 (1998)

\bibitem{zpt99}
  Zavlin~V~E, Pavlov~G~G, Tr\"{u}mper~J
 (1999a) (submitted to {\it Astrophys.\ J.})

\bibitem{zsp95}
  Zavlin~V~E, Shibanov Yu~A, Pavlov G~G
  {\it Astron.\ Lett,\ } {\bf 21} 149 (1995)

\bibitem{ztp99}
 Zavlin~V~E, Tr\"{u}mper~J, Pavlov~G~G
 (1999b) (to be submitted to {\it Astrophys.\ J.})


\end{thebibliography}
\end{document}